\documentclass[conference]{IEEEtran}
\IEEEoverridecommandlockouts
\usepackage{enumitem}

\usepackage[T1]{fontenc}
\usepackage{cite}
\usepackage{amsmath,amssymb,amsfonts}
\usepackage[printonlyused,nohyperlinks]{acronym}
\usepackage{algorithmic}
\usepackage{graphicx}
\usepackage{tabularx}
\usepackage{textcomp}
\usepackage{xcolor}
\def\BibTeX{{\rm B\kern-.05em{\sc i\kern-.025em b}\kern-.08em
    T\kern-.1667em\lower.7ex\hbox{E}\kern-.125emX}}
\usepackage{float}
\usepackage{amsfonts}
\usepackage{subcaption}
\usepackage{caption}
\usepackage{booktabs}
{}
{}
{}
\usepackage{tikz}
\usetikzlibrary{shapes, arrows}
\usepackage{pdfpages}
\usepackage{pgfplots}
\usetikzlibrary{arrows,automata}
\usepackage{pifont}
\usepackage[utf8]{inputenc}
\usepackage{multirow}
\usepackage{blindtext}
\usepackage{lipsum}% http://ctan.org/pkg/lipsum
\usepackage{pgfplots}
\usepackage{mathtools}

\usepackage{balance}
\usepackage{comment}
\usepackage{amsmath}
\usepackage{graphics}
\usepackage{pgfplots,tikz}
\usepackage{xcolor}
\usepackage{pagecolor}% http://ctan.org/pkg/{pagecolor,lipsum}
\usetikzlibrary{decorations, decorations.text,backgrounds}
\colorlet{NextBlue}{red!25!green!50!blue!75}
\usepackage[acronym]{glossaries}
\makeglossaries
\usepackage{verbatim}
\usetikzlibrary{shapes}
\usepackage{graphicx}
\usepackage{caption}
\usepackage{booktabs}
\usepackage{tabularx}
\usepackage{array}
\usepackage[hyphens]{url}
\usepackage{hyperref}
\usepackage{flushend}
\tikzstyle{line} = [draw,-latex']
\usepackage{makecell}
\usepackage{longtable}
\newcommand{\cmark}{\ding{51}} % Checkmark
\newcommand{\xmark}{\ding{55}} % Crossmark
\begin{document}

\title{Bridging Earth and Space: A Survey on HAPS for Non-Terrestrial Networks \\

\author{
    \IEEEauthorblockN{G. Svistunov\IEEEauthorrefmark{1}, 
    A. Akhtarshenas\IEEEauthorrefmark{1}, 
    D. L\'opez-P\'erez\IEEEauthorrefmark{1},
    M. Giordani\IEEEauthorrefmark{2},
    G. Geraci\IEEEauthorrefmark{3},
    H. Yanikomeroglu\IEEEauthorrefmark{4}}
    \IEEEauthorblockA{\IEEEauthorrefmark{1}Universitat Politècnica de València, Spain,
    \IEEEauthorrefmark{2}University of Padova, Italy}
    \IEEEauthorblockA{\IEEEauthorrefmark{3}Universitat Pompeu Fabra, Spain, 
    \IEEEauthorrefmark{4}Carleton University, Canada}
    \\[-5.0ex]
}

\thanks{This research was supported by the Generalitat Valenciana, Spain, through the  CIDEGENT PlaGenT,  Grant CIDEXG/2022/17, Project iTENTE, and the actions CNS2023-144333, financed by MCIN/AEI/10.13039/501100011033 and the European Union “NextGenerationEU”/PRTR. G. Geraci was supported by grants PID2021-123999OB-I00, CNS2023-145384, PID2024-156488OB-I00, and CEX2021-001195-M.}
}

\maketitle
\begin{abstract}
%\ac{HAPS} now considered as one of the key enablers in the evolution of \ac{6G} wireless networks, offering a unique capability to bridge the gap between terrestrial and non-terrestrial infrastructures. Operating in the stratosphere, \ac{HAPS} may provide wide-area coverage and energy-efficient low-latency communication, while flexible deployment makes them well-suited for a variety of critical applications. This survey presents a comprehensive overview of \ac{HAPS} technologies, use cases, and integration strategies within the broader context of \ac{6G} networks. We examine the roles of \ac{HAPS} in enhancing connectivity for underserved and remote areas, supporting dynamic backhauling, enabling massive \ac{IoT} deployments, and delivering \ac{URLLC} for emerging services such as autonomous systems and immersive applications. Key enabling technologies such as channel modeling, mobility management, \ac{AI}-driven resource allocation, interference control, and energy-efficient communication are explored in depth. In addition, the survey reviews state-of-the-art architectures for \ac{TN}-\ac{NTN} integration, highlights recent field deployments, and identifies open challenges and future research directions. By filling critical gaps in the existing literature, this work positions \ac{HAPS} as a foundational pillar of the next-generation, resilient, and sustainable global connectivity ecosystem.
\Ac{HAPS} are emerging as key enablers in the evolution of \ac{6G} wireless networks, 
bridging terrestrial and non-terrestrial infrastructures. 
Operating in the stratosphere, 
\ac{HAPS} can provide wide-area coverage, low-latency, energy-efficient broadband communications with flexible deployment options for diverse applications. 
This survey delivers a comprehensive overview of \ac{HAPS} use cases, technologies, and integration strategies within the \ac{6G} ecosystem. 
The roles of \ac{HAPS} in extending connectivity to underserved regions, enabling massive \ac{IoT}, supporting dynamic backhauling, and delivering reliable low-latency communications for autonomous and immersive services are discussed. 
The paper reviews state-of-the-art architectures for terrestrial and non-terrestrial network integration,
highlighting recent field trials.
Furthermore, key enabling technologies—such as channel modeling, \ac{AI}-driven resource allocation, interference control,  mobility management, and energy-efficient communications—are examined. 
The paper also outlines open research challenges. 
By addressing existing gaps in the literature, 
this survey positions \ac{HAPS} as a foundational component of globally integrated, resilient, and sustainable \ac{6G} networks.

\end{abstract}

\acresetall
\vspace{1em} 
\noindent\textit{This work has been submitted to the IEEE for possible publication. Copyright may be transferred without notice, after which this version may no longer be accessible.}
\section{Introduction}\label{Intro_duction}
The integration of~\acp{TN} and \acp{NTN} is widely regarded as a key enabler for achieving the ambitious goals of \ac{6G} networks --- 
to ``connect the unconnected'' and ``hyper-connect existing infrastructures.'' 
While traditional \acp{TN} are foundational, they cannot solely meet the growing demands for coverage, capacity, and low latency required by emerging applications~\cite{azari2022evolution}. To overcome these limitations, \ac{NTN} technologies have emerged as transformative solutions. \acp{NTN} leverage space-borne (satellites), airborne (\ac{HAPS}), and other non-ground-based platforms (\acp{UAV}) to extend connectivity beyond terrestrial infrastructure, enabling coverage in remote and underserved regions, as well as in mobile platforms, e.g., oceans, rural areas, or in-flight scenarios, respectively. A key distinction between conventional \acp{TN} and \acp{NTN} is the latter’s ability to decouple service coverage from the physical location of backbone infrastructure through backhauling via satellite or \ac{HAPS} links, achieved using long-distance gateway connections or multi-hop inter-platform links.

\ac{HAPS} operate in the stratosphere, complementing both \acp{TN} and other \ac{NTN} nodes such as \ac{LEO} satellites and \acp{UAV}. They combine the advantages of these systems while mitigating their limitations. \ac{HAPS} offer wide-area coverage with higher capacity and lower latency than \ac{LEO} satellites, and lower energy consumption with longer endurance than \acp{UAV}. Their quasi-stationary positioning reduces dependence on ground infrastructure and ensures reliable \ac{LoS} or near-\ac{LoS} connectivity. This capability allows \ac{HAPS} to enhance network performance in diverse environments, including urban, rural, remote, and disaster-affected areas~\cite{kurt2021vision}. Beyond bridging connectivity gaps, \ac{HAPS} enable advanced services such as massive \ac{IoT}, real-time autonomous systems, edge computing, and immersive technologies like augmented and virtual reality~\cite{mozaffari2019tutorial, kurt2021vision}. These features establish \ac{HAPS} as a key element in shaping next-generation global~connectivity.

The primary objective of this survey is to provide a comprehensive overview of the current state-of-the-art in \ac{HAPS} networks. It explores applications in wireless communication, and identifies key technologies and research directions shaping the future of this field. We begin by comparing \ac{HAPS} capabilities with \acp{TN} and other \ac{NTN} platforms. Further we highlight the diverse roles of \ac{HAPS} in several promising applications, categorized as follows:
\begin{itemize}
    \item Connectivity enhancement: \ac{HAPS} extend reliable communication to remote and underserved regions,   address connectivity challenges in natural disaster scenarios, and manage unexpected traffic surges efficiently.
    \item {Network infrastructure and management}: Acting as key enablers for robust backhaul, seamless handover management, and dynamic resource allocation, \ac{HAPS} support the growing demands of integrated \ac{TN} and \acp{NTN}.
    \item {\ac{IoT} and emerging applications}: \ac{HAPS} facilitate large-scale \ac{IoT} deployments, enable low-latency data processing through mobile edge computing, and support immersive applications such as augmented and virtual reality.
    \item {Advanced control and special applications}: Supporting \ac{URLLC} and remote control systems, \ac{HAPS} empower autonomous vehicle operations, UAV management, and other high-demand real-time applications.
    \item {Energy and sustainability}: \ac{HAPS} contribute to sustainable connectivity by reducing reliance on power-intensive terrestrial systems and optimizing energy use across integrated networks.
\end{itemize}
We conclude the application discussion with a brief overview of the \ac{HAPS} communication market.
The survey then examines \ac{TN}-\ac{NTN} integration, with particular emphasis on the role of \ac{HAPS} as a bridge between terrestrial and non-terrestrial systems. It outlines the fundamentals of the \ac{NTN} concept, describes various platform and terminal types within the integrated \ac{TN}-\ac{NTN} architecture, and reviews the main components of a \ac{HAPS} energy modeling framework. It further explores emerging frequency bands, such as \ac{mmWave} and \ac{cmWave}, highlighting their role in enhancing connectivity. Additionally, it discusses the \ac{NTN} framework in the context of 3GPP-specified architectures and analyzes past and ongoing \ac{HAPS} projects to extract key lessons learned.
Subsequently, the survey reviews key \ac{HAPS} technical enablers, including:
\begin{itemize}
    \item {Channel modeling and estimation}: 
Accurately characterize and estimate the wireless channel between \ac{HAPS} and \acp{GUE}, accounting for the unique propagation conditions of the stratosphere, such as large-scale path loss, atmospheric attenuation, and dynamic environments.
    \item {User equipment association and resource management}: 
    Dynamic \ac{UE} association techniques and efficient allocation of network resources.
    \item {Mobility management and multi-link connectivity}: 
    Techniques to ensure seamless handovers and stable connectivity in multi-link scenarios.
    \item {Massive \ac{MIMO} and \ac{ML} applications}: 
    Enhance connectivity and data processing performance through advanced antenna technologies and learning algorithms.
    \item {Interference control}: 
    Key methods to mitigate interference and optimize communication channels in dense networks.    
    \item {Energy efficiency}: 
    Optimize power usage to extend operation times and reduce the environmental impact.
    \item {Cybersecurity and Adversarial Threats}:
    Addresses jamming, spoofing and other cybersecurity issues in various communication scenarios.
\end{itemize}
Additionally, the survey addresses the critical challenges associated with the deployment and operation of \ac{HAPS}, highlighting areas that require further research and innovation to unlock their full potential. By reviewing these topics, this survey provides an in-depth understanding of technological advancements, design considerations, and open challenges shaping \ac{HAPS}-based solutions for 6G. It serves as a resource for researchers and engineers exploring the potential of \ac{TN}-\ac{NTN} integration via \ac{HAPS} communications.

The literature for this survey was compiled using a structured review technique influenced by PRISMA principles to ensure transparency and reproducibility. A comprehensive search was carried out across major academic databases, including IEEE Xplore, Scopus, and Google Scholar, encompassing papers from 2015 to 2025.

The search used keyword combinations such as "HAPS", "high-altitude platform station", "non-terrestrial networks", "6G", "TN-NTN integration", and related phrases in wireless communications and radiophysics. The search queries were refined using Boolean operators (AND and OR).

The initial search yielded approximately 450 publications. After eliminating duplicate records, about 400 papers remained for title and abstract screening. During this stage, a subset of papers was rejected because weak relevance to HAPS-based NTN communication.

The remaining publications were evaluated using full-text review. The inclusion criteria were: (i) relevance to HAPS or NTN-based communication systems, (ii) methodological rigor (analytical, simulation, or experimental validation), and (iii) contribution to crucial areas such as architectures, technologies, or integration techniques. At this stage, we excluded publications lacking substantive results, technical depth, or direct relevance, as well as non-peer-reviewed materials (except for high-quality preprints and whitepapers from trustworthy organizations strongly related to the topic). Additionally, several studies published prior to 2015 were included due to their fundamental contributions.

Following full-text evaluation, a total of 254 relevant references were selected for this survey, including peer-reviewed publications, technical reports, whitepapers, and press-releases.
This systematic methodology guarantees that the review process is transparent, reproducible, and focused on high-quality contributions to the HAPS and NTN research domains.

\begin{figure*}[!htb] 
     \centering
         \includegraphics[width=0.9\linewidth]{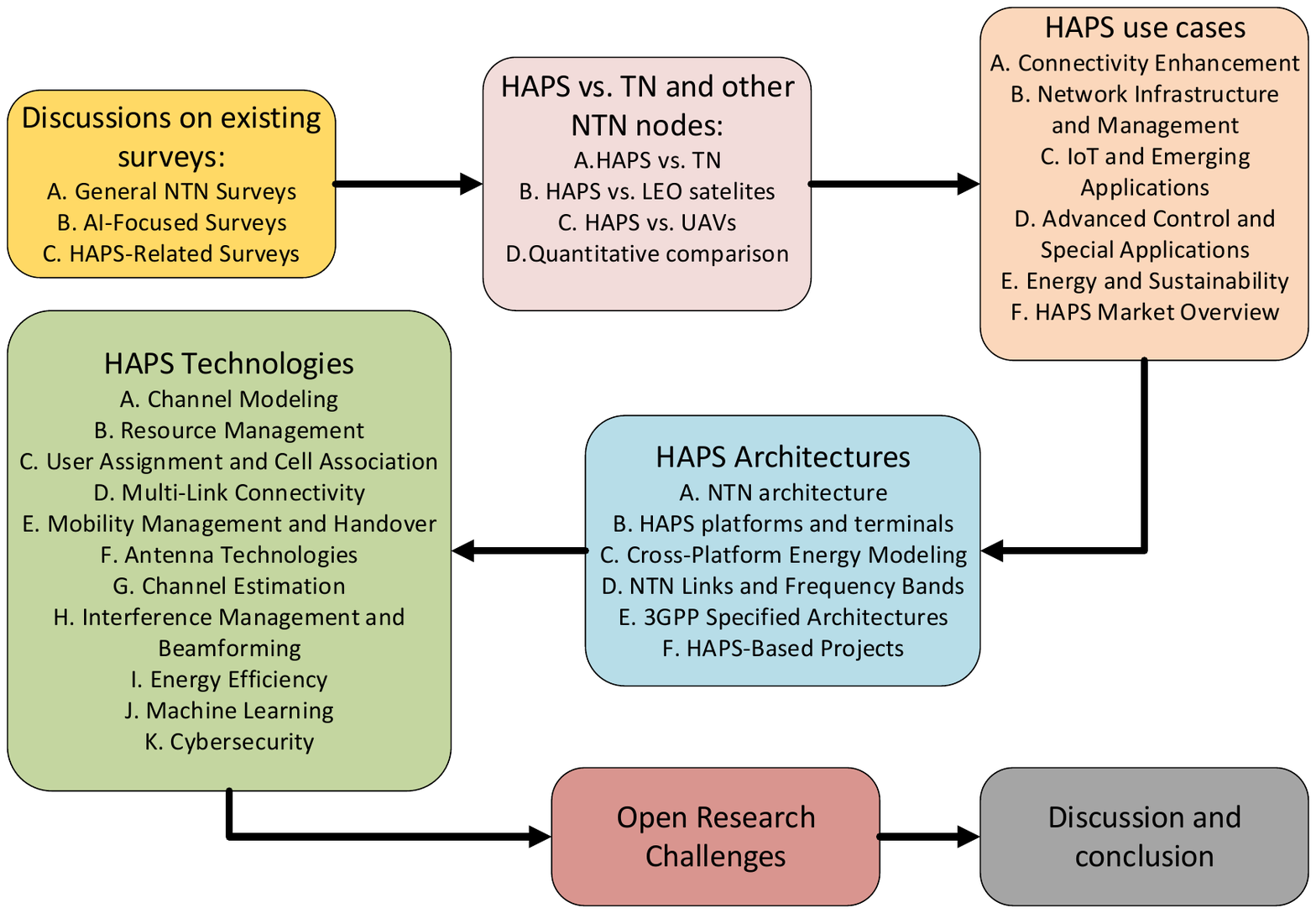}
     \caption{Survey structure}
     \label{surv_struct}
\end{figure*}

The rest of this survey is organized as follows (see Fig. \ref{surv_struct}):
Section \ref{sec: relevant_surveys} reviews the most relevant prior surveys including general studies in the area of non-terrestrial networks, studies focused on \ac{AI} and \ac{ML} applications and those specifically considering HAPS.
Section \ref{sec: haps} compares \ac{HAPS} with other \ac{NTN} nodes.
Section \ref{sec: Tech_use_case} highlights the most promising use cases for \ac{HAPS}.
Section \ref{sec:Archi} explores scenarios and architectures for \ac{HAPS} deployment.
Section \ref{sec: Techno} examines the key technical enablers for \ac{HAPS}.
Section \ref{sec:challenges} discusses the current challenges related to \ac{HAPS} communications.
Section \ref{Disc-conclu} concludes the survey with findings and insights for future research directions.

\section{Review of Previous NTN-Based Surveys}\label{sec: relevant_surveys}
Before diving into this survey, we first highlight how it differs from previous surveys on the \ac{NTN} topic.
\subsection{General NTN Surveys} 
Several surveys analyzed \acp{NTN}, but most provided no in-depth examination of \ac{HAPS}.
Baltaci~\textit{et al.}~\cite{baltaci2021survey} reviewed aerial connectivity for airplanes, vertical take-off and landing vehicles, and \acp{UAV}, providing a detailed analysis of use cases, requirements, and network architectures. It also examined aerial connectivity providers, including satellites and \ac{HAPS}, as well as communication challenges. However, its coverage of \ac{HAPS} is limited, and lacks examination of \ac{TN}-\ac{NTN} integration. Geraci~\textit{et al.}~\cite{geraci2022will} explored the role of \acp{UAV} in future communication systems, covering use cases, network architectures, and research challenges. It reviewed potential optimization opportunities in sub-6 GHz and \ac{mmWave} \acp{TN} to provide service to \acp{UAV}. Additionally, it included a performance analysis of \ac{UAV}-to-\ac{UAV} communications. 

Azari~\textit{et al.}~\cite{azari2022evolution} studied \ac{TN}-\ac{NTN} integration in \ac{5G} and \ac{6G} networks, focusing on enabling technologies such as \ac{mmWave}, \ac{IoT}/\ac{MEC} integration, and \ac{ML}. It also highlighted outcomes from field trials. However, it mostly emphasized \ac{LEO} satellites and \acp{UAV}, rather than \ac{HAPS}. \ac{TN}-\ac{NTN} integration was also examined by Geraci~\textit{et al.}~\cite{geraci2022integrating}, mostly focusing on \ac{LEO} networks, spectrum challenges, \ac{MIMO} modeling, and simulations to evaluate the performance of \acp{UAV} when served from either a massive \ac{MIMO}-based \ac{TN} or a \ac{LEO}-satellite-based operator. The study also highlighted relevant \ac{3GPP} standardization efforts and solutions, such as \ac{3D} radio access, mobility management, and network orchestration. Similarly, Rinaldi~\textit{et al.}~\cite{rinaldi2020non} explored the evolution of \ac{NTN} from second-generation to \ac{4G}, and discussed \ac{TN}-\ac{NTN} integration, mobility management, and \ac{3GPP} support for \ac{5G} \ac{NR}. The survey also highlighted the role of \ac{NTN} toward \ac{6G} for achieving global coverage and increased capacity.

Jamshed~\textit{et al.}~\cite{jamshed2025tutorial} presented a comprehensive overview of \ac{NTN} as a pivotal enabler for \ac{6G} networks, emphasizing their role in achieving seamless, intelligent, and sustainable global connectivity by integrating satellites, aerial platforms (\acp{UAV}, \ac{HAPS}), and \acp{TN} to address the digital divide and support applications like \ac{IoT}, navigation, disaster recovery, and interplanetary communication. The paper delved into enabling technologies, real-world applications illustrated through \ac{UAV}/\ac{HAPS}-assisted fronthaul/backhaul, security considerations, and future research directions, positioning \ac{NTN} as essential part of resilient \ac{6G} ecosystems.
Despite their comprehensive coverage of \acp{NTN}, these surveys only briefly mentioned \ac{HAPS} as part of a broader \ac{NTN} framework, and did not explore them in depth.

\subsection{AI-Focused Surveys}
\ac{AI} played an essential role in \acp{NTN} and \ac{TN}-\ac{NTN} integration, enabling intelligent resource management, optimization, and automation across various network layers. Several surveys analyzed the application of \ac{AI} in \acp{NTN}, focusing on key technologies, \ac{ML}-based algorithms, and network performance improvements. 
\ac{AI}-based optimization strategies for \ac{TN}-\ac{NTN} integration were analyzed by Wang~\textit{et al.}~\cite{wang2019convergence}, who discussed the potential of \ac{TN} and satellite network integration to enhance global connectivity. This study provided an in-depth analysis of convergence drivers, network architectures, and key application areas. 
Kurunathan~\textit{et al.}~\cite{kurunathan2023machine} explored the use of \ac{ML} for \ac{UAV}-related applications, including image processing, ground sensors localization, real-time \ac{UAV} control, scheduling and trajectory planning. 
Zhou, Sheng, Li and Han~\cite{zhou2023aerospace} reviewed \ac{6G} aerospace-integrated networks, analyzing architecture, satellite configurations, protocols, and key components, while also identifying future research challenges. It examined advanced technologies, with a focus on \ac{AI}-driven resource management, and \ac{NTN} research gaps. 
Iqbal~\textit{et al.}~\cite{iqbal2023empowering} also explored \ac{ML} and \ac{AI} techniques for addressing \ac{NTN}-related challenges in \ac{6G}, highlighting privacy risks and optimization techniques such as deep learning, and \ac{DRL}.
Fontanesi~\textit{et al.}~\cite{fontanesi2023artificial} provided a comprehensive overview of \ac{AI} and \ac{ML} applications in \ac{SatCom} and \ac{NTN}, focusing on \ac{GEO}, \ac{MEO}, and \ac{LEO} systems, alongside with an extensive classification of the most important use cases, its related challenges and the conventional tools used to address them. Onboard vs. on-ground \ac{AI} architectures were compared, emphasizing training paradigms (offline/online). Hardware solutions for \ac{ML} deployment were reviewed, including radiation-tolerant chipsets. Long-term challenges and insights for future research concluded the study.
Mahboob and Liu~\cite{mahboob2024revolutionizing} categorized \ac{AI} applications across current research directions covering \ac{PHY} layer, data link layer and upper layer aspects such as computational offloading, network routing and traffic prediction. Existing works, research efforts, including \ac{ML} and software defined radio testbeds, and practical challenges were reviewed, concluding with insights for future \ac{AI}-driven \ac{NTN} deployments.

Despite such a broad coverage, these surveys primarily focused on satellites and \acp{UAV}, with limited discussion of \ac{HAPS}.

\subsection{HAPS-Related Surveys}
Mozaffari~\textit{et al.}~\cite{mozaffari2019tutorial} provided a comprehensive tutorial on both \ac{HAPS} and \acp{LAPS} in wireless communications, highlighting their role as aerial \acp{BS} and cellular-connected \acp{UE}. It examined key challenges such as \ac{3D} deployment, channel modeling, and energy efficiency, while exploring potential research directions. Additionally, it reviewed analytical frameworks, including optimization theory and \ac{ML}, for addressing \ac{UAV}-related issues. However, it did not consider \ac{TN}-\ac{NTN} integration, comparing instead \ac{UAV}- and ground-based networks. 

Kodheli~\textit{et al.}~\cite{kodheli2020satellite} reviewed advancements in \ac{LEO} satellite and \ac{HAPS} communications, capturing the state of the art in \ac{SatCom}, and highlighting key developments such as new constellation architectures, on-board processing, and \ac{NTN} integration. It systematically analyzed five critical areas: system design, air interface, medium access control, networking, as well as testbeds and prototyping. While the study covered \ac{HAPS}, its primary focus remained on satellites, offering only a high-level assessment of \ac{HAPS} without detailing their unique technological challenges. Giordani and Zorzi in~\cite{giordani2020non} considered \acp{NTN} as a crucial enabler for \ac{6G}, enhancing connectivity through technologies such as \ac{mmWave} and flexible beam steering. It complemented previous works by providing simulations that compared different integration scenarios, including \ac{TN}-\ac{LEO},  as well as \ac{TN}, \ac{LEO}, and \ac{HAPS} integration. However, \ac{HAPS} was considered only as a relay, without evaluating their potential as independent network nodes or fully-integrated \ac{NTN} components.

Karabulut Kurt~\textit{et al.}~\cite{kurt2021vision} took a more focused approach on \ac{HAPS}, exploring their role in \ac{NTN} communications, particularly in the context of mega-constellation systems for next-generation wireless networks. It examined key use cases, and focused on the potential of \ac{HAPS} as \acp{SMBS}, while also addressing regulatory considerations, alongside critical technologies such as channel modeling, resource management, handoff mechanisms, and swarm control. The survey also discussed \ac{AI} applications, particularly in network topology management, resource allocation, and mobility optimization. Additionally, it reviewed emerging technologies such as reconfigurable intelligent surfaces for cost-effective deployment and faster-than-Nyquist signaling for spectral efficiency. While this survey provided a strong technological perspective, it lacked insights into \ac{TN}-\ac{NTN} interoperability and real-world deployment challenges. The role of \ac{HAPS} in \ac{NTN} was further explored by Arum, Grace, and Mitchell~\cite{arum2020review}, who examined \ac{HAPS} potential for coverage extension and capacity enhancement in \ac{TN}-\ac{HAPS} coexistence scenarios. The study focused on leveraging \ac{HAPS} as an alternative to terrestrial systems, particularly for wireless communication in rural and remote areas where broadband access remained limited. It discussed techniques such as intelligent radio resource management, interference mitigation, and spatial optimization using array antennas, radio environment map, and device-to-device communication to maximize coverage while ensuring coexistence with \ac{TN}. Additionally, the study highlighted past \ac{HAPS} projects but did not explore new applications beyond coverage expansion, such as \ac{AI}-driven optimization, network infrastructure and management, support for \ac{IoT}, or \ac{TN}-\ac{NTN} integrated architectures. Abbasi~\textit{et al.}~\cite{abbasi2024haps} outlined potential \ac{HAPS} use cases and identified particular challenges which may arise due to integration of \ac{HAPS} with conventional \acp{TN} and because of platform's physical characteristics. The authors also investigated the usefulness of HAPS-enabled \acp{vHetNet} through the estimating of cell-edge users' capacity enhancement and the benefits of the cell-free communication system of aerial \acp{BS} with \ac{HAPS} backhauling.

Among the most relevant surveys to this work, i.e., \cite{mozaffari2019tutorial, kodheli2020satellite, giordani2020non, kurt2021vision, arum2020review, abbasi2024haps}, none provided a comprehensive, up-to-date analysis of \ac{HAPS}, covering both technological advancements and \ac{TN}-\ac{NTN} integration. The surveys in~\cite{mozaffari2019tutorial, kodheli2020satellite} were still generalist \ac{NTN} studies, offering only limited exploration of \ac{HAPS}, focusing on use cases, technological capabilities, and network architectures, without detailing dedicated \ac{HAPS} advancements. Although~\cite{giordani2020non} considered \ac{HAPS} in the \ac{NTN}-enabled \ac{6G} ecosystem, it primarily focuses on relay-based NTN integration, without assessing \ac{HAPS} as independent network nodes or their role in multi-layered communication systems. The survey~\cite{kurt2021vision} provided an extensive review of \ac{HAPS} mega-constellations, regulatory aspects, and key enabling technologies, but offered little insight into \ac{TN}-\ac{NTN} interoperability, the opportunities and challenges it presents, and real-world deployment considerations. The work in~\cite{arum2020review} examined \ac{HAPS} for coverage extension and terrestrial coexistence, but did not explore their broader integration into \ac{NTN} architectures, advanced use cases beyond connectivity, or emerging technologies, especially those related to \ac{AI}. While the paper~\cite{abbasi2024haps} covered particular use cases and challenges it is more concise and less exhaustive in referencing.
To fill these gaps, this survey provided a holistic and forward-looking review of \ac{HAPS}, including \ac{TN}-\ac{NTN} integration, enabling technologies, real-world deployment challenges, and future research directions. Table~\ref{tab:haps_comparison_surveys} provided a comparative summary of this survey against the previous relevant literature. %}

\begin{table*}[htb]
    %\color{blue}
    \caption{Comparison of our survey against other HAPS-related surveys.}
    \centering
    \renewcommand{\arraystretch}{1.2}
    
    \begin{tabular}{|p{.7cm}| p{8.4cm}| p{1.2cm}| p{1.2cm}| p{1.3cm} |p{1.2cm} |p{1.2cm}|}
        \hline
        \textbf{Survey} & \textbf{Scope} & \textbf{HAPS\newline focus} & \textbf{HAPS\newline  use cases} & \textbf{\ac{HAPS} \newline technologies} & \textbf{\ac{TN}-\ac{NTN} \newline integration} & \textbf{Challenges} \\
        \hline
        
        \cite{mozaffari2019tutorial} & Examines \ac{HAPS} and \ac{LAPS} in wireless communications, identifying challenges and research directions but focusing on \ac{UAV}-ground comparisons rather than \ac{TN}-\ac{NTN} integration.  
        & Secondary & \xmark & \xmark & \xmark & \cmark \\
        
        \hline
        \cite{kodheli2020satellite} & Reviews advancements in satellites and HAPS communication, highlighting key developments such as new constellations, on-board processing, and \ac{NTN} integration, but focusing on satellite communications.  
        & Secondary & \xmark & \xmark & \xmark & \cmark \\

        \hline
        \cite{giordani2020non} & Considers \ac{NTN} for 6G, with simulations of \ac{TN}-\ac{LEO}-\ac{HAPS} integration, but focusing on \ac{HAPS} as a relay rather than independent~nodes.  
        & Secondary & \xmark & \xmark & \cmark & \cmark \\
        
        \hline
        \cite{kurt2021vision} & Focuses on HAPS mega-constellations, regulatory aspects, and key enabling technologies, including \ac{AI} and swarm control, but lacks \ac{TN}-\ac{NTN} integration insights.  
        & Primary & \cmark & \cmark \newline (partially) & \xmark & \cmark \\
        
        \hline
        \cite{arum2020review} & Examines \ac{HAPS} for coverage extension and coexistence with terrestrial networks, but does not explore broader use cases, advanced technologies, or \ac{TN}-\ac{NTN} integration.  
        & Primary  & \cmark \newline (partially) & \cmark & \xmark & \cmark \\

        \hline
        \cite{abbasi2024haps} & Outline particular HAPS use cases and challenges in 6G \acp{vHetNet} with some simulation validations.  
        & Primary  & \cmark \newline (partially) & \xmark & \cmark & \cmark \\
        
        \hline
        \textbf{Ours} & \textbf{Comprehensive review of \ac{HAPS}, covering use cases, enabling technologies, \ac{TN}-\ac{NTN} integration, current \ac{HAPS}-related projects, and future research directions.}  
        & Primary & \cmark & \cmark & \cmark & \cmark \\
        
        \hline
    \end{tabular}
    \label{tab:haps_comparison_surveys}
\end{table*}

\section*{List of Frequently Used Abbreviations}
\addcontentsline{toc}{section}{List of Abbreviations}

\begin{acronym}[AAAAAAAAAAAAAA]  % longest acronym to fix width
  \acro{3D}{three-dimensional}
  \acro{3G}{third generation}
  \acro{3GPP}{third generation partnership project}
  \acro{4G}{fourth-generation}
  \acro{5G}{fifth-generation}
  \acro{6G}{sixth-generation}
  \acro{AI}{artificial intelligence}
  \acro{AR}{augmented reality}
  \acro{AV}{aerial vehicle}
  \acro{BER}{bit error rate}
  \acro{BS}{base station}
  \acro{CFO}{carrier frequency offset}
  \acro{CNN}{convolutional neural network}
  \acro{cmWave}{centimeter wave}
  \acro{CoMP}{coordinated multi-point}
  \acro{CPU}{central processing unit}
  \acro{CSI}{channel state information}
  \acro{D2D}{Device-to-Device}
  \acro{dB}{decibel}
  \acro{DDPG}{deep deterministic policy gradient}
  \acro{DFT}{discrete Fourier transform}
  \acro{DRL}{deep reinforcement learning}
  \acro{EB}{exabytes}
  \acro{eMBB}{enhanced mobile broadband}
  \acro{FL}{federated learning}
  \acro{FR}{frequency range}
  \acro{FSO}{free space optical}
  \acro{GEO}{geostationary equatorial orbit}
  \acro{GUE}{ground user equipment}
  \acro{GNSS}{global navigation satellite system}
  \acro{GPS}{global positioning system}
  \acro{HAPS}{high-altitude platform station}
  \acro{HAPS-SMBS}{high-altitude platform station mounted super macro base station}
  % \acro{HARQ}{hybrid automatic repeat request}
  %\acro{HO}{handover}
  %\acro{IoRT}{Internet of remote things}
  \acro{IoT}{Internet of things}
  \acro{ISAC}{integrated sensing and communication}
  \acro{ISTN}{integrated space–terrestrial network}
  \acro{ITU}{International Telecommunication Union}
  \acro{ITU-R}{International Telecommunication Union Radiocommunication Sector}
  \acro{LAPS}{low-altitude platform station}
  \acro{LEO}{low Earth orbit}
  \acro{LoS}{line-of-sight}
  \acro{LoRa}{long range}
  \acro{LTE}{long term evolution}
  \acro{MBS}{macrocell base station}
  \acro{MC}{multi-connectivity}
  \acro{MDP}{Markov decision process }
  \acro{MEC}{Mobile Edge Computing}
  \acro{MEO}{medium Earth orbit}
  \acro{MIMO}{multiple-input multiple-output}
  \acro{ML}{machine learning}
  \acro{mMIMO}{massive multiple-input multiple-output}
  \acro{MMSE}{minimum mean square error}
  \acro{mMTC}{massive machine type communication}
  \acro{mmWave}{millimetre wave} 
  % \acro{MVDR}{Minimum Variance Distortionless Response}
  \acro{NB-IoT}{narrow band IoT}
  \acro{NOMA}{non-orthogonal multiple access}
  \acro{NR}{new radio}
  \acro{NTN}{non-terrestrial network}
  \acro{OFDM}{orthogonal frequency division multiplexing}
  \acro{OFDMA}{orthogonal frequency division multiple access}
  % \acro{PDCP}{packet data convergence protocol}
  \acro{PHY}{physical}
  \acro{PNT}{Positioning, Navigation, and Timing}
  \acro{PPO}{proximal policy optimization}
  \acro{QoS}{quality of service}
  \acro{RA}{resource allocation}
  \acro{RAN}{radio access network}
  \acro{RAT}{radio access technology}
  \acro{RB}{resource block}
  % \acro{REM}{radio environment map}
  \acro{RF}{radio frequency}
  \acro{RIS}{reconfigurable intelligent surface}
  \acro{RL}{reinforcement learning}
  \acro{RRC}{Radio Resource Control}
  \acro{RSRP}{reference signal received power}
  \acro{RTT}{round-trip time}
  \acro{SAGIN}{space-air-ground integrated network}
  \acro{SatCom}{satellite communication}
  \acro{SBS}{small base station}
  \acro{SIC}{Successive Interference Cancellation}
  \acro{SINR}{signal to interference plus noise ratio}
  \acro{SLNR}{signal to leakage interference and noise ratio}
  \acro{SMBS}{super macro base station}
  \acro{SNR}{signal to noise ratio}
  \acro{SVM}{support vector machine}
  \acro{TBS}{terrestrial base station}
  \acro{THz}{terahertz}
  \acro{TN}{terrestrial network}
  \acro{UAV}{unmanned aerial vehicle}
  \acro{UE}{user equipment}
  \acro{UMTS}{universal mobile telecommunications system}
  \acro{URLLC}{ultra-reliable low-latency communication}
  \acro{VEC}{Vehicular Edge Computing}
  \acro{vHetNet}{vertical heterogeneous network}
  \acro{VR}{virtual reality}
  \acro{VSAT}{very small aperture terminal}
  \acro{VTOL}{vertical take-off and landing}
\end{acronym}

\section{HAPS Comparison with Terrestrial Networks and Other NTN Nodes}
\label{sec: haps}

Global connectivity, traditionally supported by \acp{TN} and satellites, still faces significant gaps. Despite mobile broadband traffic doubling from 419 \ac{EB} in 2019 to 913 \ac{EB} in 2022, about 33\% of the world’s population remains offline~\cite{itu2023measuring}. This digital divide highlights the need for innovative solutions. \ac{HAPS}, operating in the stratosphere at 20–50 km, offer a promising approach. Acting independently or as intermediaries between \acp{TN} and other \ac{NTN} systems such as \ac{LEO} satellites and \acp{UAV}, \ac{HAPS} provide wide-area coverage, low latency, and operational flexibility. These features enable them to bridge connectivity gaps while complementing existing networks. This section examines \ac{HAPS} capabilities compared with \acp{TN} (Section~\ref{sub:comparison-tn}), \ac{LEO} satellites (Section~\ref{sub:comparison-leo}), and \acp{UAV} (Section~\ref{sub:comparison-uav}), highlighting their unique advantages. Table~\ref{tab:comparison} summarizes the distinct characteristics of \ac{HAPS} relative to \acp{TN} and other \ac{NTN} nodes in terms of altitude, performance, operational flexibility, and cost, while Section~\ref{sub:quantitative_comp} provides a comparison across \ac{NTN} architectures in terms of latency, capital and operational expenses, scalability and risks.
\begin{table*}[t!]
\centering
\def\arraystretch{1.2}
\caption{Comparison of \ac{HAPS} with TNs, \ac{LEO} satellites, and \acp{UAV}.}
\label{tab:comparison}
\begin{tabular}{|p{2.8cm}|p{3cm}|p{2.8cm}|p{2.2cm}|p{4cm}|}
\hline
\centering
\textbf{Feature}            & 
\centering
\textbf{\ac{TN}s}                &
\centering
\textbf{\ac{LEO} satellites}     & \textbf{\acp{UAV}}  
\centering
& 
\textbf{\ac{HAPS}} 
\\ \hline
\textbf{Altitude}           & Ground-based                     & 500–2,000 km                       & Up to 1.5 km                        & 20–50 km                           \\ \hline
\textbf{Coverage area}      & Local to regional                & Global (constellations)                            & Localized                           & Regional to wide-area              \\ \hline
\textbf{Latency}            & Low                              & Moderate to high                   & Low                                 & Low                                \\ \hline
\textbf{Operational duration} & Continuous                       & Years                              & Hours to days                       & Weeks to months                    \\ \hline
\textbf{Deployment flexibility} & Limited by infrastructure        & Rigid (requires launch)            & Highly flexible                     & Moderate (fixed stratospheric path) \\ \hline
\textbf{Cost}               & High for remote areas            & Very high                          & Moderate to low                     & Moderate                           \\ \hline
\textbf{Power source}       & Grid or renewable energy         & Solar or batteries                 & Batteries                           & Solar-powered                      \\ \hline
\textbf{Payload capacity}   & High                             & Moderate                           & Low                                 & Moderate to high                   \\ \hline
\textbf{Adaptability}       & Limited                          & Low                                & High                                & Moderate to high                   \\ \hline
\textbf{Emergency response} & Slow (requires rebuilding)       & Limited                            & Rapid deployment                    & Rapid deployment                   \\ \hline
\end{tabular}
\end{table*}

\subsection{Comparison with Terrestrial Networks}
\label{sub:comparison-tn}
\acp{TN} serve as the backbone of modern communication systems, providing high-capacity, low-latency connectivity in densely populated areas. However, they are constrained by geographical and infrastructural limitations. Rugged terrains, remote regions, and disaster-stricken areas often lack adequate terrestrial infrastructure, leaving significant portions of the population unconnected. Furthermore, the deployment and maintenance of \acp{TN} in such regions involve high costs, making them impractical for universal connectivity.

Recent studies on networking have also emphasized these limits. For instance, previous research~\cite{dhananjay2011wire} illustrates the challenge of extending traditional terrestrial infrastructure into rural areas, while efforts on community cellular networks~\cite{hasan2019scaling} and virtual coverage~\cite{heimerl2013expanding} reveal that operational complexity, cost, and power availability are the main obstacles to large-scale terrestrial deployment in underserved areas.

\ac{HAPS} overcome these limitations by offering wide-area coverage without reliance on extensive ground infrastructure. They can provide reliable \ac{LoS} connectivity across urban, rural, and remote areas, bridging gaps where terrestrial solutions are infeasible. Their rapid deployment capability also makes them ideal for emergency scenarios, while operational costs in remote areas are often lower than constructing terrestrial networks.

\subsection{Comparison with LEO Satellites}
\label{sub:comparison-leo}
\ac{LEO} satellites provide essential global coverage, particularly in remote and oceanic regions. However, their high operational altitudes (500 to 2,000 kilometers) result in increased signal attenuation and latency. Moreover, their large coverage areas imply that the available bandwidth is shared among a large number of \acp{UE}, leading to reduced per-\ac{UE} data rates.

These system-level limitations are further supported by recent research. In \ac{LEO} systems, downlink opportunities, contact windows, and ground-station coordination dominate end-to-end latency, according to work on distributed downlink architectures and constellation-scale traffic scheduling~\cite{vasisht2021l2d2}. Direct-to-cell designs~\cite{liu2024democratizing} highlight fundamental per-user capacity and resource-sharing limitations when large footprints serve massive user populations.
These constraints make \ac{LEO} satellites less suited for high-bandwidth or latency-sensitive applications. In contrast, \ac{HAPS} are positioned significantly closer to Earth, which reduces latency and enables better spatial reuse, resulting in higher bandwidth availability per \ac{UE}. Additionally, \ac{HAPS} offer greater operational flexibility, as they can be repositioned and upgraded without the complexity and cost of satellite launches.

\subsection{Comparison with UAVs}
\label{sub:comparison-uav}
\acp{UAV} are highly versatile, providing localized and flexible connectivity for hot-spot operations and disaster scenarios. However, their limited flight duration due to battery constraints, restricted payload capacity, and frequent maintenance requirements hinder scalability for large-area or long-term deployments.
Recent system-level investigations also show these limitations. The efficiency of \acp{UAV} for short-term, localized coverage is demonstrated by research on drone-based communication relays~\cite{ma2017drone} and integrated rural connectivity platforms~\cite{vasisht2017farmbeats}, while also highlighting endurance and sustainability constraints that preclude persistent wide-area service.%}

In contrast, \ac{HAPS} offer extended operational durations, lasting weeks or months, enabled by solar power and autonomous control. Their quasi-stationary positioning allows them to function as stable, high-capacity communication hubs, effectively bridging the gap between the localized flexibility of \acp{UAV} and the global reach of satellite systems.

\subsection{Quantitative and Structured Comparison Across NTN Architectures}
\label{sub:quantitative_comp}

    \subsubsection{Latency}
    LEO satellites experience non-negligible propagation and routing delays, on average around 25 ms~\cite{michel2022first,garcia2025detailed} of a \ac{RTT}.
    For instance, a satellite operating at an altitude of approximately 550~km has a theoretical \ac{RTT} propagation delay of about 3.6~ms; however, practical \acp{RTT} are significantly higher, ranging between 30 and 50~ms, due to routing, ground station access, and network management overheads. 
    Notably, in our experimental evaluation~\cite{figaro2026experimental}, we revealed periodic high-magnitude \ac{RTT} spikes of up to 140~ms, at regular intervals of approximately 13~s. This periodic behavior is indicative of constellation-specific control and resource management mechanisms, such as inter-beam or inter-satellite handovers, which require additional signaling for path reconfiguration and session re-initialization.
    By contrast, \ac{HAPS} platforms operating in the stratosphere at altitudes of 18–25~km have nearly negligible propagation delays, and the resulting \acp{RTT} is typically comparable to that of terrestrial 5G systems, consistently below 10~ms. This makes HAPS particularly attractive for latency-sensitive services. 
    Finally, \ac{UAV} swarms operate at very low altitudes (e.g., a few hundred meters), and achieve sub-millisecond propagation delays. However, in practice, the end-to-end latency is often dominated by multi-hop processing and coordination overhead within the swarm’s mesh~network~\cite{masaracchia2021uav,luo2022delay}.    
    
    \subsubsection{CAPEX/OPEX} Global LEO constellations require extreme upfront capital investment. Public estimates for Amazon's Project Kuiper and SpaceX's Starlink indicate a total CAPEX of approximately 10 billion USD to reach full operational capability~\cite{osoro2021techno,li2021techno}. According to~\cite{ye2024techno}, the average CAPEX per Starlink satellite is approximately 548,000~USD, for both satellite manufacturing and launch costs. Inter-satellite links (ISLs) in  second-generation satellites introduce an additional cost of several hundred thousand USD per link.
    Regarding OPEX, costs are primarily dominated by ground stations, which require an average annual cost of approximately 215,000 USD on average, vs. around 20,000 USD for each satellite. Furthermore, the relatively short operational lifetime of LEO satellites (approximately five years) necessitates frequent satellite replacement, thereby increasing long-term costs.
    In comparison, the CAPEX of HAPS is significantly lower. For example, the manufacturing cost of a Airbus Zephyr platform is estimated at approximately 4~million USD~\cite{toka2024integrating}, a number that has decreased significantly as stratospheric technology have matured. Unlike LEO satellites, HAPS platforms are typically designed to operate for around of 20 years, so that initial investments can be amortized over a much longer period. In terms of OPEX, costs are mainly due to maintenance and logistics, reaching up to 15\% of the manufacturing cost annually. However, energy costs remain minimal as HAPS platforms utilize high-efficiency solar panels for daytime flight and onboard batteries for nighttime operations, resulting in an annual energy cost of only 30,000 USD per platform.
    Finally, the individual CAPEX of large civilian heavy-lift UAVs is relatively low, typically on the order of a few thousand USD, while OPEX is dominated by energy consumption and maintenance, in particular battery replacement and charging logistics.

    \subsubsection{Scalability} While large-scale LEO satellite constellations are designed to provide ubiquitous coverage on Earth, the network capacity must be shared among a large number of UEs, which can saturate the available resources and complicate channel access~\cite{rossato20265g}.    
    In contrast, HAPS platforms only operate at a regional level, but provide high-capacity quasi-stationary connectivity via aggressive frequency reuse.
    Finally, UAV swarms provide extreme localized (though high-speed) coverage in hot-spot areas (e.g., for temporary events or emergency scenarios), even though they do not scale efficiently over large geographic regions.

    \subsubsection{Operational risk} LEO constellations face both short- and long-term operational risks, including launch failures, orbital congestion, and the increasing threat of space debris and Kessler Syndrome effects. HAPS operations are primarily challenged by stratospheric wind dynamics and seasonal power constraints, particularly during the winter solstice at latitudes above approximately 50$^\circ$, where the shorter daylight duration may limit battery recharging capability for overnight flight.
    UAV swarms face the most stringent regulatory and legal constraints on the maximum flight altitude, airspace access, and spectrum usage, especially for Beyond Visual Line-of-Sight (BVLOS) flight operations over populated areas, which typically require case-by-case authorization and continuous human supervision. Moreover, UAVs are vulnerable to adverse weather conditions, such as strong winds and precipitation.

\section{HAPS Use Cases} \label{sec: Tech_use_case}
The unique characteristics of \ac{HAPS} enable a wide range of applications across telecommunications and beyond. In telecommunications, they primarily aim to extend coverage (Section~\ref{sub:use-case-connectivity}) and enhance network efficiency (Section~\ref{sub:use-case-network}). Their rapid deployment and adaptability make them essential for bridging connectivity gaps in remote, underserved, or disaster-affected regions. Serving as mobile urban \acp{BS} or backhaul providers, \ac{HAPS} improve capacity, reliability, and energy efficiency. Beyond communications, they support advanced applications (Section~\ref{sub:use-case-iot}), including real-time \ac{IoT} sensing, immersive \ac{AR}/\ac{VR}, precision control via \ac{URLLC} (Section~\ref{sub:use-case-control}), and energy optimization (Section~\ref{sub:use-case-energy}). By bringing processing closer to \acp{UE}, \ac{HAPS} reduce latency and enable innovations across sectors such as agriculture, logistics, and disaster management. This section explores their roles in connectivity enhancement and advanced use cases shaping next-generation networks. Finally, in the Section~\ref{sub:haps-market} we provide a brief market overview of \ac{HAPS}-powered communication services.
%%%%%%%%%%%%%%%%%%%%%%%%%%%%%%%%%%%%%%%%%%%%%%%%%%%%%%%%%
\subsection{Connectivity Enhancement}
\label{sub:use-case-connectivity}
%%%%%%%%%%%%%%%%%%%%%%%%%%%%%%%%%%%%%%%%%%%%%%%%%%%%%%%%%

\subsubsection{Remote and Rural Area Coverage}
The COVID-19 pandemic highlighted global Internet access disparities, especially in remote and underserved regions, emphasizing the need for scalable connectivity solutions~\cite{chaoub20216g}. \ac{HAPS} can address these gaps by extending mobile coverage to areas lacking terrestrial infrastructure. Operating similarly to \acp{TN}, they can bridge the digital divide and deliver reliable global connectivity~\cite{9681623}.

Collaboration between \ac{HAPS} and \ac{LEO} satellites was explored by Jia~\textit{et al.}~\cite{jia2020joint}, who formulated a mixed-integer nonlinear programming problem to enhance large-scale access and backhaul. The proposed restricted trilateral matching algorithm improved coordination, connectivity, and efficiency in remote regions.
Zhao~\textit{et al.}~\cite{zhao2023backhaul} analyzed \ac{HAPS}-\ac{LAPS} integration using stochastic geometry, introducing a framework to assess platform density and altitude effects under backhaul constraints, offering design guidance for robust aerial networks in rural areas.
Arum~\textit{et al.}~\cite{arum2020review} considered \ac{HAPS} for extending rural coverage with low \ac{UE} density through integration with \acp{TN}, emphasizing coexistence via dynamic resource management, interference mitigation, and techniques such as advanced antennas, radio environment maps, and device-to-device communication.
Zhang, Kishk and Alouini~\cite{zhang2023hap} optimized resource allocation for heterogeneous services in rural \ac{HAPS} networks using \ac{FSO} backhaul. It proposed static and dynamic allocation schemes leveraging \ac{DRL}, improving age of information, data rates, and overall network efficiency.

\subsubsection{Natural Disaster Region Coverage}
In the aftermath of natural disasters, when \acp{TN} collapse, \ac{HAPS} can provide real-time aerial surveillance and rapid, reliable communication. Covering vast and remote regions, they ensure effective disaster response and emergency connectivity for first responders~\cite{gharbi2019overview}.
Yu~\textit{et al.}~\cite{yu2024research} examined \ac{UAV}-based \ac{HAPS} for disaster scenarios, highlighting their rapid deployment, flexible coverage, and communication recovery, while identifying challenges such as backhaul limitations, platform constraints, and communication capacity that require targeted solutions.
Real-time monitoring using \ac{HAPS} was demonstrated by Baraniello~\textit{et al.}~\cite{baraniello2021application}, showing persistent, high-resolution imaging from altitudes around 20\,km, enabling stable, continuous coverage of affected areas.
In the study provided by L{\"a}hdekorpi~\textit{et al.}~\cite{lahdekorpi2010replacing}, \ac{HAPS} restored services in a disrupted \ac{UMTS} network, achieving 70\% throughput recovery when terrestrial \acp{BS} were unavailable, underscoring their effectiveness in disaster recovery operations.

\subsubsection{Unexpected Traffic Events}
The inherent mobility and on-demand deployment of \ac{HAPS} make them well suited for managing dynamic communication demands, including large public events, sudden traffic surges, or routine daily traffic variations~\cite{kurt2021vision, dong2016constellation}. By adapting to these fluctuations, \ac{HAPS} alleviate congestion, enhance \ac{QoS}, and improve energy efficiency. Compared to small cell densification, which requires extensive planning, fixed infrastructure, and offers limited flexibility, \ac{HAPS} provide a faster and more cost-effective solution that can be deployed in real time without physical constraints.
A \ac{QoS}-driven deployment framework was proposed by Dong~\textit{et al.}~\cite{dong2016constellation} to optimize \ac{HAPS} operation based on coverage, capacity, and latency metrics, balancing efficiency and cost. In integrated networks where ground and aerial \acp{BS} share spectrum, interference remains a key issue. As noted by Wei~\textit{et al.}~\cite{wei2023spectrum}, advanced techniques such as dynamic resource allocation, spectrum sharing, and adaptive power control enable efficient \ac{HAPS} operation and seamless coexistence with \acp{TN}, ensuring robust performance under variable traffic conditions.
Jeon~\textit{et al.}~\cite{jeon2022energy} installed a \ac{RIS} on a \ac{HAPS} to provide extensive \ac{LoS} and full-area coverage for \ac{UAV}-\acp{BS} deployed in response to sudden urban traffic spikes. The combined optimization of \ac{RIS} phases, array-partitioning algorithms, and positioning enhanced the energy efficiency of each \ac{UAV}-\ac{BS}.
%%%%%%%%%%%%%%%%%%%%%%%%%%%%%%%%%%%%%%%%%%%%%%%%%%%%%%%%%
\subsection{Network Infrastructure and Management}
\label{sub:use-case-network}
%%%%%%%%%%%%%%%%%%%%%%%%%%%%%%%%%%%%%%%%%%%%%%%%%%%%%%%%%

\subsubsection{Backhauling}
Network backhaul is a fundamental component of communication infrastructure, responsible for high-capacity, low-latency data transmission between \acp{BS} and the core network. In dense urban areas, fiber optics remain the dominant solution, offering high throughput and minimal latency, whereas copper cables, limited in bandwidth, are largely obsolete for modern traffic demands. In contrast, rural and remote regions often rely on satellite or microwave links, as fiber deployment is economically and logistically prohibitive~\cite{dahrouj2015cost}.
However, non-geostationary satellite systems require costly ground gateways with precise placement to maintain reliable connectivity~\cite{baeza2023gateway}.
Although wireless backhauling has been studied extensively~\cite{polese2020integrated}, \ac{HAPS} are emerging as a promising enabler for next-generation networks. Positioned in the stratosphere, they provide near-\ac{LoS} connectivity with lower latency compared to satellites, supporting applications such as small cell backhauling, wireless sensor networks, and Earth observation~\cite{mohammed2011role}. In-band \ac{mmWave} backhaul has been demonstrated by Taori and Sridharan~\cite{taori2015point}, while multi-band approaches for enhanced capacity were discussed  by Tezergil and Onur~\cite{tezergil2022wireless}.

Recent studies proposed advanced architectures leveraging \ac{HAPS}. Abbasi and Yanikomeroglu~\cite{abbasi2022cell}, introduced a cell-free scheme with \ac{HAPS} as an aerial \ac{CPU} aggregating \acp{UAV} via \ac{THz} links, where large antenna arrays mitigate path loss. Hybrid \ac{RF}/\ac{FSO} solutions further enhanced flexibility: Dang, Le, Nguyen, and Pham~\cite{dang2023cooperative} presented a cooperative hybrid ARQ protocol ensuring latency fairness under varying conditions, and Yahia, Erdogan, and Kurt~\cite{yahia2023haps} explored hybrid \ac{LEO}-to-\ac{HAPS} \ac{FSO} and \ac{HAPS}-to-ground \ac{RF} links. Collectively, these advancements demonstrate the potential of \ac{HAPS}-enabled backhaul to deliver scalable, low-latency connectivity across heterogeneous environments.

\subsubsection{Handover Management}
Frequent handovers and associated failures represent a major challenge in satellite communication systems, particularly for fast-moving non-geostationary satellites that require continuous connectivity updates. \ac{HAPS} mitigate these issues in two key roles: (i) as intermediaries between satellites and \acp{UE}, handling handover requests on behalf of satellites, and (ii) as standalone communication managers, directly managing \ac{UE} connections. In both cases, \ac{HAPS} improve network stability and reduce signaling overhead.
Leveraging \ac{HAPS} as intermediaries was shown by Cianca~\textit{et al.}~\cite{cianca2005integrated} to decouple satellite mobility from ground-based handover procedures, simplifying management and enhancing continuity. Dynamic \ac{LEO}-\ac{HAPS} handover strategies proposed by Li~\textit{et al.}~\cite{li2019handover} achieved lower delays, signaling costs, and drop rates than traditional \ac{5G} methods. Prior to that, Mohammed, Mehmood, Pavlidou, and Mohorcic~\cite{mohammed2011role} explored \ac{FSO}-based \ac{LEO}-\ac{HAPS} links, demonstrating high-speed, low-latency handovers even under high satellite mobility.
When operating as quasi-stationary communication nodes, \ac{HAPS} significantly reduce handover frequency for terrestrial \acp{UE}, providing seamless service in high-mobility scenarios (e.g., vehicles, aircraft). Their stable positioning eliminates inter-platform handovers common in \ac{LEO} constellations. Combining beamforming for \ac{HAPS}-to-ground and \ac{FSO} backhaul to \ac{LEO}/\ac{MEO} satellites, Srinivasan~\textit{et al.}~\cite{srinivasan2021airplane} demonstrated efficient handover management in the airliner system even in dynamic environments. Moreover, optical \ac{HAPS}-satellite links, as analyzed be Perlot~\textit{et al.}~\cite{perlot2008system}, can further enhance reliability by ensuring robust backhaul connectivity, thereby minimizing disruptions in \ac{HAPS}-to-\ac{UE} sessions.

%%%%%%%%%%%%%%%%%%%%%%%%%%%%%%%%%%%%%%%%%%%%%%%%%%%%%%%%%
\subsection{IoT and Emerging Applications}
\label{sub:use-case-iot}
%%%%%%%%%%%%%%%%%%%%%%%%%%%%%%%%%%%%%%%%%%%%%%%%%%%%%%%%%

\subsubsection{IoT sensing}
\ac{HAPS} can significantly enhance \ac{IoT} sensing and monitoring by providing extensive, stable coverage from the stratosphere, and integrating edge computing functionalities. Edge computing ensures low latency and reliable communication by processing data closer to the source compared to cloud computing, while \ac{HAPS}' dynamic positioning allows for optimized sensor deployment and efficient data acquisition. This combination improves the efficiency and responsiveness of \ac{IoT} networks, particularly in remote and critical environments. 
A reliable \ac{IoT} architecture leveraging \ac{HAPS} was proposed by Sibiya and Olugbara~\cite{sibiya2019reliable}, integrating \ac{NB-IoT} and \ac{LoRa} to deliver scalable, cost-efficient connectivity across underserved regions. Operating at 17–35 km, \ac{HAPS} achieved coverage up to 600 km, surpassing terrestrial solutions in scalability and redundancy through overlapping cells.

Andreadis, Giambene, and Zambon~\cite{andreadis2023role} introduced a joint \ac{HAPS}-\ac{UAV} framework for \ac{IoT} applications—especially disaster scenarios. Combining umbrella coverage from \ac{HAPS} and agile \acp{UAV}, the system ensured continuous service with sub-millisecond latency (0.07–0.4 ms for \ac{HAPS}), while \ac{NB-IoT}/\ac{LoRa} access schemes minimized collisions under dense deployments.
Complementary \ac{LoRa}-based monitoring was examined by Giambene and Korre~\cite{giambene2022lora}, which optimized parameters such as altitude, transmission power, and device density under shadowing and collision constraints, confirming the viability of \ac{HAPS} for large-scale environmental \ac{IoT}.
To enhance \ac{HAPS} network capacity, Xiao,  Li, and Zhao~\cite{xiao2019lstm} introduced a \ac{5G} \ac{IoT} architecture with a uniform rectangular array antenna and a Long Short-Term Memory (LSTM) model for direction-of-arrival prediction. The approach improved beamforming precision and link stability, indirectly benefiting \ac{IoT} reliability.
Joint \ac{LEO}-\ac{HAPS} \ac{IoT} integration was explored by Ei~\textit{et al.} in~\cite{ei2023joint}, where \ac{HAPS} relayed data to ground centers via \ac{LEO} satellites, optimizing throughput and power consumption under \ac{QoS} and storage constraints. Similarly, Monzon and Alvarez~\cite{monzon2022high} presented a smart rural framework combining \ac{5G}, cloud, \ac{IoT}, and \ac{HAPS}, enabling automated remote data access and processing.

\subsubsection{Mobile Edge Computing and Fast UE Movement}
\ac{MEC} enhances network performance by positioning computing resources closer to \acp{UE}, significantly reducing latency for real-time applications. However, fast-moving \acp{UE} pose challenges for maintaining connectivity and service consistency, necessitating advanced handover techniques and predictive algorithms to ensure seamless operation in dynamic environments.
Traspadini~\textit{et al.}~\cite{traspadini2022uav} analyzed the integration of \ac{HAPS} and \acp{UAV} within \ac{VEC} systems using queuing theory to optimize task offloading in dynamic vehicular environments. The authors showed that \ac{HAPS}, with quasi-stationary operation, broad coverage, and strong \ac{LoS} links, outperform \acp{UAV} by supporting larger computing payloads and offering lower latency, making them ideal for large-scale, high-mobility \ac{VEC}/\ac{MEC} scenarios.
Later, Seid~\textit{et al.}~\cite{seid2023hdfrl}  proposed to use aerial~\ac{MEC} servers to support  \acp{GUE} and emergency communications. The authors introduced a hierarchical deep federated learning \ac{RA} scheme for intelligent aerial-enabled smart city cyber-physical system, aiming at optimizing energy consumption while preserving task offloading privacy and maintaining~\ac{QoS} in dynamic environments.
Although prior works such as~\cite{sun2020optimizing, wang2019joint} optimized \ac{UAV}-enabled \ac{MEC} for task completion and energy efficiency, these approaches can be extended to \ac{HAPS}, which offer superior payload capacity and operational stability. Yet, research on \ac{HAPS}-based \ac{MEC} remains limited, representing a promising avenue for future exploration.

\subsubsection{Aerial Data Centers and AR/VR Support}
Beyond edge computing, \ac{HAPS} can function as aerial data centers, delivering centralized cloud services for data storage, processing, and distribution from the stratosphere. With wide coverage, high capacity, and low-latency connectivity, they support resource-intensive applications such as \ac{AR} and \ac{VR}~\cite{alam2021high}, enabling immersive experiences even in remote or underserved regions~\cite{kumar2013survey}. Traspadini~\textit{et al.}~\cite{traspadini2023real} presented a \ac{HAPS}-enabled \ac{VEC} system for rural areas, showing performance gains through optimized trade-offs between computational load and latency.
A cloud-integrated \ac{HAPS} system proposed in~\cite{mershad2021cloud} achieved up to 14\% energy savings compared to terrestrial data centers, reducing offloading outages and delays in distributed server networks while supporting services such as \ac{IoT}, vehicular networks, and social media. Furthermore,~\cite{ei2023joint} introduced a collaborative \ac{LEO}-\ac{HAPS} framework in which \ac{HAPS} acted as relays, forwarding \ac{IoT} data to ground processing centers with optimized data rates, power efficiency, and \ac{QoS} compliance.

%%%%%%%%%%%%%%%%%%%%%%%%%%%%%%%%%%%%%%%%%%%%%%%%%%%%%%%%%
\subsection{Advanced Control and Special Applications}
\label{sub:use-case-control}
%%%%%%%%%%%%%%%%%%%%%%%%%%%%%%%%%%%%%%%%%%%%%%%%%%%%%%%%%

\subsubsection{URLLC Support and Remote Driving Control}
Operating at high altitudes but closer to the ground than traditional satellites, \ac{HAPS} combine the benefits of wide coverage and low-latency communication. Additionally, their significant payload capacity allows them to carry advanced antenna systems and powerful processing units, enhancing their ability to support \ac{URLLC}, e.g., for applications like remote driving control. The role of \ac{HAPS} in supporting \ac{URLLC} has been explored in several studies.
Salehi~\textit{et al.}~\cite{salehi2022ultra} analyzed a heterogeneous network integrating \ac{HAPS}, terrestrial \acp{BS}, and autonomous vehicles, showing that \ac{MC} is essential to meet stringent \ac{URLLC} requirements. \ac{HAPS} acted as auxiliary links sustaining data rates above 700\,kbps with high reliability under dynamic conditions. Building on this, the authors in~\cite{salehi2024reliability} extended the model with \ac{LEO} satellites, optimizing \ac{MC} path selection to satisfy \ac{QoS} constraints with minimal link usage. 
Results revealed that \ac{LoS} interference is a key performance-limiting factor in radio access links, which was further addressed by Manzoor, Ozger, Schupke, and Cavdar~\cite{manzoor2024combined} through refined \ac{MC} strategies for enhanced reliability and efficiency.
Beyond \ac{URLLC}, \ac{HAPS} can improve Global Navigation Satellite System (GNSS) performance due to their quasi-stationary position and strong signal power. Zheng~\textit{et al.}~\cite{zheng2023analysis} demonstrated a \ac{HAPS}-aided GPS architecture that improves \ac{3D} positioning accuracy, enhancing navigation and control functions. This capability is particularly valuable for remote or underserved regions—such as mountainous, coastal, or oceanic areas—where \ac{HAPS} can act as super macro base stations (\ac{HAPS}-SMBS) to backhaul isolated nodes and extend coverage~\cite{kurt2021vision}.

\subsubsection{UAV Control and Aerial Highways}
The cellular community has been actively working to enable \ac{UAV} connectivity and control, focusing on leveraging existing ground cellular networks while ensuring minimal impact on ground \acp{UE} performance. Current solutions, such as separating time and frequency resources, offer a temporary improvement by assigning specific slots/bands to \acp{UAV}. However, these methods lack the scalability required to manage the anticipated surge in \acp{UAV} operations in the skies~\cite{3GPP36777, NguAmoWig2018}. To address these limitations, various strategies have been explored, including increasing the density of network infrastructure, developing specialized aerial communication systems~\cite{garcia2019essential,d2020analysis,mozaffari2021toward,kim2022non,pan2023resource}, or using \ac{LEO} satellites to enhance \ac{UAV} connectivity~\cite{geraci2022integrating}.
A promising concept in this domain is the establishment of aerial highways, i.e., designated \ac{UAV} corridors to facilitate organized and secure \ac{UAV} operations. These corridors, much like terrestrial road networks, are envisioned as regulated airspace where \acp{UAV} follow predefined safe and efficient routes~\cite{cherif20213d, bhuyan2021secure}. To support connectivity within these corridors, researchers have investigated adjustments to cellular network configurations. Efforts include network-level optimizations tailored to \ac{UAV} traffic in these aerial highways~\cite{bernabe2022optimization, bernabe2023novel, maeng2023base, chowdhury2021ensuring, singh2021placement}, and theoretical studies focused on developing frameworks for \ac{UAV} operations in structured airspace~\cite{karimi2023optimizing, karimi2023analysis}. Leveraging their ability to provide reliable, low-latency communication and wide-area coverage, \ac{HAPS} are well-suited to support \ac{UAV} connectivity and control, whether in open airspace or within the structured framework of aerial highways. As discussed in Section~\ref{sub:use-case-network}, this extensive coverage significantly reduces the need for handovers between terrestrial \acp{BS}, ensuring seamless connectivity for high-speed \acp{UAV}. Additionally, their quasi-static positioning in the stratosphere eliminates the need for frequent inter-platform communication, a challenge often encountered with \ac{LEO} satellite systems due to their high mobility. This combination of stability, wide-area coverage, and operational simplicity positions \ac{HAPS} as an ideal and efficient solution for enabling robust and reliable \ac{UAV} control.

%%%%%%%%%%%%%%%%%%%%%%%%%%%%%%%%%%%%%%%%%%%%%%%%%%%%%%%%%
\subsection{Energy and Sustainability}
\label{sub:use-case-energy}
%%%%%%%%%%%%%%%%%%%%%%%%%%%%%%%%%%%%%%%%%%%%%%%%%%%%%%%%%
Energy efficiency is a critical consideration for future integrated \acp{TN} and \acp{NTN}. Traditional \acp{TN}  rely on continuous operation of macro \acp{BS} to provide ubiquitous coverage, leading to substantial energy consumption even during low-traffic periods. This ''zero bit, non-zero watt'' phenomenon reflects the inefficiency of maintaining coverage when demand is minimal. To address this issue, a promising solution involves offloading traffic from terrestrial macro \acp{BS} to \ac{NTN}-powered hypercells, such as those supported by \ac{LEO} satellites, \ac{HAPS}, or tethered \acp{UAV}. During nighttime or periods of low demand, this approach allows power-intensive terrestrial macro \acp{BS} to enter low-power states or shut down entirely, significantly reducing energy consumption. In this context, \ac{LEO} satellites are well-suited for offloading outdoor \acp{UE} due to their extensive coverage and ability to serve low-capacity traffic efficiently during nighttime. In contrast, \ac{HAPS}, with their superior link budget and better signal penetration, are better suited for offloading indoor \acp{UE}, especially in urban environments where high spatial reuse is essential.
Song~\textit{et al.}~\cite{song2024high} developed a heuristic algorithm to optimize this offloading process. By prioritizing the least-utilized \ac{BS} for offloading, and leveraging \ac{HAPS}'s self-sufficient energy capabilities, the proposed approach minimized hourly energy consumption across the network. Simulation results indicated that this method could reduce energy consumption by up to 29\% over a week, with savings reaching 41\% during nighttime hours. These results highlight the complementary roles of \ac{HAPS} in advancing sustainable and energy-efficient connectivity within integrated \acp{TN} and \acp{NTN}.

%%%%%%%%%%%%%%%%%%%%%%%%%%%%%%%%%%%%%%%%%%%%%
\subsection{HAPS Market Overview}
\label{sub:haps-market}
%%%%%%%%%%%%%%%%%%%%%%%%%%%%%%%%%%%%%%%%%%%%%
HAPS offer a cost-effective complement to terrestrial and satellite networks in NTN architectures, particularly for rural connectivity and emergency services~\cite{rousseau2023haps}. This subsection addresses economic viability through cost breakdowns, sustainable business models, and market analysis of key providers.
As mentioned above, the CAPEX/OPEX comparison between LEO satellites and \ac{HAPS} is generally more favorable for \ac{HAPS}, and the overall environmental burden is typically lower.
Meanwhile, sustainable models for HAPS operators emphasize multi-layer integration with satellites and \ac{TN} infrastructure, industry partnership, and green energy to achieve long-term viability~\cite{kuikel2025triplec}. Reliance on solar power supports UN Sustainable Development Goals 7 and 13, reducing carbon emissions and enabling eco-efficient operations in remote, disaster, and maritime scenarios. 
Collaborative models, such as those between Airbus, NTT DOCOMO, and SKY Perfect JSAT~\cite{NTT2022HAPSStudy} or SoftBank and Sceye~\cite{softbank2025pressrelease}, focus on direct-to-device services via HAPS-NTN hybrids, targeting air/sea connectivity and disaster response. These ensure sustainability through spectrum sharing, regulatory alignment, and revenue from government/defense contracts (e.g., surveillance, environmental monitoring).

While the global \ac{HAPS} market is clearly evolving and expanding across a broad range of applications, it remains nascent and fragmented. The limited availability of long-term historical data makes market projections highly sensitive to assumptions regarding adoption rates and regulatory developments.
According to MarkNtel Advisors, the \ac{HAPS} market is projected to grow from \$99 million in 2024 to \$240 million by 2030, driven by a diverse set of industrial applications~\cite{MarkNtelAdvisors2025HAPSMarket}. Published in March 2025, this report focuses primarily on core \ac{HAPS} platforms, excluding ancillary services, ground infrastructure, and broader ecosystem components from its scope.
In contrast, Global Insight Services estimates a significantly larger market expansion—from \$1.7 billion in 2025 to \$3.3 billion by 2035—anticipating growth supported by advances in materials science, solar power technologies, and regulatory facilitation~\cite{gis2026haps}. This assessment adopts a broader ecosystem perspective, encompassing not only the platforms themselves but also associated products, subsystems, and supporting equipment.
Despite differences in scope and valuation methodology, both reports converge on the conclusion that the communications segment constitutes the dominant share of the \ac{HAPS} market (approximately 35–45\%), followed by surveillance and navigation applications, with the Asia-Pacific region emerging as a leading growth area.
\section{HAPS Architectures}
\label{sec:Archi}
The integration of \acp{TN} and \acp{NTN} is critical to addressing the diverse connectivity challenges identified in Section~\ref{sec: Tech_use_case}, such as extending coverage to remote areas, enabling robust disaster communication, managing unexpected traffic surges, supporting \ac{IoT} applications, and facilitating advanced services such as mobile edge computing and \ac{AR}/\ac{VR}. To this end, \ac{NTN} platforms must efficiently transmit/receive communication signals between \acp{UE}, gateways, \ac{BS}, and core networks. While the recent progression of \ac{NTN} standardization is focusing satellites~\cite{kyung2025standardization}, we expect, by serving as dynamic platforms and connectors, \ac{HAPS} can play a key role in bridging \ac{NTN} platforms and creating a resilient, high-capacity, global communication infrastructure.

\subsection{NTN Architecture}
\begin{figure*}[!tb]
\centerline{\includegraphics[width=0.85\linewidth]{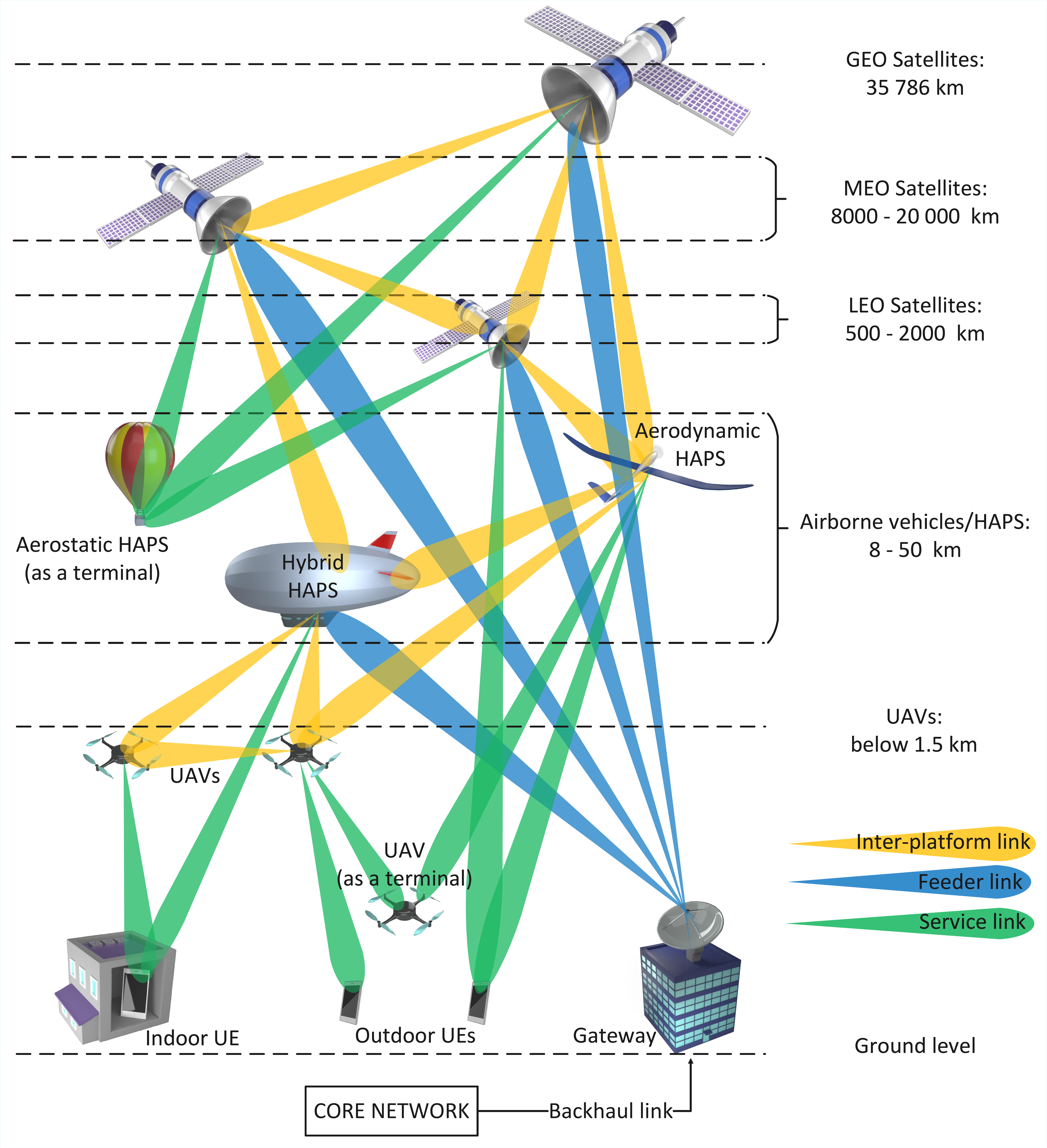}}
\caption{General NTN representation.}
\label{fig:general_NTN}
\end{figure*}

\acp{NTN} refers to the networks which are using satellites or uncrewed aerial platforms in various constellations as relay nodes or \acp{BS}. In general, \acp{NTN} (see Fig. \ref{fig:general_NTN}) include several key components, each playing a specific role in ensuring global, reliable communication. These components can be categorized into three main layers, namely space, aerial, and ground layers~\cite{hoyhtya2022sustainable}. The space layer comprises \ac{LEO}, \ac{MEO}, \ac{GEO} and even highly elliptical orbiting satellites, while the aerial layer consists of \ac{HAPS} and low-altitude \acp{UAV}. Satellites, \ac{HAPS}, and \acp{UAV} in this case are considered as network platforms, which carry the node able to facilitate communications. However, in certain use cases (for example, Earth observation satellites and \ac{HAPS} or delivery drones), they may serve as network terminals, i.e., connectivity consumers, providing no communication functionality.
The ground layer typically includes infrastructure that interacts with the space and aerial layers, such as gateway stations, core networks, and data centers. Considering \ac{TN}-\ac{NTN} integration, the different layers can be combined together to provide better coverage, capacity, resilience, and flexibility compared to standalone deployments~\cite{wang2021potential}.
The \ac{NTN} architecture, as defined by the \ac{3GPP} in \cite{TR38821}, supports two fundamental payload designs: transparent (or bent-pipe) and regenerative. In the transparent approach, the \ac{NTN} platform carries communication equipment that functions as a simple relay, forwarding signals without modification. In contrast, the regenerative architecture enables onboard signal processing at the \ac{NTN} platform, allowing the digital processing of the --- from packet data convergence protocol (PDCP) to \ac{PHY} layer --- to be performed directly within the platform. These architectural choices impact key factors such as spectral efficiency, rate, latency, and overall system complexity, shaping how \acp{NTN} integrate with \acp{TN}. Building on this foundation, we can distinguish three main network components within the \ac{NTN} architecture: 
(i) platforms, which carry communication equipment and implement either a transparent or regenerative payload to relay or generate communication signals; 
(ii) terminals, including direct \acp{UE} or specialized devices designed to interact with \ac{NTN} platforms; 
(iii) gateways, which connect \acp{NTN}, e.g., satellites and \ac{HAPS}, to core networks, facilitating data flow between \acp{TN} and~\acp{NTN}.

In Section~\ref{sub:ntn-components},
we will describe the different NTN components, including platforms, terminals, gateways, and their interactions, focusing on  \ac{HAPS}.
Further, in Section~\ref{sub:energy_modeling} we discuss the platform-level metrics for HAPS energy modeling.
At the same time, the \ac{NTN} architecture incorporates the following types of links~\cite{3gpp-tr-38.811}: service links, feeder links, backhaul links,  and inter-platform links, which we discuss is subsection~\ref{sub:ntn-bands}, as well as the promising frequency bands for \acp{NTN} based on the link types, with considerations related to range, throughput, latency, and interference sensitivity.
In Section~\ref{sub:3gg-architectures},
we will compare \ac{3GPP} bent-pipe and regenerative architectures, focusing on the role of \ac{HAPS}.
Finally, in Section~\ref{sub:trials},
we will discuss ongoing and already closed \ac{HAPS}-related projects, challenges they are facing and lessons which must be learned from the progress done so far.

%%%%%%%%%%%%%%%%%%%%%%%%%%%%%%%%%%%%%%%%%%%%%%%%%%%%%%%%%%
\subsection{HAPS Platforms and Terminals}
\label{sub:ntn-components}
%%%%%%%%%%%%%%%%%%%%%%%%%%%%%%%%%%%%%%%%%%%%%%%%%%%%%%%%%%

\begin{table*}[t]
\centering
\def\arraystretch{1.2}
\caption{Comparison of \ac{HAPS} platforms.}
\label{tab:haps_comparison}
\setlength{\tabcolsep}{6pt} % Adjusts column spacing
\begin{tabular}{|p{3cm}|p{4.3cm}|p{4.3cm}|p{4.3cm}|}
\hline
\textbf{Characteristic} & \textbf{Aerostatic} & \textbf{Aerodynamic} & \textbf{Hybrid} \\  
\hline
\textbf{Lift mechanism} &  
Buoyant force (helium/hot air) &  
Aerodynamic lift (fixed/rotary wings) &  
Combination of buoyant and aerodynamic forces \\  
\hline
\textbf{Mobility} &  
Low when operating stationary or drifting slowly; high with favorable wind patterns &  
High, regardless of wind direction; can reposition dynamically &  
Moderate (repositioning capability) \\  
\hline
\textbf{Payload capacity} &  
High &  
Low &  
Moderate (payload/mobility trade-off) \\  
\hline
\textbf{Energy efficiency} &  
High (minimal energy to stay~aloft) &  
Low (continuous energy for flight) &  
Moderate (depends on lift mechanics) \\  
\hline
\textbf{Weather resilience} &  
Sensitive to wind and turbulence &  
More resistant to adverse weather, better operational stability &  
Balanced \\  
\hline
\textbf{Deployment flexibility} &  
Requires ground infrastructure (hangars, airfields) in close proximity to the serving region &  
Can take off and land outside serving region (deployment flexibility) &  
Moderate (limited by mobility) \\  
\hline
\textbf{Operational duration} &  
Long  and persistent (weeks/months) &  
Variable (days/months), and may be solar-powered to extend flight life &  
Potentially long, depending on design and hybrid lift use \\  
\hline
\textbf{Development cost} &  
Low, due to simpler design and manufacturing &  
High, requiring advanced materials and engineering &  
Highest, due to the complexity of integrating multiple lift systems \\  
\hline
\textbf{Notable example} &  
Loon Balloons~\cite{loonproject} %Tactical \ac{HAPS} by CIRA~\cite{eurohaps2023pressrelease} 
&  
SoftBank’s ``Sunglider''~\cite{softbank2024pressrelease} &  
Tactical \ac{HAPS} by CIRA~\cite{eurohaps2023pressrelease} \\  
\hline
\end{tabular}
\end{table*}

Now, we shift our focus to the aerial layer, with a particular emphasis on \ac{HAPS}, the subject of this survey. We distinguish between NTN platforms, terminals, and gateways.

\subsubsection{Platforms}
\ac{HAPS} can be classified depending on the physical principle providing the lifting force: (i) aerostatic platforms, which rely on buoyant force to stay aloft, and include airships and balloons; (ii) aerodynamic platforms, which generate lift through fixed or rotary wings to maintain flight; (iii) hybrid platforms, which rely on a combination of aerostatic and aerodynamic effects at different stages of flight, such as the tactical \ac{HAPS} by CIRA~\cite{eurohaps2023pressrelease}.
All these types of \ac{HAPS} have their own benefits and drawbacks,
which we discuss in the following and are summarized in Table~\ref{tab:haps_comparison}. 

\paragraph{Aerostatic platforms}
Aerostatic platforms are less expensive to manufacture, launch, and operate compared to aerodynamic and hybrid systems. While prototyping and refining their products, the Loon project team was able to fill and launch a balloon into the stratosphere every 30 minutes with their designed and custom-built auto-launchers and keep it in the air for more than 300 days~\cite{loonproject}. Thus, by hovering over a specific region, aerostatic platforms can provide consistent coverage and are well-suited for persistent surveillance and localized communication needs. Since this platforms do not require forward speed to maintain flight, they only need power to hold or adjust their position. This results in potentially better power efficiency, and allows for heavier payloads compared to aerodynamic platforms. Furthermore, their large volume capacity enables them to carry more complex communication payloads, making aerostatic platforms particularly suitable for both short-term and long-term missions over specific regions. However, these platforms are more sensitive to atmospheric conditions, which may impact their operational stability and reliability. Additionally, launching and landing aerostatic platforms require specific facilities such as hangars and airfields located near the target region, which may limit their flexibility in deployment. However, the Loon project team demonstrated a method to navigate the balloons using dominant winds at different altitudes. This approach requires advanced predictive models to identify favorable wind patterns and intelligent algorithms to determine the most efficient flight paths through varying wind layers~\cite{loonproject}. This allows the launch facilities to be located away from the target region, but makes deployment reliant on wind patterns.

\begin{table*}[t!]
\centering
\def\arraystretch{1.2}
\caption{Comparison of \ac{NTN} terminals.}
\label{tab:ntn_terminals_comparison}
\setlength{\tabcolsep}{5pt} % Adjusts column spacing
\begin{tabular}{|p{3cm}|p{4.3cm}|p{4.3cm}|p{4.3cm}|}
\hline
\textbf{Characteristic} & \textbf{Portable consumer terminals} & \textbf{Vehicular terminals} & \textbf{Fixed \ac{UE} terminals} \\  
\hline
\textbf{Mobility} &  
Medium (handheld/wearable devices) &  
High (cars, trains, aircraft, \acp{UAV}) &  
Low (stationary gateways/sensors) \\  
\hline
\textbf{Power availability} &  
Limited (battery-powered) &  
Moderate to high (power sources) &  
High if fixed power supply available; limited for low-power \ac{IoT} sensors \\  
\hline
\textbf{Antenna} &  
Small, integrated, limited gain &  
Large, more powerful (e.g., VSAT, phased arrays) &  
High-gain dish or directional antennas for long-range communication; limited gain for low-power IoT sensors \\  
\hline
\textbf{Coverage \newline Handover requirements} &  
Seamless transition between \acp{TN} and \acp{NTN} for uninterrupted connectivity &  
Multi-band, multi-standard support for high-speed mobility across regions &  
Persistent connectivity with high-capacity backhaul support; limited data rates for low-power \ac{IoT} sensors\\  
\hline
\textbf{Use cases} &  
Wearables, first responder devices, personal communication (phones) &  
Connected vehicles, public transport, autonomous robots, \acp{UAV} &  
Broadband Internet, industrial \ac{IoT}, remote sensors \\  
\hline
\textbf{Network requirements} &  
Low latency for applications like gaming and video streaming &  
Low latency for safety-critical applications, high throughput for real-time~data &  
High throughput for broadband applications, low power for \ac{IoT} sensors \\  
\hline
\textbf{Deployment constraints} &  
Consumer-driven must be cost-effective and compact & 
Requires vehicle integration and external antennas &  
Fixed installation with higher infrastructure requirements \\  
\hline
\textbf{Example devices} &  
Smartphones, tablets, smartwatches, disaster relief radios &  
Connected cars, aircraft communication systems, surveillance drones &  
Satellite broadband dishes, industrial \ac{IoT} gateways, environmental sensors \\  
\hline
\end{tabular}
\end{table*}

\paragraph{Aerodynamic platforms}
Aerodynamic platforms offer significant advantages given their high flexibility,
allowing them to move easily between regions in response to communication demands or environmental changes. 
This adaptability reduces the need for multiple airfields 
and enhances their capability for precise positioning through specific flight patterns. 
Modern aerodynamic designs are generally more robust against adverse weather conditions than aerostatic platforms,
increasing their reliability in various operational scenarios.
Additionally, many modern aerodynamic platforms, 
such as SoftBank's ``Sunglider''~\cite{softbank2024pressrelease}, 
can be entirely solar-powered, 
enabling them to remain operational for extended periods, 
ranging from weeks to months. 
However, these platforms come with some limitations.
Their payload capacity was generally smaller compared to large aerostatic platforms,
which can affect the scale and complexity of their communication systems.
Moreover, the development and manufacturing costs of aerodynamic platforms are higher,
reflecting their advanced engineering and specialized materials.

\paragraph{Hybrid platforms}
Hybrid designs aim to combine the strengths of both aerostatic and aerodynamic platforms,
offering a balance between payload capacity, stability, and mobility. 
They can carry significant payloads like aerostatic platforms, 
while also benefiting from the mobility and weather resilience typically associated with aerodynamic platforms. 
This combination makes hybrid solutions suitable for a wide range of applications,
including scenarios requiring both high payload capacity and the ability to reposition dynamically.
However, these advantages come with notable challenges.
The complexity of hybrid designs makes them costly to develop and manufacture.
Sophisticated control systems are required to manage the dual mechanisms for lift and propulsion, 
adding to the operational complexity. 
Moreover, while hybrid platforms mitigate some disadvantages of pure aerostatic or aerodynamic designs, 
they often inherit certain drawbacks, 
such as higher sensitivity to adverse weather compared to fully aerodynamic platforms and increased infrastructure requirements compared to simpler aerostatic systems.

\subsubsection{Terminals}

Platforms ultimately aim at delivering connectivity to network terminals, a.k.a, \acp{UE}.
Given the broad spectrum of application scenarios outlined in Section~\ref{sec: Tech_use_case}, 
it is evident that \acp{UE} extends far beyond conventional smartphones, and also includes wearable devices, \ac{IoT} sensors, robots, and \acp{UAV}.
Terminals can be classified in terms of features like form factor, operational environment, and communication capabilities.
In the following paragraphs,
we will describe  the main groups of terminals, namely portable consumer terminals, vehicular terminals, and fixed terminals, as well as their features, and specific requirements, as summarized in Table~\ref{tab:ntn_terminals_comparison}.

\paragraph{Portable consumer terminals}
Probably the largest group of \acp{UE} consists of portable consumer terminals, 
including smartphones, smartwatches, tablets, and laptops. 
These devices are designed for everyday use across urban, suburban, and rural areas,
and require seamless handover between \acp{TN} and \acp{NTN} to ensure uninterrupted connectivity. 
The diverse range of capacity-demanding and latency-sensitive  applications, 
such as gaming and video streaming, 
necessitates support for multiple frequency bands, 
compatibility with existing standards (e.g., \ac{5G}), 
and \ac{NTN}-specific adaptations for satellite and \ac{HAPS} links.
Conventional mobile devices are further complemented by \textit{wearable and implantable terminals}, 
such as fitness trackers, health monitors, and medical implants. 
These devices are typically ultra-portable, low-power, and designed for continuous operation. 
They often require only short-distance, low-data-rate communication with a \ac{UE}’s hub device, 
such as a smartphone, 
relying on the ubiquitous coverage provided by the hub.
Another significant group within portable terminals includes communication devices for first responders, disaster relief teams, and military personnel. 
These devices are specifically designed to operate in extreme conditions, 
providing reliable connectivity in areas where terrestrial infrastructure is damaged or nonexistent.
They often require enhanced security and encryption features for sensitive communications, 
as well as high-capacity links to enable rapid data exchange in critical scenarios.

\paragraph{Vehicular terminals}
Another distinct category of mobile devices comprises vehicular terminals, installed in cars, buses, trains, ships, and aircraft. These terminals operate under challenging conditions due to high mobility and diverse environments, including remote and maritime regions. Unlike handheld devices, they can accommodate more efficient power sources and advanced antennas, such as \ac{VSAT} parabolic reflectors or phased array antennas.
Typically, \ac{VSAT} systems feature parabolic dishes ranging from 0.75 m to 2.4 m in diameter, with some larger models, providing high-gain, narrow-beamwidth links ideal for point-to-point communications. The antenna size depends on required gain, operating frequency, and environmental factors. In contrast, phased arrays, such as those employed in Starlink or BGAN systems, are more compact and rely on electronic beamforming through controlled phase shifts across multiple radiating elements. This enables rapid, precise beam steering without mechanical movement, offering enhanced flexibility and robustness under vibration or harsh conditions. Modern designs integrate features like auto-acquisition, self-alignment, and automatic aiming, simplifying deployment and ensuring reliability during high-speed operations.
A related subset of vehicular devices includes unmanned vehicles, such as UAVs, self-driving cars, autonomous robots, and surveillance balloons. These systems combine high mobility with operational versatility across urban and remote areas. They require reliable, low-latency command-and-control links and high-capacity channels for real-time data transmission (e.g., video or telemetry). To maintain continuous connectivity beyond terrestrial coverage, NTN support—via platforms like LEO satellites or HAPS—is essential for ensuring global reach and service continuity.

\paragraph{Fixed UE terminals}
Despite the widespread use of mobile \acp{UE}, fixed~\acp{UE} remains an essential category of network terminals. These devices are primarily designed for stationary broadband access in rural and remote areas. Their antenna systems are similar to those used in vehicular \ac{VSAT} terminals. Unlike mobile devices, however, they are not constrained by strict form-factor limitations, allowing for larger, more efficient antenna designs. Such equipment allows high throughput for broadband applications, the ability to support multiple simultaneous connections, and resilience to adverse weather conditions, particularly for satellite-based links.

A distinctly different subset of fixed terminals comprises \ac{IoT} devices, including sensors, indoor consumer smart household appliances and industrial equipment. While industrial equipment and household appliances can be connected through industrial gateways or indoor access points, remote sensors rely on their own connectivity solutions. These sensors are typically compact, low-power, and deployed in large numbers. They require \ac{mMTC} connectivity, often in remote or hard-to-reach locations, and energy-efficient communication protocols to maximize battery~life.
This diverse range of \ac{UE} requirements in integrated \ac{TN}-\ac{NTN} systems highlights the critical role of platform adaptability and technology integration. 
Meeting these diverse needs necessitates a collaborative ecosystem of terrestrial,  satellite platforms, and \ac{HAPS}, each uniquely suited to specific terminal categories and operating conditions. Specifically, the \ac{TN} provides the best capacity and latency for densely populated areas, but is constrained in coverage and mobility support in remote regions, while satellites are indispensable for remote and hard-to-reach areas, but suffer from higher latency and lower rates. At the same time, \ac{HAPS} serves as a versatile middle ground, offering good coverage, capacity, latency, and mobility, especially in semi-remote areas or regions requiring temporary connectivity.

\subsubsection{Gateways}
Aside from \ac{NTN} platforms, 
several essential components constitute the network infrastructure,
including gateways and associated network elements that ensure connectivity, reliability, and efficient communication~\cite{wang2022performance}. 
Gateways,
often referred to as ground stations, 
serve as critical interfaces between \acp{TN} and \acp{NTN}.
Strategically located to optimize coverage and capacity and minimize latency, 
gateways are typically unconstrained in terms of form factor and power consumption.
They are equipped with a large antenna structure designed for higher frequency bands, and are capable of maintaining efficient uplink connections with \ac{NTN} platforms.
Depending on the overall architecture,
gateways may perform functions such as signal processing, routing, and resource management.
They also aggregate traffic, and facilitate integration of \ac{NTN} components with the terrestrial core network.

%%%%%%%%%%%%%%%%%%%%%%%%%%%%%%%%%%%%%%%%%%%%%%%%%%%%%
\subsection{Cross-Platform Energy Modeling for HAPS}
\label{sub:energy_modeling}
%%%%%%%%%%%%%%%%%%%%%%%%%%%%%%%%%%%%%%%%%%%%%%%%%%%%%
A single energy-model framework that separates (i) solar energy harvesting, (ii) storage restrictions, and (iii) platform/payload power budgets is necessary for a consistent comparison of \ac{HAPS} types (such as fixed-wing solar aircraft and aerostatic/airship systems). The reported energy efficiency and endurance findings are not directly comparable across research due to the lack of established measures and assumptions.

Solar harvesting is commonly represented as a function of solar irradiance $G(t)$ (W/m$^2$), conversion efficiency $\eta_{\mathrm{PV}}$ (\%), and effective photovoltaic area $A_{\mathrm{PV}}$ (m$^2$). The harvested power can be expressed as
\begin{equation}
P_{\mathrm{harv}}(t) = \eta_{\mathrm{PV}} \, A_{\mathrm{PV}} \, G(t)
\end{equation}
using temporally integrated daily gathered energy. To present harvested energy in kWh/day and facilitate cross-platform normalization, fixed-wing solar \ac{HAPS} experiments explicitly parameterize these values under latitude- and season-dependent irradiance circumstances \cite{arum2020energy}. Similar irradiance-driven formulations are used in system-level \ac{HAPS} energy studies, confirming their applicability as a platform-independent reference model \cite{abderrahim2023data}.

Storage limits are characterized using the storage capacity $E_{\mathrm{sto}}$ (kWh), the usable fraction (depth-of-discharge and round-trip constraints), and the equivalent sustainment time under a defined load. The storage sustainment time is given by

\begin{equation}
T_{\mathrm{sto}} = \frac{E_{\mathrm{use}}}{P_{\mathrm{tot}}}
\end{equation}
where $P_{\mathrm{tot}}$ denotes the total platform power demand. Studies on the viability and power budget of high-altitude airships show how storage sizing relates to propulsion or station-keeping requirements and payload loads. These works also provide guidance for consistent reporting of storage capacity and endurance across platform types \cite{colozza2004initial, choi2006power}.

Payload power budgets are most comparable when decomposed into communications payload power $P_{\mathrm{pl}}$, propulsion or station-keeping power $P_{\mathrm{prop}}$, and auxiliary/avionics power $P_{\mathrm{aux}}$. The total platform power consumption can therefore be written as

\begin{equation}
P_{\mathrm{tot}} = P_{\mathrm{pl}} + P_{\mathrm{prop}} + P_{\mathrm{aux}}
\end{equation}

This decomposition enables comparison between propulsion-dominant fixed-wing platforms and station-keeping-dominant aerostatic systems. In particular, the payload power share $\left(P_{\mathrm{pl}}/P_{\mathrm{tot}}\right)$ can be used as a normalized efficiency indicator \cite{arum2020energy, choi2006power}.

To ensure fair interpretation for TN--NTN integration research, these platform-level metrics should be reported alongside altitude, endurance targets, and irradiance assumptions (latitude and season). Furthermore, linking platform power budgets to network-level effects—such as reducing terrestrial densification requirements or supporting traffic spikes—strengthens the relevance of energy comparisons in integrated architectures \cite{kement2023sustaining}. Consequently, a transparent and repeatable basis for cross-platform evaluation can be achieved by presenting solar harvesting characteristics, storage capacity and sustainment, and complete power budget breakdowns within a unified comparison table.

%%%%%%%%%%%%%%%%%%%%%%%%%%%%%%%%%%%%%%%%%%%%%%%%%%%%%%%%%%
\subsection{NTN Links and Frequency Bands}
\label{sub:ntn-bands}
%%%%%%%%%%%%%%%%%%%%%%%%%%%%%%%%%%%%%%%%%%%%%%%%%%%%%%%%%%

\begin{table*}[t!]
    \centering
    \renewcommand{\arraystretch}{1.2}
        \caption{Comparison of frequency bands for NTN, in terms of range, throughput, latency, and interference sensitivity.}
    \label{tab:frequency_band_comparison}
    \begin{tabular}{|p{4cm}|p{2.5cm}|p{2.7cm}|p{2cm}|p{4.5cm}|}
        \hline
        \textbf{Band} & \textbf{Typical range} & \textbf{Peak throughput} & \textbf{Latency profile} & \textbf{Interference sensitivity} \\
        \hline
        \textbf{Sub-7 GHz} (410 - 7125 MHz) & Hundreds of km & Low ($<$1 Gbps) & Low & Low (robust propagation) \\
        \hline
        \textbf{cm-Wave} (7.125 - 24.25 GHz) & Tens–hundreds of km & Medium (1–10 Gbps) & Medium & Moderate (rain fade in Ku-band) \\
        \hline
        \textbf{mm-Wave} (24.25 - 52.6 GHz), \textbf{FR2-2} (52,6 - 71 GHz) & $<$10 km & High (10–100 Gbps) & Medium & High (severe rain attenuation) \\
        \hline
        \textbf{Sub-THz} (100 - 300 GHz) & $<$5 km & Ultra-high (100+ Gbps) & Low & Very high (strong atmospheric loss) \\
        \hline
        \textbf{THz} (0.1 - 10 THz) & $<$1 km & Tbps-class & Very low & Extreme (highly fragile links) \\
        \hline
        \textbf{FSO} (3 THz - 30 PHz) & Up to hundreds of km & Ultra-high (Tbps) & Very low & Weather-dependent (fog/clouds) \\
        \hline
    \end{tabular}
\end{table*}

To establish reliable \ac{NTN} communication,  
various types of links facilitate data exchange between different network components, namely feeder, service, inter-platform, and backhaul links.
These links differ in terms of function, propagation characteristics, and performance requirements,  influencing the choice of frequency bands and transmission technologies.
\begin{itemize}
\item 
{Service links} provide connectivity between \ac{NTN} platforms and \acp{UE}.  
These links must accommodate a diverse types of devices, ranging from stationary to highly mobile terminals, as described in Section~\ref{sub:ntn-components}, 
posing challenges such as signal power constraints, Doppler effects, and interference management.  
\item {Feeder links} connect \ac{NTN} platforms directly to gateways,  
typically operating at high frequencies to enable high-capacity transmission.  
To ensure reliability, these links incorporate advanced error correction and signal processing to mitigate atmospheric disruptions.  

\item {Inter-platform links} facilitate direct communication between \ac{NTN} platforms,  
either within the same layer (e.g., satellite-to-satellite) or across different layers  
(e.g., satellite-to-\ac{HAPS} or \ac{HAPS}-to-\ac{UAV}).  
By supporting direct routing and handover, these links reduce dependency on terrestrial infrastructure.  
\item {Backhaul links} interconnect \ac{NTN} gateways with core network facilities,  
typically utilizing fiber-optic infrastructure to ensure high-speed, low-latency data transport.  
\end{itemize}

%\subsubsection{Mapping \ac{NTN} Links to Frequency Bands}

The performance of these links is heavily influenced by the choice of operating frequency bands.  
Each frequency range offers distinct propagation characteristics,  
affecting factors such as coverage, capacity, and resilience to atmospheric conditions.  
The introduction of \ac{5G} marked a significant shift in spectrum allocation,  
expanding the use of high-frequency bands such as in the \ac{mmWave} spectrum to enhance network capacity.  
In parallel, \acp{NTN} operates in highly heterogeneous environments,  
requiring careful spectrum planning to address diverse communication needs.  
As \ac{6G} and future \ac{TN}-\ac{NTN} integration evolves,  
the selection of appropriate frequency bands will be critical in optimizing system performance.  

%David: Redo this paragraph indicating the bands specfied so far for NTNs. https://www.everythingrf.com/community/ntn-frequency-bands
Notably, in Release 17
the \ac{3GPP} formally integrated \ac{NTN} into the 5G \ac{NR} framework,  
defining use cases within the sub-7 GHz spectrum,  
particularly for massive access and narrowband \ac{IoT} applications.  
Release 18 further expanded \ac{NTN} support,  
introducing enhanced 5G \ac{NR} operations for handheld terminals in the 10 GHz and above range.  
These updates focused on improved spectrum management and coexistence strategies  
to ensure seamless integration between \acp{TN} and \acp{NTN}~\cite{3gpp-tr-38.811, TR38821, TR38741}.  

The following subsections examine the key frequency bands currently used in \ac{NTN} communications and those with potential for future adoption depending on the type of link, as summarized in  
Table~\ref{tab:frequency_band_comparison}.

\subsubsection{Sub-7 GHz}
Also known as \ac{3GPP} \ac{FR} 1, 
this well-established frequency range, 
spanning from 410\,MHz to 7125\,MHz~\cite{3gpp-tr-38.104},
has been widely used in \ac{LTE} and \ac{NR} networks.
\ac{FR}1 is valued for its excellent propagation characteristics,
resilience to atmospheric conditions,
and an extensive ecosystem of mature telecommunication equipment. 
An important advantage of this band is the possibility to establish direct satellite-to-\ac{UE} links,
as it was recently demonstrated by companies like AST SpaceMobile, Starlink and Lynk Global.
However, its limited bandwidth and high congestion restrict the achievable data rates,
making it less suitable for high-capacity applications.

\subsubsection{mmWave}
Referred to \ac{3GPP} \ac{FR}2-1,
\ac{mmWave} spans frequencies from 24.25 GHz to 52.6 GHz while \ac{3GPP} \ac{FR}2-2 designated for frequencies from 52.6 GHz to 71 GHz~\cite{3gpp-tr-38.104}.
Given its large bandwidth and short \ac{OFDM} symbols,  
this spectrum enables ultra-high data rates and extremely low latency,  
making it ideal for bandwidth-intensive applications such as \ac{eMBB}.
\ac{mmWave} also offers fiber-like speeds in fronthaul and backhaul communications,
supporting network densification through small cells and high-directivity antennas.

Despite its advantages, \ac{mmWave} has seen limited adoption in \acp{TN} due to its short range and high infrastructure costs.
The severe free-space path loss, combined with high susceptibility to blockage from buildings, foliage, and human movement,
requires a dense network of \acp{BS},
making large-scale deployment expensive.
While small cell and advanced beamforming techniques based on \ac{MIMO} helps to mitigate these limitations,
deployment remains challenging,
particularly in rural and remote areas~\cite{chiaraviglio2017bringing}.
The use of \ac{mmWave} in \acp{NTN} for broadband connectivity has recently gained significant research attention.
The studies~\cite{giordani2020non,giordani2020satellite} explored \ac{NTN} applications in \ac{6G},
evaluating aerial- and space-based \ac{mmWave} links to provide connectivity for ground terminals under various network configurations.
However, expanding \ac{NTN} into \ac{mmWave} bands requires effective spectrum-sharing techniques to balance isolation, licensing costs, and coexistence with existing \acp{TN}.
Moreover, large-scale propagation models must be re-evaluated,
considering factors such as path loss, penetration loss (also through the atmosphere), and shadow fading,
which significantly impact \ac{NTN} protocol design~\cite{rappaport2017overview, shafi2018microwave}.

\subsubsection{cmWave}  
Also known as \ac{3GPP} \ac{FR}3,  
\ac{cmWave} spans 7.125\,GHz to 24.25\,GHz,  
bridging the gap between sub-7\,GHz and \ac{mmWave} bands~\cite{3gpp-tr-38.820}.  
Unlike \ac{mmWave} band, 
which suffer from severe propagation characteristics,  
\ac{FR}3 offers improved coverage and efficiency,  
with channel properties closer to sub-7\,GHz in the sub-12\,GHz region than to \ac{mmWave}~\cite{cui20236g, kang2024cellular, meyne2024antenna, shakya2024wideband}.
Propagation studies~\cite{Huawei2024cmWave} indicate that \ac{FR}3 communication at 10\,GHz experiences approximately 9.5\,dB higher path loss in \ac{LoS},  
and 9.7\,dB in non-\ac{LoS}, compared to 3.5\,GHz;  
yet these losses remain manageable with advanced beamforming. 
This band also exhibits lower penetration loss than at \acp{mmWave},  
making it a strong candidate for outdoor-to-indoor coverage~\cite{kang2024cellular}. 
Additionally, rain attenuation is negligible below 10\,GHz,  
and reduced delay spread improves signal quality,  
while angular spreads remain consistent with sub-6\,GHz deployments.  
A key challenge in \ac{FR}3 networks is the coexistence with satellite services,  
particularly in the 12\,GHz band,  
which is heavily utilized by \ac{LEO} constellations such as Starlink~\cite{cui20236g}.  
Interference mitigation techniques,  
including advanced beamforming and ephemeris-based coordination,  
have been proposed to ensure harmonious spectrum sharing between terrestrial and satellite networks~\cite{kang2024cellular}.  
Dynamic spectrum access mechanisms and interference nulling strategies are also being explored to optimize spectral efficiency  
while maintaining reliable satellite communications~\cite{kang2024terrestrial}.  

\subsubsection{Beyond-100\,GHz: sub-THz and THz}
\label{subsec:beyond100}
The beyond-100\,GHz region is covering sub-\ac{THz} typically defined between 100 and 300\,GHz~\cite{atzeni2025subthz, chukhno2024models} and even wider \ac{THz} range (0.1–10\,\ac{THz}).
It offers extreme bandwidth for immersive communications and sensing \cite{saeed2021terahertz,jiang2024terahertz}.
Measurements at 140\,GHz report 2.3\,Gbps over 120\,m,
highlighting the rate–range trade-off and equalization needs \cite{molisch2024properties}. 
Simulation studies also show that \ac{ISAC} in these bands can outperform 5G-spectrum baselines \cite{katwe2024cmwave}.
For NTN topologies, sub-\ac{THz} has been proposed for backhaul between \ac{UAV}-\acp{BS} and \acp{CPU}, 
with \ac{HAPS} acting as the \ac{CPU} to preserve \ac{LoS} \cite{abbasi2024uxnb}. 
Feasibility analyses indicate that \ac{THz} feeder/backhaul links around 10\,Gbps are possible in light rainfall given careful link budgets and signal processing \cite{tamesue2024thz}
Reconfigurable intelligent surfaces (\acp{RIS}) can further alleviate blockage \cite{amodu2024technical}.
Despite this potential, 
beyond-100\,GHz faces severe constraints for NTNs:
high free-space path loss, strong water-vapor absorption, weak diffraction, and hardware limits (low output power, thermal stability) restrict range and robustness \cite{babay2022sub,song2021terahertz}.
Consequently, sub-\ac{THz}/\ac{THz} should be viewed as complementary, 
beyond-5G options suited mainly to short-range, \ac{LoS} inter-platform or backhaul links under controlled conditions—not for handheld \textit{service} or long-range \textit{feeder} links. 
Practical deployment will require multi-link coordination (e.g., fallbacks to FR2/FR3 or \ac{FSO}), highly directive arrays with null steering, site/spatial diversity, and aggressive modulation/coding adaptation \cite{saeed2021terahertz,jiang2024terahertz}.

\subsubsection{FSO}
\ac{FSO} frequency region covering an ultra-wide range of  spectrum, roughly, between 3\,THz and 30\,PHz, which include infrared, visible light, and ultraviolet categories and basically overlap with \ac{THz} region. It is worth noting, however, that unlike \ac{RF} bands, no single bounded range is allocated for \ac{FSO} by regulations, and discrepancies in the reported values may therefore occur.
\ac{FSO} systems offers a broad range of indoor and outdoor applications including cyber security, surveillance and monitoring, inter-building communication, high-capacity backhauling, etc.~\cite{JAHID20221contemporary}.
\ac{FSO} technology is a valuable approach for future wireless networks,
offering cost-effectiveness, easy deployment, high bandwidth, and strong security.
Samir~\textit{et al.}~\cite{al2020survey} provided an in-depth analysis of \ac{FSO} systems,  
highlighting their role in next-generation optical communication.  
However, \ac{FSO} propagation is challenged by atmospheric turbulence and alignment issues,  
necessitating mitigation strategies to maintain link stability.  
Shang~\textit{et al.}~\cite{shang2024enhancing} evaluated a ground-to-\ac{HAPS}-to-\ac{RIS}-to-\ac{UE} communication system,  
where \ac{FSO} is used for the ground-to-\ac{HAPS} link, while \ac{RF} communication is established for \ac{HAPS}-to-\ac{RIS}-to-\ac{UE} transmissions.
Key performance metrics, such as ergodic capacity, average \ac{BER}, and outage probability, are analyzed using the bivariate Fox-H function.  
Results indicate that pointing errors and shadowing significantly affect system performance,  
with larger zenith angles and severe shadowing leading to degradation.  
Heterodyne detection improves efficiency by enhancing capacity and reducing \ac{BER} and outage probability.

Based on the above discussion, 
it is important to recognize that power limitations, operational altitude, and environmental factors play a crucial role in \ac{HAPS} communication.  
Selecting the appropriate frequency band for each link requires balancing multiple considerations:
\begin{itemize}  
    \item \text{Power constraints:}  
    Higher frequency bands typically require more power for transmission,  
    which can be challenging given the strict power budgets of \ac{HAPS}.  

    \item \text{Weather sensitivity:}  
    Atmospheric conditions, e.g., rain and turbulence,  
    impact high-frequency bands more significantly than lower ones,  
    affecting link reliability and~\ac{QoS}.  

    \item \text{Coverage vs. capacity:}  
    Lower frequencies (e.g., sub-7 GHz and \ac{cmWave}) offer wider coverage (with indoor penetration) with lower throughput,  
    while higher frequencies (e.g., \acp{mmWave}, \ac{THz}, \ac{FSO}) provide higher capacity but require precise beamforming.  
\end{itemize}  

The design of efficient \ac{HAPS}-based \acp{NTN} requires balancing trade-offs between coverage, capacity, and resilience.  
A hybrid multi-band approach can enhance adaptability by leveraging different frequency bands for specific link requirements. For example, (i) service links may operate at lower frequencies (sub-7 GHz, \ac{cmWave}), to provide broad coverage and reliable signal propagation~\cite{3gpp-tr-38.811, TR38821, TR38741}; (ii) feeder links may operate at \ac{cmWave} and \ac{mmWave} bands, to enable high-capacity, long-range transmission~\cite{lou2023haps}; (iii) Inter-platform and backhaul links require ultra-high-speed data transfer, and may operate at sub-\ac{THz}, \ac{THz}, and \ac{FSO} frequencies for seamless network integration.
This multi-band, hybrid spectrum strategy ensures a balance between coverage, capacity, and system resilience, supporting the development of robust \ac{HAPS}-assisted \ac{NTN} architectures.

\begin{figure}[!tb]
\centerline{\includegraphics[width=0.5\textwidth]{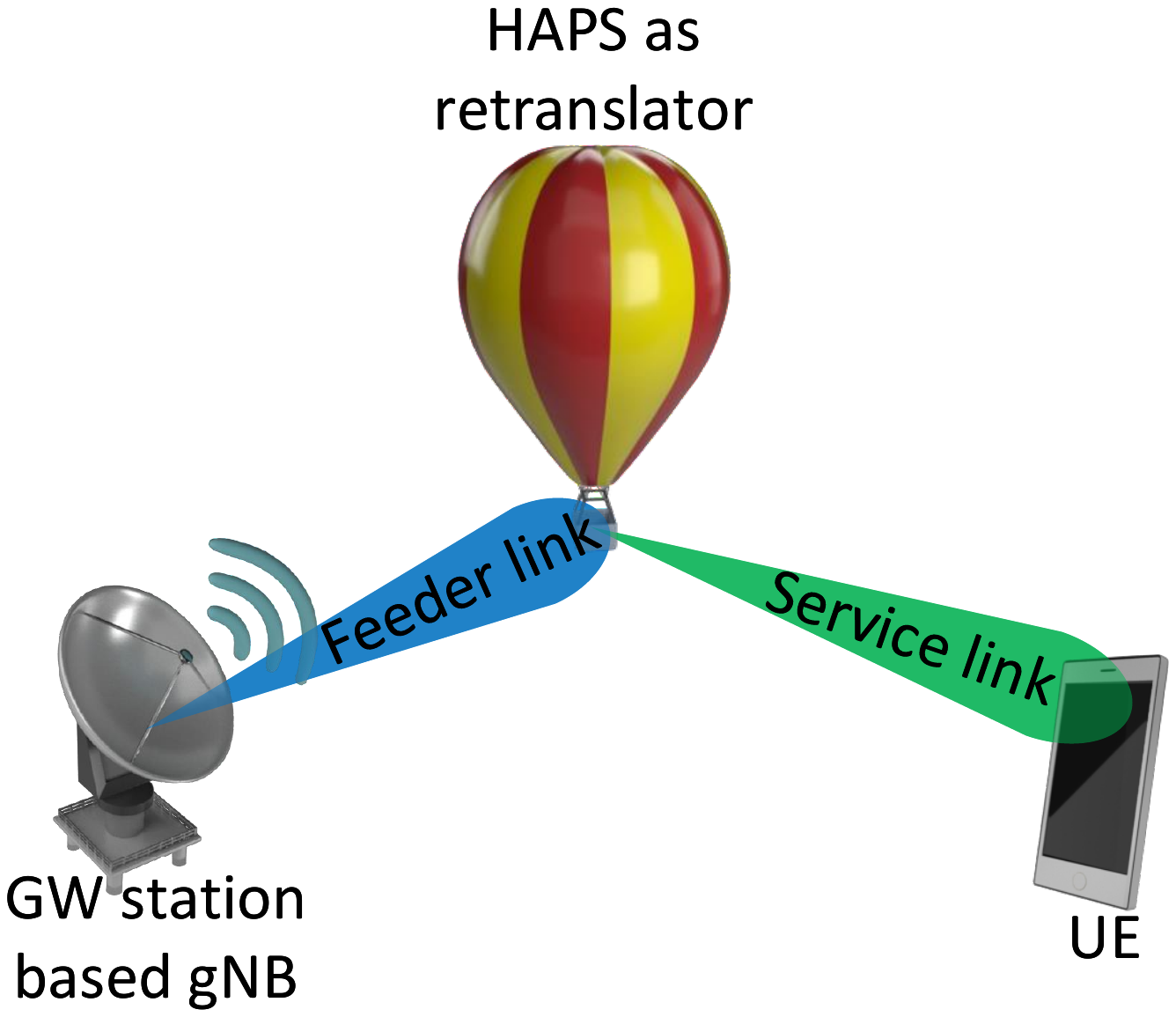}}
\caption{Transparent NTN architecture.}
\label{fig:transparentNTN}
\end{figure}
%%%%%%%%%%%%%%%%%%%%%%%%%%%%%%%%%%%%%%%%%%%%%%%%%%%%%%%%%%
\subsection{3GPP Specified Architectures}
\label{sub:3gg-architectures}
%%%%%%%%%%%%%%%%%%%%%%%%%%%%%%%%%%%%%%%%%%%%%%%%%%%%%%%%%%
As introduced earlier, in the \ac{3GPP} \ac{NTN} framework~\cite{TR38821}, two primary platform payload configurations are defined: transparent (or bent-pipe) and regenerative. The choice of payload configuration determines the \ac{NTN} architecture at a fundamental level, shaping its operational behavior. 
In the transparent architecture (see Fig. \ref{fig:transparentNTN}), 
the platform acts as a transparent relay, forwarding signals directly without significant processing.
A bent-pipe payload typically implements frequency conversion, analog filtering, and signal amplification. This straightforward design results in reduced payload complexity, lower power consumption on the platform, and generally lower development and deployment costs. However, the absence of onboard signal processing limits the system's ability to handle dynamic traffic or execute advanced processing tasks. These functions are offloaded to the ground station, leading to a round-trip signal path (from the \ac{UE} to the \ac{NTN} platform and then to the ground station) which inherently increases latency. 

\begin{figure}[!tb]
    \centerline{\includegraphics[width=0.5\textwidth]{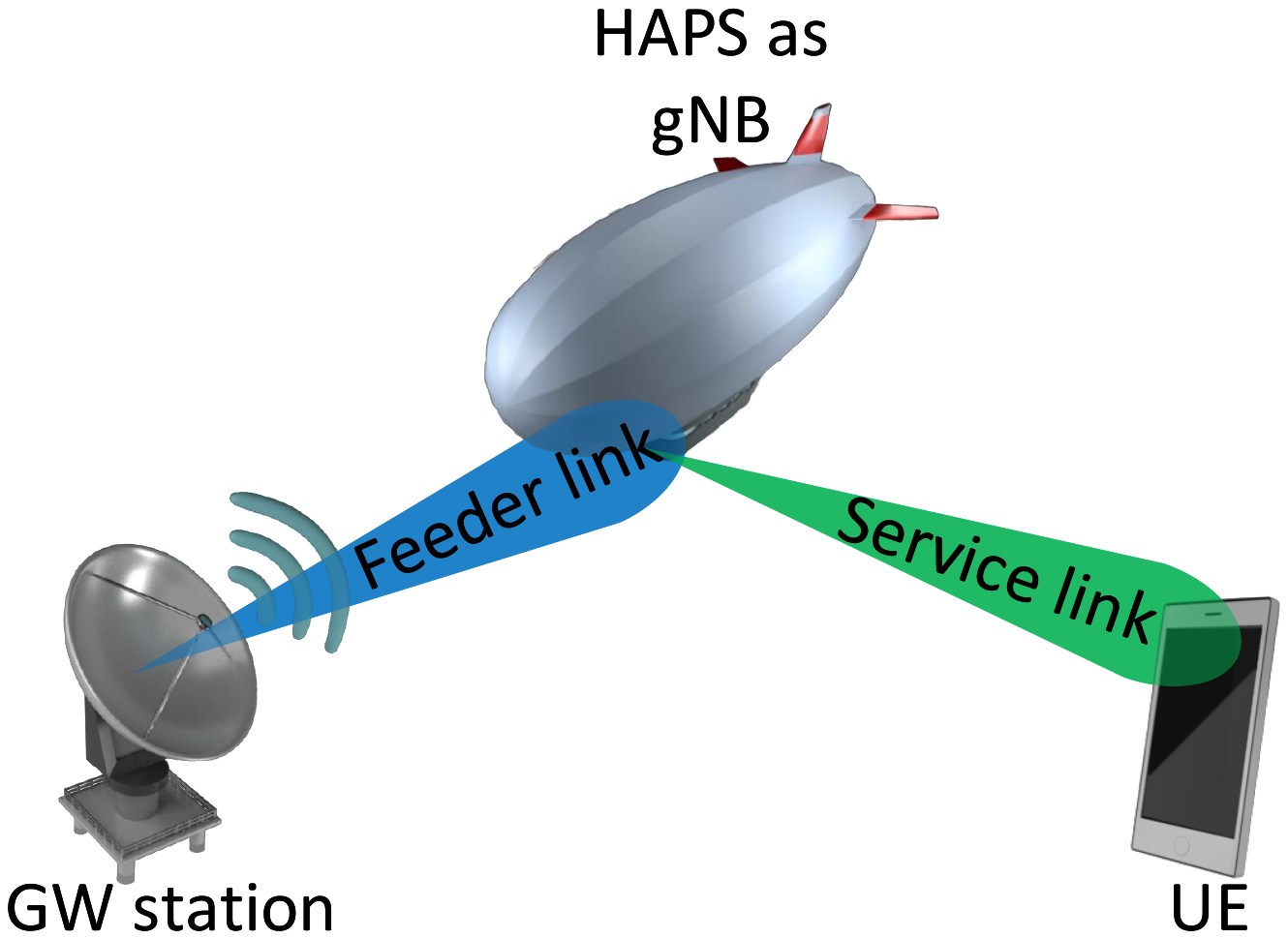}}
    \caption{Regenerative NTN architecture.}
    \label{fig:regenNTN}
\end{figure}

The regenerative architecture (see Fig. \ref{fig:regenNTN}) 
involves onboard signal processing, including tasks such as demodulation, decoding, and routing, in addition to the functions performed in the bent-pipe configuration. This onboard processing capability enables dynamic traffic management, such as load balancing based on \ac{QoS} requirements. Platforms with sufficient computational resources can execute some \ac{UE} tasks independently of the ground station, significantly reducing latency for time-sensitive applications. However, this capability comes at the cost of higher payload complexity and power consumption. The advanced technology required for regenerative systems also increases development and operational expenses.

Both transparent and regenerative \ac{NTN} configurations can be implemented in multiple ways, depending on the functionalities assigned to the platform. In a \emph{transparent} (bent-pipe) architecture---used in 3GPP Rel-17 and Rel-18---the \ac{NTN} platform acts as a radio repeater (with or without beam control) while the gNB remains on the ground; this limits payload complexity and enabled early deployment. Notably, the Rel-17 “Satellite Access Node” definition in 3GPP RAN4 still allowed some distribution of gNB functions between Earth and space while keeping RAN4 RF/performance specifications applicable. In contrast, Rel-19 introduces a \emph{regenerative} (packet-processing) architecture in which a complete gNB is placed on the platform~\cite{xing2021high}. This enables native use of the $X_n$ interface over inter-satellite links (inter-gNB mobility, resource coordination, energy saving), on-payload packet switching and routing/handovers directly between satellites with minimal ground involvement, and reduced $U_u$ round-trip time (e.g., for random access and hybrid automatic repeat request). Regenerative payloads also provide the building block for Rel-19 “store-and-forward” (S\&F) for IoT \acp{UE}, allowing continued operation if the feeder link is severed and even supporting UE-to-UE termination in space. These gains come at the expense of higher payload power and complexity. Earlier alternatives with only a gNB-distributed unit onboard (gNB-central unit on the ground) were found suboptimal due to heavy F1 signaling over the feeder link, lack of a standardized direct inter-satellite interface, and limited support for feeder-link switching and S\&F. Consensus therefore converged on the full gNB-on-board approach.

Regardless of the chosen architecture, \ac{NTN} platforms can be integrated into a multi-connectivity (MC) framework for uplink, downlink, or bidirectional communication~\cite{TR38821}. In this context, a \ac{UE} can connect simultaneously to both an \ac{NTN} platform and a terrestrial access point, or to multiple \ac{NTN} platforms. This approach improves throughput, enhances reliability, and ensures seamless service continuity in heterogeneous service areas.

The role of \ac{MC} in integrated \ac{TN}-\ac{NTN} systems depends on the operational objective. Beyond conventional dual- or multi-node connectivity, secondary MC nodes may be activated on demand to boost throughput. For example, Majamaa, Martikainen, Sormunen, and Puttonen~\cite{majamaa2022multiconnectivity} proposed a secondary-node activation algorithm, demonstrating a 9.1\% increase in maximum average per-\ac{UE} throughput compared to systems without MC, and a 5.3\% improvement over baseline algorithms.

In addition, MC enables dynamic traffic splitting between terrestrial and non-terrestrial nodes, allowing the network to balance load across layers and offload traffic from congested terrestrial nodes to \ac{NTN} links. Alam~\textit{et al.} \cite{alam2024ontherole} introduced a framework for UE association, BS transmit power and activation, and bandwidth allocation between terrestrial and non-terrestrial tiers. Simulation results showed a 45\% reduction in total power consumption and an approximately 250\% increase in average throughput compared to standard \ac{3GPP}-based network models.

Within integrated \ac{NTN} systems, \ac{HAPS} can assume diverse roles: from serving as terrestrial network extensions or satellite signal repeaters to operating as full-scale network centers equipped with baseband units and edge data processing capabilities. These roles directly impact latency, energy efficiency, and coverage performance. For instance, Erdogan~\textit{et al.}~\cite{erdogan2022cooperation} proposed a quasi-stationary HAPS to enhance inter-satellite connectivity via \ac{FSO} links with \ac{LEO} satellites. Simulations under extreme conditions, such as volcanic activity, demonstrated the feasibility of supporting optical inter-satellite communication.
Experimental evaluations further confirm the versatility of HAPS-based systems. Hasegawa, Konishi, Ohta, and Nagate~\cite{hasegawa2023experimental} used an automatic gain control system, which enabled a HAPS wireless repeater compatible with 5G NR to maintain stable transmission at maximum power, ensuring wide-area coverage. Similarly, Lou~\textit{et al.}~\cite{lou2023haps} analyzed different HAPS-based link architectures and showed that HAPS can reduce latency in UAV ad-hoc networks, enhance energy efficiency in UAV cell-free architectures, improve coverage probability in satellite cell-free networks, and expand the service area of macro base stations through integrated access and backhaul designs.

Nevertheless, real-world data highlight operational limitations. Empirical analysis of Loon project flight tracking data revealed that while stratospheric balloons offer a cost-effective, delay-tolerant connectivity service, outages can last several hours, limiting the achievable service stability compared to planned terrestrial networks. Such constraints may explain the current trend toward fixed-wing HAPS designs in modern projects, which promise greater reliability and sustained performance~\cite{serrano2023balloons}.

%%%%%%%%%%%%%%%%%%%%%%%%%%%%%%%%%%%%%%%%%%%%%%%%%%%%%%%%%%
\subsection{HAPS-Based Projects}
\label{sub:trials}
%%%%%%%%%%%%%%%%%%%%%%%%%%%%%%%%%%%%%%%%%%%%%%%%%%%%%%%%%%
Over the past decades, numerous projects have explored various applications of \ac{HAPS}, particularly in communication, but also in fields like surveillance, disaster management, and environmental monitoring. While some of these projects achieved promising results, many encountered significant challenges that led to their closure, even at advanced stages of development. These challenges included technological difficulties, unsustainable financial models, or regulatory hurdles.

One notable example is the Loon project~\cite{loonproject}, initiated by X Development LLC, formerly part of Google's innovation-focused division. The project aimed at extending Internet connectivity to underserved and remote areas using stratospheric balloons. Loon achieved several milestones, showcasing the potential of \ac{HAPS}-based solutions. For instance, in 2017, the team successfully navigated balloons to disaster-stricken regions, providing emergency connectivity in Peru during severe flooding and in Puerto Rico after a hurricane destroyed critical ground infrastructure. In 2020, Loon partnered with Telkom Kenya to launch a commercial Internet-via-balloon service for Telkom Kenya subscribers, marking a significant step toward scalable deployment. The balloons, following a regenerative architecture, were equipped with an e-NodeB and various antennas to transmit \ac{LTE} signals to ground \acp{UE} and communicate with each other and with gateway stations~\cite{serrano2023balloons}.

However, despite an important role and a number of delivered breakthroughs in both academia and industry, the Loon project was ended in 2021. Its shutdown stemmed from many reasons including economical (high capital expenditure and operating costs), regulatory (spectrum allocation restrictions, international airspace coordination and data sovereignty issues), technological (channel modeling and network planning challenges, complicated station-keeping) and environmental/resource (helium scarcity, power limitations from photo-voltaic panels, occasional environmental pollution caused by polyethylene balloon wreckage)~\cite{zhang2022lessons}.

Another ambitious project, Aquila, initiated by Facebook in 2014, aimed to develop a \ac{HAPS} system based on solar-powered unmanned aircraft to deliver broadband connectivity to underserved and remote regions. Operating in the stratosphere, Aquila was designed as a quasi-stationary relay providing line-of-sight coverage over areas of several thousand square kilometers, complementing terrestrial networks where infrastructure deployment is economically unviable. The project emphasized long-endurance flights (target: 60–90 days), solar energy harvesting, and \ac{mmWave} communication payloads for high-capacity air-to-ground links. Despite promising early demonstrations and due to difficulties with in-house aerospace expertise led to insufficient solar energy harvesting and battery efficiency, inadequate flight endurance and, above all, structural integrity issues, their own aircraft development was ended in 2018 in favor of ecosystem partnerships with aerospace firms such as Airbus~\cite{harris2018facebook}.

Zephyr, a series of \ac{HAPS} aircraft initially designed by Qinetiq and produced by Airbus, effectively became a successor of Aquila~\cite{palmer2019facebook}. Now it is an ongoing fixed-wing solar-powered \ac{HAPS} project by Airbus, is designed as a payload-agnostic platform capable of extended operation in the stratosphere~\cite{zephyr}. It provides persistent surveillance and communication services with various types of payloads. Zephyr supports both transparent and regenerative network architectures, utilizing either its default in-house connectivity payload designed to extend \ac{4G} and \ac{5G} networks, or by integrating other compatible technologies, thereby functioning as platform-as-a-service. Combining a wide range of monitoring, tracking, sensing and detection applications with a communication payload, this platform can offer great opportunities for intelligent network optimization and \ac{QoS} improvement.

Similarly, HAPSMobile, a collaborative project between AeroVironment and SoftBank, aims at establishing a network of Sungliders~\cite{softbank2024pressrelease} to provide connectivity to billions of people lacking basic wireless communication, and to supply additional bandwidth for emerging \ac{5G} and \ac{IoT} networks~\cite{haps_avinc}. SoftBank also announced that, in partnership with the U.S.-based company Sceye Inc., it plans to launch pre-commercial \ac{HAPS} stratospheric telecommunications services in Japan in 2026.~\cite{softbank2025pressrelease}. In these systems, where the \ac{HAPS} acts as a flying \ac{BS}, a regenerative payload architecture is the likely choice, although details are hard to find at this stage.

Failures of the Loon and Aquila projects highlighted several major lessons that must be learned to enable more successful future development of \ac{HAPS}-based solutions.
From a technological perspective, it is crucial to ensure platform maturity through iterative prototyping and proper alignment of expertise. Aquila’s structural integrity issues and limited endurance stemmed from insufficient aerospace-domain knowledge. 
It is also important to consider economic sustainability in addition to technical feasibility. Technological proof-of-concepts (e.g., Loon’s over 300-day flights or Aquila’s mmWave links) are insufficient without a viable business model. Early-stage techno-economic analyses, including sensitivity to subsidies or public–private partnerships, are required to avoid over-optimistic assumptions.

Similarly, environmental and resource constraints must be accounted for in the design phase. Helium scarcity limited Loon’s scalability, while Aquila’s solar–battery efficiency proved inadequate under variable conditions. Consequently, it is essential to adopt sustainable alternatives aligned with \ac{6G} sustainability goals, emphasizing energy-efficient \ac{NTN} solutions to reduce ecological footprints.

Regulatory foresight and spectrum allocation strategies should also be embedded into project roadmaps and aligned with \ac{ITU} and \ac{3GPP} frameworks in order to avoid conflicts with local regulations and international coordination challenges.
These issues may be mitigated through closer academia–industry collaboration and effective partnerships across multiple industrial sectors, including aerospace, communications, energy harvesting, and materials engineering. For this purpose, future \ac{HAPS} research should prioritize system-level prototyping using open-source tools, demand-driven deployment strategies, and technological cooperation. 

As an example of such collaboration, the \ac{HAPS} Alliance, founded in 2020, brings together globally recognized companies from telecommunications, technology, aviation, and aerospace, as well as academic research groups. Moreover, shifting from competition for market share among terrestrial providers, \ac{SatCom} operators, and emerging \ac{HAPS}-based projects toward mutually beneficial collaboration may represent a key step toward a fully integrated \ac{TN}-\ac{NTN} architecture incorporating multiple layers to achieve truly ubiquitous and seamless connectivity.

\section{HAPS Technologies}  \label{sec: Techno}

Technology achievements enable \ac{HAPS} to enhance connectivity, optimize performance, and ensure sustainable operations. As an example, Karabulut Kurt~\textit{et al.}~\cite{kurt2021vision} explored the application of advanced techniques such as massive MIMO and \ac{mmWave} communication into \ac{HAPS} systems, identifying these as promising directions for future research. 
Similarly, d’Oliveira, Melo, and Devezas~\cite{d2016high} discussed key technological developments, trends, and challenges associated with \ac{HAPS}. Nonetheless, this work primarily adopts an aeronautical engineering perspective, and lacks a thorough analysis from a telecommunications point of view. In addition, the study was published in April 2016, and does not reflect the most recent developments in the field.

In this section, we address these limitations, and present an updated and comprehensive overview of recent technological trends that support the large-scale adoption of \ac{HAPS} in the context of 6G networks.
We focus on key enablers such as channel modeling (Sec.~\ref{sub:channel-modeling}), resource management (Sec.~\ref{sub:resource-management}), user assignment and cell association (Sec.~\ref{sub:user-assignment}), multi-link connectivity (Sec.~\ref{sub:multi-link}), mobility management and handover (Sec.~\ref{sub:mobility-management}), antenna (Sec.~\ref{sub:mimo}), channel estimation (Sec.~\ref{sub:channel-estimation}), interference management and beamforming (Sec.~\ref{sub:interference-management}), energy efficiency (Sec.~\ref{sub:energy}), \ac{ML} (Sec.~\ref{sub:machine-learning}) 
and Cybersecurity (Sec.~\ref{sub:security}) techniques for \ac{HAPS}. 

In order to review technology achievements systematically we classify studies using a multi-dimensional taxonomy based on following dimensions: research methodology (analytical, simulation-based, experimental); network integration level (standalone HAPS, HAPS-TN, HAPS-NTN, HAPS-TN-NTN, no HAPS); and related frequency bands (Sub-7~GHz, mmWave/FR2, Sub-THz/THz, and FSO).

From Table VIII in Appendix summarizing the classification, it can be observed that analytical research methodology dominates the body of \ac{HAPS}-related studies. This prevalence reflects the current maturity level of \ac{HAPS} technology, where theoretical frameworks enable rapid feasibility assessments, while simulation-based investigations are limited by the availability of validated channel and system models. Experimental studies remain scarce due to regulatory, financial, and technical barriers. Consequently, researchers often rely on self-developed simulation models with limited capabilities in order to validate their analytical findings. Nevertheless, collaborative initiatives, such as those led by the HAPS Alliance, may facilitate the collection of empirical data and contribute to the standardization of datasets for more reliable and reproducible simulation studies. It is also worth noting that the majority of existing studies focus on the sub-7~GHz and mmWave/FR2 frequency ranges, while the emerging cmWave and THz bands remain comparatively underexplored.

\subsection{Channel Modeling}
\label{sub:channel-modeling}
Channel modeling for~\ac{HAPS} is evolving with new technologies, improving accuracy in predicting signal propagation, especially based on advanced simulation techniques and real-time data. 
However, existing \ac{HAPS}-related channel modeling studies predominantly focus on \ac{HAPS}-to-ground links, whereas inter-platform channel characteristics remain comparatively underrepresented.

Network simulators often lack realistic path loss and fading modules. To address this, models for rain, cloud, gaseous absorption, and Ricean fading have been developed. Using QualNet, Zheng, Wang, and Meng~\cite{zheng2013modeling}  showed how different attenuation types affect the packet delivery rate of \ac{HAPS}-to-ground links on 10, 15, 20 and 25 GHz communication  frequency, improving simulation realism and accuracy.
Also, Zain~\textit{et al.}~\cite{zain2013improving} showed Ka-band rain attenuation is about 10× higher than in C-band and strongly location-dependent, and presented an analytical model based on the \ac{ITU-R} empirical model to predict Earth–space attenuation for fully and partially rain-affected paths. For a 1-km cell, attenuation reaches 43~dB at 0.01\% availability and roughly doubles as cell size increases.

To address the complexity of propagation models in suburban and urban environments, Hsieh~\textit{et al.}~\cite{hsieh2019propagation} applied ray tracing simulations to study ~\ac{HAPS}-to-ground signal propagation in Chicago downtown. They developed a path loss model in order to estimate total propagation loss by considering terrestrial clutter and free-space propagation. The model adapts to different aerial vehicle heights and elevation angles using two environment-specific factors, and the study also analyzes multipath characteristics to support system design and fast fading modeling.
To examine the inter-platform propagation, Yılmaz~\textit{et al.}~\cite{yilmaz2023path} analyzed a channel model of \ac{HAPS}-to-\ac{HAPS} links in order to determine its capacity while considering polarization mismatches, atmospheric gases, rain, and clouds/fog. Their simulations revealed the path gain characteristics and channel capacity across different antenna types, platform heights, and carrier frequency scenarios.
Meanwhile Palanci~\textit{et al.}~\cite{palanci2022high} modeled ~\ac{HAPS}-to-satellite communication channels with a focus on ionospheric effects. It showed that plasma density variations influence the channel while magnetic field effects on the refractive index can be neglected. Using the Appleton-Hartree model and an adapted Chapman model, the authors analyzed how Total Electron Content impacts path loss, supported by extensive numerical results.

A multi-state statistical channel model for \ac{HAPS} communication in the Ka-band was proposed by Zhao~\textit{et al.}~\cite{zhao2020ka} to account for tropospheric weather effects and large-area user mobility. The model uses three fading states—\ac{LoS}, moderate shadowing, and deep shadowing—represented by a Markov chain. It also provides the probability density function of the received signal power, a method for generating time-series, and evaluates \ac{BER} performance using the proposed model.

Guan~\textit{et al.}~\cite{guan2020channel} show that \ac{HAPS} mobile channels invalidate the terrestrial plane-wave assumption, requiring consideration of both elevation and azimuth spreads. Large delay differences, Doppler shifts, and strong spatial correlation further complicate analysis. By characterizing dispersion across time, frequency, and space, the authors develop a wideband tapped-delay-line \ac{HAPS} channel model with environment-specific parameters to support simulation and verification of broadband \ac{HAPS} systems.

\subsection{Resource Management}
\label{sub:resource-management}
Given the recent relevance of \ac{HAPS} in wireless communication, \ac{RA} has become one of the key areas of investigation, including the positioning of the aerial tier units within the network. While multiple parameters must be optimized to address the \ac{RA} problem comprehensively, more practical solutions operate under specific circumstances and constraints, defining optimization objectives accordingly. 
In this subsection, we will consider various system models, from single-tier (based on \ac{HAPS}) to multi-tier (also known as \acp{SAGIN}) architectures.

\subsubsection{Single-tier architecture}
Ibrahim and Alfa~\cite{ibrahim2015using} formulated an optimization problem aimed at maximizing the total number of \ac{UE}s capable of receiving multicast session transmissions within a given \ac{OFDMA} frame, while satisfying specific \ac{QoS} requirements. This optimization problem was decomposed into two linear programming sub-problems, which were solved separately but sequentially, with the solution of one informing the other in an iterative process. The first sub-problem focused on \ac{UE} selection for a multicast session group and frequency-time slot allocation. To address this, the authors proposed three methods based on Lagrangian relaxation. The second sub-problem involved power allocation, modeled as a continuous knapsack problem, for which a greedy approach was proposed to find the optimal solution. The considered scenario involved a single \ac{HAPS} providing cellular coverage for \acp{GUE} in a specified area. 

In a multi-platform scenario, where several \ac{HAPS} systems serve a coverage area with multiple beams, improved bandwidth efficiency can be achieved. Grace~\textit{et al.}~\cite{grace2005improving} explored this capacity by leveraging the orientation of fixed user dish antennas with relatively narrow beamwidths, which are assumed to point directly at the \ac{HAPS} with which they communicate (i.e., the main \ac{HAPS}). The authors considered two scenarios: one with a single beam per \ac{HAPS}, and another with a multi-beam layout from each platform. Results demonstrated that the system capacity increases when using multiple \ac{HAPS} within the same allocated spectrum. Notably, this approach is particularly scalable as the number of users increases.

In multi-cell settings, both inter- and intra-beam interference are significant. In integrated networks, overlapping coverage between \acp{TBS} and \ac{HAPS}-based \ac{BS}s, especially with co- or adjacent-channel operation, induces inter-system interference, making effective interference coordination essential for future integration (see Sec.~\ref{sub:interference-management}).
In this context, Liu~\textit{et al.}~\cite{liu2022interference} proposed a dynamic spatial–time–frequency coordination method between \ac{HAPS}-mounted \acp{BS} and \acp{TBS}. Two service modes are considered: direct \ac{HAPS}–\ac{GUE} links and indirect access where \ac{HAPS} serve \acp{TBS} that, in turn, serve \acp{GUE}. \acp{UE} associate to \ac{HAPS} (direct or indirect) or \acp{TBS} based on maximum received power. The scheme distributes traffic load and employs an ergodic search over allocation patterns across \ac{HAPS} and terrestrial networks. Simulations show the best average per-cell throughput for user groups at any distance from \ac{HAPS}.
Unlike the single-\ac{HAPS} model in~\cite{liu2022interference}, Wei~\textit{et al.}~\cite{wei2023spectrum} studies a \ac{HAPS} network via stochastic geometry, modeling \acp{TBS} and \ac{HAPS} as Poisson point processes and deriving a closed-form coverage probability for spectrum sharing. Results show \ac{HAPS} cause minimal interference to \acp{TBS}, supporting spectrum sharing in integrated networks.

\subsubsection{Multi-tier architecture}

In the space domain, \ac{LEO} satellite systems like Starlink by SpaceX are already extending communication services to areas beyond the reach of terrestrial networks. In January 2024, Starlink launched its first six satellites equipped with direct-to-cell capabilities, enabling device-to-device connections with unmodified \ac{LTE} phones, and providing text messaging services integrated with mobile operators' core networks~\cite{starlink2024directtocell}. 
To achieve this, Starlink developed large, advanced phased arrays, measuring 2.7 m x 2.3 m, which feature highly sensitive receivers and powerful transmitters. These arrays allow the satellites to function as cell towers in space, capable of communicating with cellphones equipped with low-gain antennas and limited transmit power.
The goal faces significant technological challenges, particularly in densely populated areas with many users. Addressing these requires advancements in both network architecture and \ac{RA}. An integrated multi-tier \ac{SAGIN} can provide major performance benefits through deployment across diverse nodes and links.

Li~\textit{et al.} ~\cite{li2020hierarchical} focused on \ac{RA} optimization within a multi-tier architecture where \ac{LEO} satellites provide wide-area coverage, while \ac{HAPS} and full-duplex terrestrial relays cover hot-spot regions to enhance capacity. Additionally, \ac{GEO} satellites were considered as routing nodes. The authors designed a hierarchical resource management framework for this system, and proposed an optimization method to maximize \ac{LEO}-to-ground and \ac{HAPS}-to-ground downlink throughput based on predictions derived from the regular orbit information of \ac{LEO} satellites.
In the proposed architecture, mobile \ac{UE}s in hot-spot regions connect via \ac{HAPS} or terrestrial relays, while other areas use direct \ac{LEO} links. This approach boosts system capacity in hot-spots and lowers handover rates. Different frequency bands are allocated for space/air-to-ground, terrestrial, space-to-air, and inter-satellite links. Capacity analysis and simulations show that the multi-tier design greatly outperforms satellite-only systems. Additionally, dynamic radio resource optimization improves the average throughput of \ac{HAPS} and \ac{LEO} beams by over 10\%. In this scenario, spectrum sharing is required if \ac{HAPS} and satellites are designed to transmit on the same or adjacent frequency bands, or when deploying \ac{HAPS} in a region already covered by a separate satellite system.

Meanwhile, a \ac{LEO}-\ac{HAPS} spectrum sharing technique under non-ideal spectrum sensing was studied by Wang~\textit{et al.} in~\cite{wang2019dynamic}, where a multi-beam \ac{LEO} served as the primary system and a single-beam \ac{HAPS} supported \ac{GUE}s without direct \ac{LEO} access. A dynamic spectrum and power allocation strategy was proposed to maximize \ac{HAPS}-\ac{GUE} throughput with minimal \ac{QoS} degradation for \ac{LEO}-associated \acp{GUE}. Results showed that \ac{HAPS} interference to \ac{LEO}-associated \acp{GUE} is negligible and affects only a small fraction.

Later, \ac{RA} for a multi-tier cognitive satellite-\ac{HAPS} network with \ac{NOMA} was studied by Liu~\textit{et al.}~\cite{liu2023resource}. The model included a primary satellite and a secondary \ac{HAPS} network sharing spectrum via underlay cognitive radio to serve terrestrial users. The sum-rate maximization problem for the secondary network was addressed under \ac{NOMA} group connectivity limits, power constraints, and \ac{QoS} requirements. The proposed joint \ac{RA} algorithm (sub-channel and power allocation) outperformed orthogonal multiple access, especially as user or sub-channel numbers increased.

Introducing smaller aerial components, such as \acp{LAPS} like drones, into the network architecture adds an additional degree of freedom in deployment planning. While this approach can further improve coverage and system capacity, it also complicates \ac{RA} due to the need for dynamic load balancing and \ac{3D} placement of mobile network nodes.
An aerial heterogeneous cellular network with \ac{HAPS}-based macrocells and \ac{LAPS}-based femtocells was studied by Helmy and Arslan~\cite{helmy2018utilization}. Using stochastic geometry, the authors analyzed cell loads, outage probability, optimal \ac{LAPS} placement, and downlink \ac{SINR}. Results showed that users associated with an \ac{LAPS} \ac{BS} experienced much lower outage probability, and the proposed \ac{LAPS} positioning algorithm further reduced outages.

Later, Qin~\textit{et al.}~\cite{qin2021joint} investigated joint \ac{RA} and \ac{UAV} \ac{3D}-location optimization in a cluster-\ac{NOMA} network with a \ac{HAPS} macro \ac{BS} and multiple \ac{UAV} small \acp{BS}. The goal was to maximize IoT uplink sum rate by decomposing the problem into spectrum/power allocation and \ac{UAV} positioning (horizontal and altitude), solved iteratively. Results showed improved uplink sum rate and spectrum efficiency over benchmark \ac{RA} schemes.

As previously mentioned, the \ac{SAGIN} architecture is likely the most promising option for future communication networks, including~\ac{6G}. Therefore, it is crucial to clarify the features that enable the coexistence and collaboration of \ac{HAPS} and terrestrial systems within this integrated framework. 
To provide high-throughput connectivity to ground users within an integrated satellite-\ac{HAPS}-\ac{TBS} network, a \ac{RA} framework was proposed in~\cite{alsharoa2020improvement}. This framework aimed at enhancing \ac{GUE} downlink throughput while operating under power, access, and backhauling constraints. Two distinct problems were formulated based on the frequency of parameter optimization. The short-term problem addressed the optimization of access, backhauling, and power allocation while maintaining a fixed \ac{HAPS} location. In contrast, the long-term problem focused on optimizing \ac{HAPS} location based on the average distribution of users.
The studied system model comprised three tiers: a satellite in the space tier, multiple \ac{HAPS} in the air tier, and several \acp{TBS} in the ground tier, with gateway feeder stations for backhauling. Based on \ac{OFDMA}, spectrum was divided into \acp{RB}, with each user served by one \ac{BS} using one \ac{RB}, ensuring no intra-cell interference but allowing inter-cell interference among \acp{TBS}. No inter-cell interference was assumed for satellites and \ac{HAPS} due to distinct frequency sets. Simulations showed a notable average data rate gain compared to two baselines: (i) joint optimization with uniform power, and (ii) optimized \ac{HAPS} placement with random access/backhaul under uniform power.

From the literature, it is clear that aerial network mobility boosts capacity and flexibility but complicates optimization due to \ac{3D} positioning, dynamic load balancing, and spectrum diversity. Effective interference management is vital for stability, while a well-designed \ac{RA} strategy can mitigate inter-tier interference and enable smooth integration with \ac{TN}.

\subsection{User Assignment and Cell Association}
\label{sub:user-assignment}
Unlike stationary \acp{TBS}, \ac{HAPS} platforms enable flexible deployment that can be dynamically optimized according to user distribution.
Consequently, the \ac{HAPS} deployment problem has been investigated using various analytical models across different application scenarios and architectural configurations.
For example, Zong~\textit{et al.}~\cite{zong2012deployment} examined \ac{HAPS} deployment for maximizing network transmission rate while meeting service requirements, where each \ac{GUE} connects to the \ac{HAPS} offering the best rate. A game-theoretic, self-organized optimization using a restricted spatial adaptive play algorithm was proposed and shown to converge in simulations. Although simplified compared to full \ac{SAGIN} architectures, the results highlighted the potential of \ac{HAPS} for self-organized deployment optimization.

To meet \ac{6G} data rate demands, Cumali~\textit{et al.}~\cite{cumali2023user} studied user selection in a downlink \ac{NOMA}-based multi-user \ac{HAPS} system. A joint user selection and correlation-based pairing algorithm was proposed to maximize beamforming gain, along with a polar-cap codebook design for limited-feedback \ac{HAPS} channels under Rician fading. Simulations showed superior data rates over existing \ac{NOMA} schemes. However, the model did not consider integration with terrestrial/satellite networks or support for aerial users, which are essential for \ac{6G}.

While satellite deployment is less flexible than \ac{HAPS}, its inherently wide coverage area provides opportunities for optimizing user association strategies, especially within multi-tier integrated architecture.
Accordingly, the user association problem in the context of computational task offloading has been investigated by Zhang~\textit{et al.}~\cite{zhang2020satelite}. The authors considered a disaster scenario where \acp{TBS} have been destroyed, and two-tier \acp{UE}, including both \acp{GUE} and \acp{UAV},  with limited computational capabilities offload disaster information to \ac{HAPS} and/or LEO satellite servers for further processing. 
The study aimed to jointly optimize uplink fronthaul capacity and computation performance by modeling the problem as a weighted 3-uniform hypergraph, where each hyperedge weight represents the combined fronthaul uplink and computing rates. A local search-based algorithm was developed to find a near-optimal \ac{3D} hypergraph matching with maximum total weight.
The model considered a \ac{LEO} satellite with an enhanced \ac{MEC} server providing backhaul to multiple \ac{HAPS}, which cover disaster areas and offload computation, while \ac{UE}s use orthogonal sub-channels within shared Ka-band \acp{RB}, avoiding uplink interference. In simulations with 6 \ac{HAPS} and 20 \acp{UAV} over a disaster area, the proposed hypergraph search algorithm achieved significant system sum rate gains over greedy search, showing strong practical potential.

Later, a \ac{vHetNet} comprising one \ac{GEO} satellite, one \ac{HAPS}, and several \acp{TBS} was examined by Liu, Dahrouj and Alouini~\cite{liu2024joint}. In this setup, both the \ac{HAPS} and \acp{TBS} are equipped with multiple antennas, enabling them to simultaneously serve multiple \acp{GUE} through \ac{RF} links. The connection between the satellite and the \ac{HAPS} is established via an \ac{FSO} link. In this scenario, when a \ac{GUE} is served by the \ac{HAPS}, the required data is first transmitted from the satellite to the \ac{HAPS}, and then relayed to the \ac{GUE}.The system model employs a space division multiplexing scheme, where all \ac{RF} links share the same central frequency. Consequently, the performance of the proposed architecture is highly dependent on the intra-cell and inter-cell interference among \ac{TBS}s, intra-cell interference within the \ac{HAPS}, and the interference between \ac{HAPS} and \ac{TBS}s.
The study formulated a joint optimization problem for associating \ac{GUE}s with either \ac{TBS}s or the \ac{HAPS} and designing corresponding beamforming vectors, aiming to maximize sum-rate under constraints of \ac{HAPS} payload connectivity, \ac{FSO} backhaul capacity, and power limits of \ac{HAPS}/\ac{TBS}s. An iterative algorithm optimized parameters step by step, and simulations demonstrated that integrated satellite-\ac{HAPS}-terrestrial networks can significantly enhance the sum-rate.

As heterogeneous integrated networks face load imbalance from uneven user distribution and mobility, Tu~\textit{et al.}~\cite{tu2024priority} proposed a priority-based load-balancing scheme for a two-layer downlink \ac{SAGIN} using network slicing to mitigate underutilization and overload across cells.
The proposed two-layer \ac{RAN} includes \acp{TBS} in the terrestrial layer (priority for user association) and \acp{UAV}/\ac{LEO} satellites in the non-terrestrial layer. Overloaded \acp{TBS} offload traffic to the non-terrestrial layer via \ac{RAN} slicing into high-throughput, low-delay, and wide-coverage slices, with lower-priority slices preferentially offloaded. The optimization problem jointly considered throughput, delay, coverage, user association, resource allocation, and \ac{UAV} deployment. Results showed improved slice throughput and average delay. While \ac{HAPS} were not included, the study tackled the critical load-balancing challenge in \ac{SAGIN}.

Prior to that, Shahid~\textit{et al.}~\cite{shahid2020load} explored load balancing for \ac{5G} integrated satellite-terrestrial networks using multiple \acp{RAT}. The approach handled both intra- and inter-\ac{RAT} \ac{UE} offloading from overloaded cells by introducing the radio resource utilization ratio as a common load metric, enabling efficient traffic redistribution without overloading target cells. Simulations showed improved throughput and \ac{QoS}. While limited to a \ac{GEO} satellite, the metric can be generalized to other \ac{NTN} components, including \ac{HAPS}.

Alam~\textit{et al.}~\cite{alam2023throughput} studied load balancing as a joint radio resource management problem—covering bandwidth allocation, \ac{UE} association, and power control—in an integrated \ac{TN}-\ac{NTN} rural downlink network with multiple macro \acp{TBS} and \ac{LEO} satellites. The proposed framework maximized the sum of log throughput across all \acp{UE}, improving overall network capacity.
The framework tackled the optimization sequentially—first user association and bandwidth allocation under fixed power, then transmit power with the first two fixed. Simulations showed notable throughput gains over the maximum \ac{RSRP} baseline and confirmed that integrating \acp{TN} and \acp{NTN} in rural areas can greatly reduce out-of-coverage \acp{UE}.
In their subsequent study~\cite{alam2024ontherole}, the authors proposed a framework for bandwidth allocation, user association, and power control that can balance network capacity and energy consumption based on the current traffic conditions. By adapting the network behavior according to traffic demands, this approach significantly enhances mean network throughput during high-traffic periods and reduces power consumption during low-traffic periods.

Based on the reviewed studies, we can conclude that user association in \acp{SAGIN} that incorporate \ac{HAPS}-mounted \acp{BS} and are fully integrated with \acp{TN}, still requires further research. It is particularly important to consider scenarios involving a realistically sufficient number of mobile ground and aerial \acp{UE}. Additionally, there is a need to investigate the user association challenge under the practical constraints imposed by interference management solutions and load balancing.

\subsection{Multi-Link Connectivity}
\label{sub:multi-link}
As mentioned earlier, some emerging 6G applications, such as aerial vehicle control and autonomous driving, impose stringent requirements in terms of both latency and connection reliability, a paradigm referred to as \ac{URLLC}.
Meanwhile, the Stage-2 architectural framework for \ac{MC} across E-UTRA and \ac{NR} is defined by 3GPP TS~37.340~\cite{3GPP37340}. It outlines the functions of the master node and secondary node, bearer splitting and duplication mechanisms, and the related signaling architecture that permits simultaneous connectivity to multiple radio nodes. In order to make clear how these processes might be used to enhance service continuity and resilience in \ac{HAPS}-enabled \ac{NTN} deployments, we have broadened the conversation.

In this context, multi-link connectivity (or \ac{MC}) has shown the potential to meet the \ac{URLLC} service requirements~\cite{salehi2022ultra}.
Specifically, \ac{MC} architectures can be categorized at the \ac{BS} level (i.e., the user connects to multiple \acp{BS} within the same network, usually managed by the same operator), at the network level (i.e., the user is connected to different networks, possibly from different operators), or at the technology level (i.e., the user connects via different \acp{RAT})~\cite{salehi2024reliability}. 
In this paper, we focus on the first category, which also includes techniques such as \ac{CoMP}, which gained significant importance in \ac{LTE}-advanced and~\ac{5G}, where multiple \acp{BS} within the same network simultaneously serve the same~user.

Li~\textit{et al.}~\cite{li2020airtoair} studied air-to-air communication using a \ac{3D} model integrating aerial \acp{BS}, aerial \acp{UE}, \acp{TN}, and \acp{GUE}. Two \ac{UAV} types were considered: aerial \acp{BS} connecting via \ac{HAPS}/\ac{TBS}, and aerial users relaying data to \acp{GUE}. A joint-transmission \ac{CoMP} scheme was proposed where each aerial user is served by four aerial \acp{BS} forming a tetrahedral cell. The study derived data rates and coverage probabilities, and introduced a frequency allocation scheme to reduce inter-cell interference. Compared to the binomial-Voronoi dynamic cooperation set, the approach achieved similar performance with much lower complexity and signaling overhead.

Meanwhile, Amer, Saad, and Marchetti~\cite{amer2020mobility} studied aerial \ac{UE} connectivity using \ac{CoMP} transmissions for \acp{UAV} served by \acp{TBS}. Two scenarios were analyzed—static hovering and mobile \acp{UAV}—with coverage probability bounds derived and handover rates evaluated via a custom mobility model. Results showed that \ac{CoMP} transmission greatly improves achievable coverage probability.
Prior to that, Liu, Zhang, and Zhang~\cite{liu2019comp} considered an uplink \ac{CoMP} scenario where \acp{UAV} collect and forward signals from \acp{GUE}. A UAV placement and movement scheme was proposed to maximize average throughput for mobile \acp{GUE}, though UAV energy constraints remain a critical challenge.

Recently, Salehi, Ozger, and Cavdar~\cite{salehi2024reliability} proposed an \ac{MC} \ac{SAGIN} algorithm for remote \ac{AV} piloting beyond \ac{LoS}. The system model included \acp{LEO}, \ac{HAPS}, and \acp{TBS} for \ac{AV} communication, along with direct inter-\ac{AV} links. However, it excluded \ac{LEO}-\ac{HAPS} and \ac{LEO}/\ac{HAPS}-\ac{TBS} links, assuming both \ac{LEO} and \ac{HAPS} connect to the core network via their ground stations.
The transmission strategy in~\cite{salehi2024reliability} used message cloning, sending multiple copies via all available links for independent decoding at the \ac{AV}. The model incorporated practical antenna setups, unreliable backhaul, buffer queues, and fading effects. Reliability and availability of different \ac{MC} configurations were evaluated through simulations, with brute-force optimization used to find the optimal \ac{MC} path ensuring \ac{QoS}. Results showed that \ac{LoS} interference strongly impacts link performance, and \ac{MC} is vital for safe operation of certain \acp{AV}.

At thesame time, latency-sensitive scenario was explored by Sadovaya~\textit{et al.}~\cite{sadovaya2024enhancing}, which addressed the issue of \ac{MC} computational task offloading in \ac{NTN}. The system model featured a \ac{HAPS} and several \acp{UAV} providing on-demand \ac{MEC} services to randomly distributed \acp{GUE}. The simulation results showed that \ac{MC} offloading strategies can significantly reduce task computation delay.

\subsection{Mobility Management and Handover}
\label{sub:mobility-management}
Mobility management involves significant challenges in maintaining seamless connectivity during handovers between network cells without service interruption. Issues include preserving connection quality during handovers, and adapting to rapid movements or changing signal conditions. The design of new handovers techniques, specifically tailored to \ac{LEO} satellite and \ac{HAPS} systems, is essential to ensuring reliable service continuity and a positive user experience.
Meanwhile the architectural and functional principles of connected-mode mobility and handover in \ac{NR} and \ac{LTE} are specifically described in 3GPP TS~38.300 and TS~36.300~\cite{3GPP38300,3GPP36300}, respectively. Referencing the specific handover signaling and timing protocols outlined in the related Radio Resource Control \ac{RRC} standards, these specifications specify the functions of radio access network (RAN) nodes, mobility statuses, and high-level handover concepts. In the context of \ac{HAPS}-assisted mobility, where wide coverage regions and quasi-stationary platforms produce different handover dynamics than terrestrial networks, its applicability to \ac{NTN} scenarios is examined.

In this context, Juan~\textit{et al.}~\cite{juan20205g} evaluated the standard \ac{5G} \ac{NR} handover procedure for a high-speed train in a \ac{LEO}-based deployment versus a \ac{3GPP} urban macro \ac{TN}. Simulations showed radio connection failures during handovers were 10× higher and outages 5× longer in the \ac{NTN} scenario.
Further, an intra-satellite mobility solution was proposed~\cite{juan2022handover} that avoids \ac{UE} radio measurements by exploiting predictable satellite trajectories and antenna gains. Simulations in rural and urban scenarios showed reduced service failures and unnecessary handovers, with improved downlink \ac{SINR}.
Zhang~\textit{et al.}~\cite{zhang2020ai} introduced a \acp{CNN}-based handover approach for \ac{LEO} satellite network. The authors modeled the handover process as a directed graph, where a user with (partial) knowledge of its future signal strength can identify the optimal handover sequence. The \ac{CNN} was trained to learn near-optimal handover strategies from historical signal strength data.
Voicu, Bhattacharya, and Petrova~\cite{voicu2024handover} compared three inter-satellite handover methods for space-to-ground performance: Closest Satellite, Max Visibility, and \ac{SINR}-based. The first two improved spectral efficiency and reduced delays, while the \ac{SINR}-based approach minimized the number of handovers.
Warrier~\textit{et al.}~\cite{warrier2024future} introduced a graph theory-based algorithm for \ac{UAV} handovers between terrestrial and satellite networks. By adapting in real time to \ac{UAV} mobility and network changes, it minimizes latency and disruptions. The algorithm considers \ac{RSRP}, \ac{SNR}, and elevation angle to optimize handover timing and ensure smooth connectivity.

The quasi-stationary nature of \ac{HAPS} in the stratosphere introduces additional challenges for conventional handover algorithms, which rely on fixed thresholds. He, Cheng and Ni~\cite{he2016improved} introduced an adaptive handover algorithm, addressing the limitations of fixed-threshold methods, in order to handle possible \ac{HAPS} disturbance. Using a least mean square prediction-based power model, the scheme adjusts handover parameters according to terminal speed and the confidence levels of expected signal strength, improving reliability under \ac{HAPS} mobility conditions.

Further~\cite{he2017adaptive} a cooperative transmissions scheme was proposed for \ac{HAPS} systems to mitigate outage probability and long service interruptions. The algorithm adaptively selects the platform with the highest channel gain based on user mobility and real-time channel conditions, improving reliability and service continuity.
Later~\cite{wang2019effect} Wang, Li, and Zhou analyzed the impact of \ac{HAPS} vertical and swing movements on handover performance. A refined ground coverage model and analytical expressions were developed, showing that path loss strongly affects coverage, with swing movement exerting a greater influence on handover probability than vertical movement.
Menwhile, Li~\textit{et al.}~\cite{li2019handover} developed various handover techniques for an integrated \ac{LEO}-\ac{HAPS} network, and evaluated the handover delay and signaling overhead. Specifically, they formulated the problem of determining the optimal handover triggering time, which was analytically solved using Lagrange duality.
Finally, Almuallim~\textit{et al.}~\cite{almuallim2023handover} studied a mobility management solution for integrated \ac{HAPS}-ground networks, allowing aerial users to connect to either terrestrial \acp{BS} or a \ac{HAPS}. The approach focused on minimizing unnecessary handovers while ensuring \ac{QoS}.

\subsection{Antenna Technologies}
\label{sub:mimo}
Integrating advanced antenna technologies, such as \ac{mMIMO}, spatial multiplexing, digital beamforming, and \acp{RIS}, into \ac{HAPS} systems can improve network capacity and efficiency across a range of diverse and challenging environments. 
For example, Dicandia and Genovesi~\cite{dicandia2021spectral} evaluated the advantages of using a triangular (rather than square) lattice for phased arrays in \ac{HAPS} systems operating in the~\ac{5G} NR n257 and n258 frequency bands. By considering a 64-element planar array, the results demonstrated that the triangular lattice array offers superior performance in terms of array gain, average sidelobe level, and mutual coupling, making it a more effective choice for \ac{5G} \ac{mMIMO} systems.
Meanwhile, Tashiro, Hoshino, and Nagate~\cite{tashiro2021cylindrical} proposed a cylindrical \ac{mMIMO} system for \ac{HAPS}. They showed that the cylindrical \ac{mMIMO} shape provides up to 2.1 times higher capacity compared to traditional planar arrays, with an extended coverage of up to 100 km in radius, and improves the \ac{SINR} by approximately 10 dB at the cell edge.
Further, in order to consider the coexistence with terrestrial systems, the authors introduced a two-stage precoding system for \ac{TN}-\ac{HAPS} spectrum sharing using phased array antennas~\cite{tashiro2022nullforming}. The first stage forms nulls toward \ac{TN} cells, and the second directs beams to \ac{HAPS} users. With a 196-element cylindrical array, the approach improved \ac{SINR} for \ac{TN} users by up to 10 dB over a conventional \ac{MMSE} precoder.

Cylindrical antenna configuration was later considerd by Shafie~\textit{et al.}~\cite{shafie2024mimo}. The authors proposed a vertical uniform linear array sector-based cylindrical antenna for \ac{HAPS}, combined with a \ac{NOMA} clustering method to mitigate strong geographic user correlation. An optimization algorithm for power distribution was introduced to maximize energy efficiency, spectral efficiency, and \ac{QoS} using successive interference cancellation. Simulations showed that spatial correlation strongly affects spectrum and energy efficiency in multi-antenna \ac{HAPS} systems.

Alternative antenna configuration was studied by Abbasi, Yanikomeroglu, and Kaddoum~\cite{abbasi2024hemispherical} to address the suboptimal gain for users located beneath the \ac{HAPS}. The authors introduced a hemispherical antenna array for \ac{HAPS} to maximize gain by aligning users with specific antenna elements. Using analog beamforming, user data rates were estimated from steering vectors, while an antenna selection algorithm and Bisection-based power allocation optimized performance. Results showed cumulative data rates up to 14 Gbps.
The authors also proposed a cell-free scheme to mitigate inter-cell interference between \acp{GUE} and neighboring \ac{UAV}-\acp{BS}~\cite{abbasi2024uxnb}. The scheme operates in two stages: \acp{GUE} send messages to \acp{UAV} in the sub-6 GHz band, and \ac{UAV}s then allocate resources and beamform signals to a \ac{HAPS} in the sub-\ac{THz} band acting as a \ac{CPU}. An optimization problem was formulated to maximize the minimum \ac{SINR} by optimizing power allocation and \ac{UAV} positions.

Regardin the \ac{UE} antenna systems, Kim~\textit{et al.}~\cite{kim2023dual}, proposed a dual-polarized shared aperture antenna for Ku- and Ka-bands, developed through a four-step derivation process using slot aperture coupling to achieve broadside radiation in both bands. Based on array factor theory, the optimal inter-element spacing was determined, resolving the trade-off between beam steering angles and array spacing in dual-band systems.

The use of \acp{RIS} is another approach to improve signal quality and coverage. In fact, low-cost \ac{RIS} components enable customizable communication channels, \ac{LoS} connectivity, and obstruction avoidance~\cite{ye2022nonterrestrial}. 
When combined with traditional \ac{mMIMO} techniques, \acp{RIS} can also boost data rates and reliability. In a system where \ac{HAPS} relay data via \acp{RIS} due to transmission blockages, the \ac{RIS} acts as an amplify-and-forward relay between source and ground users. This approach was evaluated by Odeyemi, Owolawi, and Olakanmi~\cite{odeyemi2022reconfigurable} through metrics like \ac{BER}, capacity, and outage probability, with analytical and asymptotic results showing significant performance improvements for \ac{HAPS} and \acp{RIS} integration.

\subsection{Channel Estimation}
\label{sub:channel-estimation}
\ac{6G} is expected to use high-frequency bands such as \ac{cmWave}, \ac{mmWave} or even sub-~\ac{THz}, to support new scenarios for seamless ubiquitous \ac{3D} connectivity (see Sec.~\ref{sub:ntn-bands}). However, the channel conditions in these bands will become more variable, suffering from significant path loss, fading, and interference. Thus, accurate channel estimation is necessary to adapt transmission parameters dynamically, ensure reliable connectivity, and maximize data throughput while minimizing latency. Advanced techniques like beamforming, \ac{mMIMO}, and \ac{ISAC}, also require precise \ac{CSI} via channel estimation, to provide reliable communication services.

In this context, Iskandar and Aziz~\cite{iskandar2014study} evaluated \ac{LTE} downlink performance over \ac{HAPS}, comparing least square and \ac{MMSE} channel estimation under a Ricean fading model, and found that \acp{GUE} at low elevation angles suffer degradation from multipath and Doppler. Later Hidayat and Iskandar~\cite{hidayat2015pilot} analyzed a single-carrier \ac{LTE} \ac{HAPS} channel, showing that elevation angles above 40° yield optimal conditions, with performance influenced by bandwidth, modulation, and Doppler shifts. More recently, Rahayu and Iskandar~\cite{rahayu2023study} compared least square, \ac{MMSE}, and \ac{MMSE}+\ac{DFT} estimation methods for \ac{HAPS}-based \ac{OFDM} systems.

Also, Nawaz~\textit{et al.}~\cite{nawaz2017location} estimated sparse \ac{HAPS} channels using a superimposed pilot sequence method that leverages prior knowledge of \acp{UE}' locations. The model featured a \ac{HAPS}-based beamforming multi-antenna \ac{BS} serving mobile \acp{GUE}, with backhaul via ground stations and satellites. Numerical results examined the influence of antenna configuration, channel sparsity, \ac{SNR}, and interferer density.
Later, Tekbıyık~\textit{et al.}~\cite{tekbiyik2021channel} studied a \ac{vHetNet} combining \ac{GEO}/\ac{LEO} satellites, \ac{HAPS}, and \ac{TN} with full-duplex \ac{RIS}-assisted backhauling over \ac{HAPS} channels. A graph attention network-based channel estimator was proposed to address channel variability and hardware impairments, requiring estimation at only one transmitting node without on-off control or half-duplex switching. Simulations confirmed strong performance across diverse channel and hardware conditions.

Meanwhile, Güven and Karabulut Kurt~\cite{guven2022cnn} studied a \ac{HAPS}-\ac{LEO} architecture to address channel estimation and synchronization challenges. A \ac{CNN}-based estimator was proposed to mitigate carrier frequency offset from residual Doppler effects, with results showing improved mean square error and \ac{BER} performance over benchmark methods.
And recently, Guo~\textit{et al.}~\cite{guo2024power} analyzed the effect of channel estimation errors in an integrated satellite-\ac{HAPS}-terrestrial communication system, and proposed a novel NOMA-based power allocation scheme to satisfy \ac{QoS} requirements.

\subsection{Interference Management and Beamforming}
\label{sub:interference-management}
Interference management and beamforming technologies enhance \ac{HAPS} communication by jointly improving signal strength and reducing interference in remote and demanding areas. A comprehensive analysis of stratosphere-to-Earth co-channel interference caused by \ac{HAPS} was presented by Milas and Constantinou~\cite{milas2006simulations}, who introduced a methodology for assessing the impact of this interference component on \acp{TN} by evaluating performance degradation. The methodology considered various factors, including the mobility of \ac{HAPS}, the realistic distribution of azimuth and elevation angles of terrestrial microwave links, and the gradual loading of \ac{HAPS} structures.

A \ac{3D} \ac{mMIMO} system was examined by Hu, Hong and Evans~\cite{hu2016modelling}, who considered an uplink scenario with pure \ac{LoS} propagation from multiple \acp{GUE} to a \ac{HAPS} mounting a horizontal planar antenna array. The study modeled intra-cell interference using Beta distributions, and demonstrated that adjusting the altitude of the array can significantly reduce interference, thereby improving key performance metrics such as signal to interference ratio, coverage, and throughput.

Later, Guan~\textit{et al.}~\cite{guan2019multi} introduced a space-division multiple-access scheme with a multi-beam antenna and beamforming algorithm that adapts array weight vectors to improve performance and reduce interference. An improved elliptical cell design balanced \ac{SNR} by broadening inner beams and narrowing outer ones, expanding inner cell coverage while shrinking outer areas. This reduced the number of required cells while maintaining coverage quality. The integration of space-time block coding with multi-beamforming further enhanced coding and diversity gains without extra bandwidth.

Fujii, Ohta and Fujii~\cite{fujii2020study} analyzed a multi-cell \ac{HAPS} system, where capacity is limited by feeder link bandwidth. To enhance spectrum efficiency, a multi-gateway approach was proposed, allowing gateways to share feeder link frequencies to reduce interference and increase capacity. Building on this, the authors~\cite{fujii2020astudyon} introduced interference cancellation techniques for the multi-gateway feeder link, further improving transmission quality and overall communication performance.
To further reduce interference, Ishikawa~\textit{et al.}~\cite{ishikawa2023spectrum} introduced a null-sweeping technique where the \ac{HAPS} dynamically adjusts null points, while terrestrial \acp{BS} schedule \acp{GUE} accordingly. This coordinated approach improved the \ac{SINR} for \acp{GUE} without degrading the \ac{HAPS} link quality, resulting in up to a 15\% increase in ground system capacity.

Considering \ac{HAPS}-aided \ac{SAGIN} architecture, Kawamoto~\textit{et al.}~\cite{kawamoto2024interference} aimed to mitigate the interference between \ac{HAPS}-to-\acp{GUE} links and back lobes from \ac{HAPS}-to-satellite links. Unlike traditional zero forcing, which allocates random nulls, the proposed method generates targeted nulls based on \ac{GUE} locations using a codebook of antenna patterns tied to weight matrices, enabling dynamic adaptation to reduce interference and improve communication quality.

\subsection{Energy Efficiency}
\label{sub:energy}
A key challenge for emergency communication is ensuring energy-efficient transmissions due to the limited power resources available during large-scale disasters. Integrated~\ac{HAPS}/satellite networks can help overcome this issue by optimizing power consumption while maintaining robust connectivity. In this context, Dong~\textit{et al.}~\cite{dong2015energy} proposed an adaptive link-state advertisement routing algorithm designed for a slow-flat Rician fading channel to estimate link information. This approach can reduce energy consumption at the terminal by selecting the most energy-efficient path.

In order to improve energy efficiency, Rallage and Kandeepan~\cite{rallage2022improving} proposed a low-power-consumption decoding and forward relay system via~\ac{HAPS}.The results demonstrated that, in certain channel conditions, a relay-based system can outperform a non-relay-based system in terms of energy efficiency.
Prior to that, Xing~\textit{et al.}~\cite{xing2021high} studied \ac{HAPS} as repeaters and aerial \acp{BS}, with energy efficiency assessed via consumption factor theory in single- and multi-cell scenarios using Monte Carlo simulations. Results showed both architectures achieve good downlink spectral efficiency, though coverage is constrained by \ac{GUE} transmit power and antenna gain. Performance in uplink and downlink can be further improved with directed \ac{GUE} antennas on \ac{LTE} band 1.
Also, Alfattani~\textit{et al.}~\cite{alfattani2022multi} proposed an energy-optimization algorithm for \ac{HAPS}, enabling mode switching to reduce consumption and extend loitering duration. The modes include a low-power \ac{HAPS}-\ac{RIS} mode for passive communication and a \ac{HAPS}-\ac{SMBS} mode for enhanced computing, caching, and communication. The system prioritizes the passive mode, activating higher-energy modes only when additional capacity is required.
Finally, Salamatmoghadasi, Mehrabian, and Yanikomeroglu~\cite{salamatmoghadasi2023energy} studied a vertical HetNet with a \ac{HAPS} as an \ac{SMBS}, alongside an \ac{MBS} and multiple \acp{SBS}, to improve energy efficiency. An algorithm was proposed to minimize consumption by leveraging the \ac{HAPS} and dynamically managing \ac{SBS} sleep modes based on real-time traffic load, reducing power use by 30\% while maintaining \ac{QoS}.

Li~\textit{et al.}~\cite{li2019energy} studied energy-efficient \ac{RA} in~\acp{SAGIN} for remote~\ac{IoT} using~\ac{UAV} relays. The goal was to improve energy efficiency by optimizing sub-channels selection and the uplink transmission power using Lagrangian dual decomposition, and \acp{UAV} deployment via sequential convex approximation. The proposed approach achieved more than 21.9\% improvement in terms of energy efficiency. 
Later, Zhu~\textit{et al.}~\cite{zhu2021optimal} explored the use of~\ac{HAPS} in~\acp{SAGIN} to support data transmission between~\ac{LEO} satellites and smart devices for remote~\ac{IoT} applications. The proposed approach jointly optimized \ac{RA} and~\ac{HAPS} deployment to minimize the power consumption. By breaking down the complex problem into two sub-problems, the authors found a near-optimal solution. Simulation results showed a significant reduction in power consumption compared to conventional \ac{RA} strategies, while still satisfying smart device rate requirements.
Finally, Zhao and Chang~\cite{zhao2023energy} optimized energy efficiency for data collection in a \ac{UAV} network supported by a \ac{HAPS} aerial \ac{BS}. The approach dynamically clustered the target area, selected sink sensing devices, adjusted transmit power, and managed \ac{UAV} operations. A centralized \ac{RL} framework based on \ac{PPO} jointly optimized \ac{UAV} trajectories and power, while an Affinity Propagation clustering algorithm identified optimal sink devices for device-to-UAV communication.

Recently, Song~\textit{et al.}~\cite{song2024high} evaluated the energy efficiency of a \ac{HAPS}-based \ac{RAN} against a terrestrial \ac{RAN} using a numerical model based on real traffic trends and showed that leveraging \ac{HAPS} features—self-sustainability, high altitude, and wide coverage—cuts energy consumption by up to 30\%.

\subsection{Machine Learning}
\label{sub:machine-learning}
\ac{AI} has been utilized to tackle several critical challenges facing current and future wireless networks~\cite{bakambekova2024interplay}. 
Notably, ML-based algorithms can be applied to \ac{HAPS} to optimize key functions such as topology management, \ac{RA}, handover management, and improve interference mitigation, ensuring more efficient, reliable, and adaptive communication networks~\cite{kurt2021vision}. These algorithms can dynamically adjust to changing conditions, reducing disruptions and maximizing service quality.

\Ac{RL} is increasingly applied to \acp{NTN} to optimize resource allocation, platform positioning, and trajectory control. By leveraging \ac{MDP} frameworks, \ac{RL} enables aerial platforms like \ac{HAPS} and \acp{UAV} to autonomously adapt and enhance network performance under real-world constraints.
In their survey, Naous~\textit{et al.}~\cite{naous2023reinforcement} reviewed \ac{RL} applications for \acp{NTN}—including \ac{LEO} satellites, \ac{HAPS}, and \acp{UAV}—focusing on resource allocation, platform positioning, and trajectory control under NTN-specific constraints. Using \ac{MDP} frameworks, it provided a taxonomy of current research, assessed study realism, and identified open challenges to narrow the simulation-to-reality gap.

In this context, Nguyen~\textit{et al.}~\cite{nguyen2022deep} developed a deep \ac{RL}-based offloading strategy for \ac{HAPS}-assisted vehicular networks. The partial task offloading problem was formulated as a \ac{MDP} and solved using the \ac{DDPG} algorithm. Among all evaluated schemes, the \ac{DDPG}-based approach achieves the highest task success rate, with only a 0.7\% failure rate when the task size is 1.2 Mbits.

Dahrouj, Liu, and Alouini~\cite{dahrouj2023machine} highlighted the advantages of optimizing user scheduling in space-\ac{HAPS}-ground networks by ensembling deep neural networks. In particular, it was demonstrated that, when the ensemble size increases, the suggested approach performs better than all existing offline optimization strategies, with a sum-rate improvement of up to 26\%.
Also, Sharifi~\textit{et al.}~\cite{sharifi2023deep} proposed a deep SARSA-based \ac{RL} algorithm for user scheduling in a wireless network with a \ac{HAPS} backup. Modeled as an \ac{MDP}, it operated with outdated \ac{CSI} to maximize sum-rate while minimizing active antennas at the \ac{HAPS} for energy savings. Results showed near-optimal sum-rate performance, and a heuristic method was also introduced for solving the scheduling problem.

Jo~\textit{et al.}~\cite{jo2022deep} introduced a multi-agent deep Q-learning algorithm for \ac{HAPS} power control to reduce outage probability while satisfying interference constraints with \acp{TN}. A double deep Q-learning variant was also introduced to avoid action-value overestimation. Simulations showed that agents collaboratively learned a near-optimal power control policy through effective state, reward, and training design.

Later, a dynamic coverage path planning algorithm was proposed by Liu~\textit{et al.}~\cite{liu2023dynamic} for multi-airship formations to overcome the limitations of static approaches for stratospheric airship formations in dynamic coverage scenarios. Using \ac{RL}, the method stored experiences in a pool for rapid updates via semi-centralized exploration and centralized playback, with task-specific rewards improving network coverage.

Meanwhile, Arani, Hu, and Zhu~\cite{arani2023haps} considered a heterogeneous network with \acp{UAV} and \ac{HAPS} as aerial \acp{BS}, aiming at optimizing fairness and \ac{QoS} for  \acp{GUE}. The proposed approach combines \ac{DRL} with fixed-point iteration to determine the optimal \acp{UAV} positions, \ac{3D} trajectories, and channel allocation strategies based on real-time traffic~load.

Minani~\textit{et al.}~\cite{minani2023fundamental} introduced an \ac{RA} method for uplink \ac{HAPS}-\ac{LEO} communication using multi-carrier \ac{NOMA}. The problem was formulated as a mixed-integer nonlinear program under power and channel constraints, and solved with a multi-agent double deep Q-network employing prioritized experience replay and priority sampling to handle dynamic resources and short satellite visibility.
Later, Le~\textit{et al.}~\cite{le2024resource} studied \ac{HAPS} as flying \acp{BS} serving \acp{GUE} with \ac{NOMA} in a downlink \ac{THz} system. The method optimized power allocation, bandwidth, antenna beamwidth, and \ac{HAPS} altitude to improve transmission rates while meeting data requirements. A \ac{DRL}-based \ac{DDPG} algorithm was applied to manage network dynamics and complexity, maximizing long-term rewards with continuous decision variables.

Recent work shows that \ac{FL} can strengthen \ac{HAPS}-enabled networks by enhancing privacy and lowering communication overhead. Approaches like \ac{SVM}-based \ac{FL} for task offloading and hierarchical \ac{FL} with \ac{HAPS} as distributed servers improve model accuracy, energy efficiency, and reduce training time for large-scale applications.
For example, Wang~\textit{et al.}\cite{wang2021federated} addressed the minimization of energy and time consumption in mobile-edge computing-enabled \ac{HAPS} networks. It formulates an optimization problem for user association, service sequencing, and task allocation, and proposes a \ac{SVM}-\ac{FL} algorithm to predict user associations without sharing private data. Simulations demonstrate up to 15.4\% improvement in energy and time efficiency compared to terrestrial centralized methods.

Addressing the limitations of traditional \ac{ML} in satellite communications,  Elmahallawy and Luo \cite{elmahallawy2022fedhap} introduced a \ac{FL} framework that employs \ac{HAPS} as distributed parameter servers for \ac{LEO} satellite constellations. The proposed framework integrates a hierarchical communication architecture along with model dissemination and aggregation algorithms, significantly reducing training time from days to hours while improving model accuracy.
To enable global-scale \ac{FL} in \acp{NTN}, a distributed hierarchical \ac{FL} framework was proposed by Farajzadeh, Yadav and Yanikomeroglu~\cite{farajzadeh2025federated}, with \ac{HAPS} as intermediate servers. Integrating \ac{LEO} satellites and ground clients while using \ac{GEO} and \ac{MEO} satellites for model relaying, the framework enhances privacy, improves model accuracy, manages latency, and boosts scalability. Numerical results confirm its efficiency in training loss reduction and network optimization.

Güven and Karabulut Kurt~\cite{guven2022cnn} proposed a CNN-aided receiver for \ac{HAPS}--\ac{LEO} links that jointly addresses channel estimation and \ac{CFO} estimation under high Doppler and rapid channel variations. Supervised CNN models are trained to learn the nonlinear mapping from received pilot observations to channel and \ac{CFO} estimates, reducing estimation errors compared to conventional pilot-based methods. Simulation results demonstrated improved synchronization accuracy and detection performance in representative HAPS--LEO scenarios.

Mahyastuty~\textit{et al.}~\cite{mahyastuty2022massive} investigated CNN-assisted and CNN-based \ac{SIC} replacement for PD-\ac{NOMA} in \ac{HAPS}-enabled mMTC and WSN scenarios, where the HAPS acts as an aggregation node. The proposed approach targets low-complexity massive access by leveraging CNNs for multiuser detection, and simulation results showed improved \ac{BER} performance compared with conventional PD-\ac{NOMA} receivers under the considered system configurations.
Later, Mahyastuty, Pamukti, and  Oktaviana~\cite{mahyastuty2025deep} introduced a CNN-based decoding framework for power-domain \ac{NOMA} in \ac{HAPS} systems to overcome the limitations of successive interference cancellation (\ac{SIC}), particularly error propagation and sensitivity to user dynamics. The study analyzed the impact of user location and mobility on decoding performance, showing that users closer to the HAPS achieve target \ac{BER} at lower \ac{SNR}, while increasing user density raises the required \ac{SNR} for neighboring users.

\begin{table*}[t]
\centering
\caption{Mapping between learning paradigms, HAPS-specific constraints, representative applications, and open challenges in HAPS-enabled NTN.}
\label{tab:ai_haps_mapping}
\renewcommand{\arraystretch}{1.2}
\setlength{\tabcolsep}{6pt}
\begin{tabular}{|p{2.5cm}|p{3.7cm}|p{4.5cm}|p{5.2cm}|}
\hline
\textbf{Learning Paradigm} &
\textbf{HAPS-Specific Constraints Addressed} &
\textbf{Representative Applications in HAPS/NTN} &
\textbf{Pending Challenges and Open Issues} \\
\hline

\textbf{DRL / RL}~\cite{naous2023reinforcement, arani2023haps, nguyen2022deep, sharifi2023deep, liu2023dynamic, jo2022deep, minani2023fundamental, le2024resource} &
Mobility and non-stationarity (platform/user dynamics, intermittent visibility).
Real-time decision latency (online control and scheduling).
Energy budget for onboard inference/training.
 &
Platform positioning and 3D trajectory control (HAPS/UAV).
Dynamic \ac{RA}, power control, and interference mitigation.
User scheduling under outdated/partial \ac{CSI}.
Joint beamforming and \ac{RIS} configuration.
&
Sim-to-real gap; robustness to channel/traffic model mismatch.
Sample inefficiency and training instability in non-stationary NTN.
Multi-agent scalability and coordination overhead.
Safety/constraint satisfaction (hard QoS, interference, energy).
\\
\hline

\textbf{FL }~\cite{wang2021federated,farajzadeh2025federated,elmahallawy2022fedhap} 
&
Privacy constraints (no raw-data sharing).
Limited backhaul / feeder link capacity.
Intermittent connectivity and long RTTs in NTN.
&
User association and task offloading prediction at the edge.
HAPS-assisted distributed training (HAPS as aggregator/parameter server).
Global-scale hierarchical learning across ground--HAPS--satellite tiers.
&
Communication overhead of model updates; compression/quantization needs.
Stragglers and synchronization under intermittent links.
Statistical heterogeneity (non-IID data) and personalization requirements.
Secure aggregation and adversarial/poisoning robustness.
\\
\hline
\textbf{CNN}~\cite{guven2022cnn, mahyastuty2022massive, mahyastuty2025deep} 
&
High Doppler, residual \ac{CFO}, and fast channel variations in HAPS/LEO links.
Strong multiuser interference and imperfect \ac{SIC} in NOMA-based HAPS systems.
Non-stationary propagation, traffic dynamics, and domain shift across scenarios.
&
CNN-based channel and \ac{CFO} estimation for HAPS-assisted and HAPS--LEO links.
CNN-based multiuser detection/decoding for PD-NOMA in HAPS and mMTC scenarios.
Improved BER, throughput, and robustness compared to conventional receivers.
&
Dependence on representative training datasets; sensitivity to domain mismatch.
Increased computational and energy cost at HAPS/UE receivers.
Limited interoperability and deployment trade-offs (accuracy vs. latency).
 \\
\hline
\end{tabular}
\end{table*}

\subsection{Cybersecurity and Adversarial Threats in HAPS-Enabled NTN}
\label{sub:security}

Cybersecurity threats can put \ac{NTN} systems at significant risk. Due to exposed wireless links and strong reliance on \ac{PNT}, platforms such as \ac{HAPS} are particularly vulnerable, making robust security mechanisms essential.
%%{Cybersecurity of High-Altitude Platform Stations: Threat Taxonomy, Attacks and Defenses with Standards Mapping --- DDoS Attack Use Case}
By mapping threats to \ac{HAPS} subsystems (such as communication payload, power management, telemetry, tracking, and flight control, as well as ground station interfaces), Hjaiji, Ouni and Alouini~\cite{hjaiji2026cybersecurity} provided a \ac{HAPS}-focused cybersecurity analysis.
The authors specifically highlighted jamming, replay, and system intrusion attacks, data manipulation and spoofing, DoS/DDoS, adversarial \ac{ML} attacks, and supply-chain risks; and extensively reviewed the resent advancements in defense mechanisms applicable to \ac{HAPS} subsystems.
A simulation-based distributed-denial-of-service case study that quantifies service/control-plane availability deterioration under assault was also provided as a significant technical component of the study.

% %%% Jamming %%%
%%% Cooperative Beamforming for UAV Anti-jamming (IEEE IoT-J 2022)
Yu~\textit{et al.}~\cite{yu2022let} examined the communication between several transmitters and a multi-antenna \ac{UAV} when jamming is present. They suggested cooperative beamforming techniques that reduce deliberate jamming and inter-user interference. The analysis showed how anti-jamming performance is affected by geometry, \ac{UAV} motion, and beam design. In addition to \ac{HAPS}-assisted \ac{NTN} systems, it offers practical techniques for aerial-layer resiliency.

%%{SATCOM Earth Station Arrays Anti-Jamming Based on MVDR Algorithm}
Using the minimum variance distortionless response (MVDR) beamforming approach, Xi, Liu and Zhao~\cite{xi2023satcom} investigated anti-jamming for satellite ground-station arrays. After creating a signal model that takes direction-dependent array/reflector effects into consideration, proposed approach calculates MVDR weights to reduce interference. Simulations verified the improvements in output \ac{SINR} in relation to jammer angle, power, and snapshot availability. Aiming array-level interference reduction in \ac{SatCom}/\ac{NTN} feeder lines, the study may be potentially generalized to \ac{HAPS}-related application.

%%{Dark Side of HAPS Systems: Jamming Threats towards Satellites}
Otay, Humadi, and Karabulut Kurt~\cite{otay2023dark} investigates a security-relevant scenario in which a \ac{HAPS} prevents \ac{LEO}-to-ground communication by acting as a \emph{downlink jammer}. A direct satellite-ground link with a jamming \ac{HAPS} and a cooperative two-satellite path (direct+relay) under the same \ac{HAPS} jammer are the two scenarios that are examined. The authors derive analytical formulas based on outages that quantify the degradation of link dependability caused by jammer power and shape. One important lesson is that satellite cooperation and relaying can lower the likelihood of a serious disruption because, in the worst scenario, both pathways must be jammed at the same time.

%%{A Multi-Agent Deep Reinforcement Learning Anti-Jamming Spectrum-Access Method in LEO Satellites}
Cao~\textit{et al.}~\cite{cao2025multi} adapts spectrum-access choices for \ac{LEO} satellite downlinks under jamming by using multi-agent deep reinforcement learning. Agents choose channels and actions that minimize the impact of jamming while maintaining throughput by learning policies from observation histories. Performance was compared to less adaptable techniques using a jammer model with channel-switching dynamics. For \ac{NTN} resource management, including \ac{HAPS}-based architectures, it offers a practical illustration of autonomy-driven anti-jamming.

% %%% Spoofing/data manipulation %%%
%%{GNSS Spoofing Detection Based on Coupled Visual/Inertial/GNSS Navigation System}
Gu, Xing, and You~\cite{gu2021gnss} proposed a \acp{GNSS} spoofing detection approach based on coupled visual/inertial/\ac{GNSS} positioning algorithm. To lower false alarms, it incorporates confidence handling and residual/consistency testing in the fused navigation solution. Datasets and experimental settings that mimic spoofing effects are used to validate the method. It is pertinent to aerial platforms where secure control and safe flying depend on strong~\ac{PNT}.

%%{GPS Spoofing Detection Method for Small UAVs Using 1D Convolution Neural Network}
A lightweight 1D-\ac{CNN} detector targeted at small \acp{UAV} with an emphasis on early detection of \ac{GPS} spoofing was proposed by Sung~\textit{et al.}~\cite{sung2022gps}. It reports practical runtime/energy feasibility on embedded hardware and assesses detection performance against classical \ac{ML} baselines. The technique supports defensive actions (such as safe landing logic) and was shown under practical \ac{UAV} operational restrictions. In layered \ac{NTN} ecosystems, it is a powerful applicable reference for \acp{UAV} security.

%%{Secure and Efficient Group Handover Protocol in 5G Non-Terrestrial Networks}
The issue of huge, nearly simultaneous handovers in \ac{NTN} (e.g., \ac{LEO} footprints moving quickly), which might overwhelm control signaling and create security gaps, was addressed by Zhang~\textit{et al.}~\cite{zhang2024secure}. It suggests a safe group-handover protocol that combines steps to lower signaling costs while maintaining authorization and authentication objectives. The study compares cost and delay to baseline methods and assesses performance using discrete-event simulation.

%%{LEO-Range: Physical Layer Design for Secure Ranging with Low Earth Orbiting Satellites}
For \ac{LEO}-to-device links, Coppola~\textit{et al.}~\cite{coppola2025leo} develop a physical-layer ranging technique that can withstand spoofing and relay-style manipulation. It examines attack surfaces and presents a safe range waveform/processing chain that works with realistic \ac{OFDM}-style systems. The authors assess ranging accuracy and resilience under representative satellite channel effects and offer security justifications. The end product is a tangible foundation for \ac{NTN} systems' secure timing/ranging (\ac{PNT}-adjacent) services.

%%%Covert communication%%%
%%{Covert Communication in Integrated High Altitude Platform Station Terrestrial Networks}
Wu~\textit{et al.}~\cite{wu2023covert} examine covert communication in an integrated \ac{HAPS}—\ac{TN}, approaching security from a \emph{low likelihood of detection} perspective. To conceal the signal from a warden or eavesdropper, it employs a constant-power auxiliary node that sends artificial noise. The authors present closed-form formulas for outage probability and effective covert rate after deriving the covert constraint connecting transmitter and auxiliary node powers. These findings apply to \ac{HAPS} deployments where a design objective under surveillance is to conceal transmission behavior (not just encryption).

\section{Open Research Challenges}
\label{sec:challenges}

A key direction for future \ac{6G} systems is the integration of \acp{TN} and \acp{NTN} through \ac{HAPS}-enabled connectivity.
Although several related challenges have been partially studied in terrestrial, satellite, and standalone \ac{HAPS} networks, many issues remain unresolved—or become more critical—under large-scale, multi-tier \ac{TN}/\ac{NTN} integration.
In particular, heterogeneous propagation conditions, wide coverage areas, platform mobility, and stringent latency and reliability constraints introduce new cross-layer coupling effects spanning the physical, MAC, and network layers.
The following summarizes the key open research challenges (see Table~\ref{tab:haps_open_challenges_summary}), clearly distinguishing \ac{HAPS}-specific open issues from those already identified in prior work.

\begin{table*}[htb]
\centering
\caption{Summary of Open Research Challenges in \ac{HAPS}-Enabled \ac{TN}/\ac{NTN} Integration}
\label{tab:haps_open_challenges_summary}
\renewcommand{\arraystretch}{1.05}
\setlength{\tabcolsep}{6pt}
\begin{tabular}{| p{6cm} | p{11.0cm} |}
\hline
\textbf{Research Challenge (Time Horizon, Type)} & \textbf{Key Open Issues} \\
\hline
Channel Modeling (Short-term, Technical) &
Especially at mmWave and higher frequencies, joint consistent modeling spanning terrestrial--\ac{HAPS}--satellite links under non-stationarity, large coverage, and weather-/terrain-dependent effects. \\
\hline
Spectrum Management (Short- to Long-term, Technical/Regulatory) &
Interference cooperation and scalable spectrum sharing across diverse tiers with dense deployments and dynamic beams.\\
\hline
Mobility Management and Handover (Short-term, Technical) &
Robust signaling and precise handover time in the face of diverse propagation delays, quasi-stationary platforms, and large coverage areas. \\
\hline
Multi-Connectivity (Short-term, Technical) &
Bearer splitting/duplication across heterogeneous networks under dynamic mobility: control-plane overhead, synchronization, and scalability. \\
\hline
Interoperability and Standardization (Long-term, Technical/Standardization) &
Protocol development beyond existing norms to facilitate real-time multi-tier coordination, huge connection, and large delays. \\
\hline
Energy Efficiency (Short- to Long-term, Technical) &
Energy harvesting, storage, and energy-conscious scheduling are all components of sustainable HAPS operation under stringent energy limitations. \\
\hline
Legal and Regulatory Issues (Long-term, Regulatory) &
Large-scale \ac{HAPS} deployment requires international cooperation, safety, spectrum control, and airspace management.\\
\hline
Deployment Costs (Long-term, Economic) &
For widespread adoption, public-private collaborations, sustainable business models, and economic viability are required. \\
\hline
Security, Jamming, and Adversarial \ac{ML} Threats (Short- to Long-term, Technical) &
Protection in wide-area \ac{LoS} situations from hostile \ac{ML} attacks, spoofing, cyberthreats, and jamming. \\
\hline
\end{tabular}
\end{table*}

 \textbf{Channel Modeling:} As previously examined in the literature (e.g., \cite{zheng2013modeling,hsieh2019propagation,yilmaz2023path,palanci2022high,zain2013improving}), Doppler effects, delay spread, and atmospheric attenuation are among the important propagation characteristics of \ac{HAPS}-enabled \acp{NTN} that have been examined under a variety of assumptions.
However, the requirement for jointly consistent representations across diverse links and large geographic areas keeps channel modeling open in \ac{HAPS}-enabled \ac{TN}/\ac{NTN} integration.
Channel coherence is affected by non-stationarity brought on by platform motion, altitude changes, and atmospheric dynamics, while spatial heterogeneity is introduced by effects that rely on the topography and the weather.
At mmWave and higher frequencies, when attenuation and beam misalignment can significantly impair link reliability, these difficulties are made worse.

\textbf{Spectrum:} As reported in earlier studies on spectrum sharing and interference management (e.g., \cite{grace2005improving,liu2022interference,wei2023spectrum,li2020hierarchical}), spectrum coexistence between \ac{HAPS}, satellite, and terrestrial systems is a well-known challenge.
This problem is exacerbated by diverse coverage scales, dynamic beamforming, and dense node distributions in \ac{HAPS}-enabled multi-tier deployments.
Scalable coordination methods, scheduling that considers interference, and adaptive spectrum access techniques that function well in the face of mobility and quickly shifting interference situations are among the unresolved difficulties.

\textbf{Multi-connectivity:} It has been demonstrated that multi-connectivity can support strict \ac{URLLC} requirements, including in circumstances related to \ac{NTN}~\cite{salehi2022ultra,salehi2024reliability}.
But even with the Stage-2 architecture outlined in 3GPP TS~37.340~\cite{3GPP37340}, there is still room for multi-connectivity advantages to be practically realized in \ac{HAPS}-enabled \ac{NTN} deployments.
Syncing across heterogeneous links, scalability under dynamic mobility situations, and higher signaling overhead are among the main obstacles~\cite{sadovaya2024enhancing}.

\textbf{Mobility Management:} Numerous studies have been conducted on mobility and handover in \ac{HAPS}-assisted \acp{NTN}~\cite{juan20205g,juan2022handover,li2019handover}.
However, mobility management is still a challenge in \ac{HAPS}-enabled \ac{TN}/\ac{NTN} integration because of diverse propagation delays, quasi-stationary aerial platforms, and vast coverage regions.
The terrestrial assumptions that underlie current \ac{LTE}/\ac{NR} mobility procedures~\cite{3GPP36300,3GPP38300} may result in inefficient handover timing and signaling, especially for services that are sensitive to latency and reliability under high mobility and limited signal strength gradients~\cite{yu2022performance}.

\textbf{Interoperability:} Interoperability is still a major obstacle, as stated in current \ac{TN}/\ac{NTN} standardization initiatives (e.g., \cite{TR38821,3gpp-tr-38.811}).
Although the present standards offer a starting point, more protocol development is needed to accommodate huge connectivity, substantial propagation delays, and real-time coordination between satellites, \ac{HAPS}, and terrestrial nodes.

 \textbf{Energy:} Energy efficiency is still a significant challenge, as was previously mentioned in the context of energy-efficient \ac{HAPS} operation (e.g., \cite{rallage2022improving,bolandhemmat2019energy}).
In order to guarantee continued operation under changing lighting and traffic situations, there are still unresolved challenges with effective energy harvesting, long-term storage, dynamic power allocation, and energy-aware scheduling.

 \textbf{Legal Regulations:} Large-scale \ac{HAPS} deployment presents unresolved issues with airspace management, safety, spectrum regulation, and international cooperation, despite the fact that regulatory features of aerial and non-terrestrial platforms have been taken into account in previous work.

 \textbf{Deployment Costs:} The economic viability of large-scale \ac{HAPS} deployment is still up for debate, despite the fact that techno-economic elements have been examined in previous research.
To justify long-term investment, viable revenue models and public-private partnerships are necessary due to high deployment and maintenance costs.

 \textbf{Security, Jamming, and Adversarial Machine Learning Threats:} As was previously mentioned in the context of \ac{HAPS}-enabled \ac{NTN} security~\cite{coppola2025leo,zhang2024secure,ren2023novel,xi2023satcom,yang2020intelligent,yu2022let,gu2021gnss,sung2022gps,cao2025multi,otay2023dark,hjaiji2026cybersecurity,wu2023covert}, \ac{HAPS}-enabled systems are susceptible to cyberattacks, spoofing, and jamming due to their exposed wireless configuration.
Additionally, the growing dependence on machine learning for optimization and control raises the possibility of adversarial attacks, underscoring the necessity of strong and reliable security measures.

\section{Discussion and conclusion}\label{Disc-conclu}
%\begin{comment}
The evolution toward \ac{6G} networks demands a seamless integration of \acp{TN} with \acp{NTN} to achieve ubiquitous, high-capacity, and low-latency connectivity. \ac{HAPS} operating in the stratosphere, emerge as a pivotal component in this vision, bridging the gap between \acp{TN}, \ac{LEO} satellites, and \acp{UAV}. This survey provided a comprehensive and up-to-date review of \ac{HAPS} networks, spanning their architectures, frequency operations, enabling technologies, and emerging applications.

In terms of architecture, \ac{HAPS} offer flexible deployment models, functioning as either  transparent bent-pipe relays or regenerative platforms with onboard processing, each suited to specific application and system complexity requirements. \ac{HAPS} networks incorporate key infrastructure elements, including \acp{UE}---ranging from portable devices to vehicular and \ac{IoT} terminals---and gateway stations that enable high-capacity feeder and service links. The aerial platform design, whether aerostatic, aerodynamic, or hybrid, offers trade-offs between payload capacity, mobility, and operational resilience, allowing tailored solutions depending on mission requirements.

This study also explored the critical role of frequency band selection in \ac{HAPS} communications. Sub-7 GHz bands provide robust coverage but limited data rates, while cmWave and mmWave bands offer higher throughput with manageable trade-offs in range and propagation challenges. Emerging sub-THz and \ac{FSO} links promise ultra-high-capacity backhaul and inter-platform connectivity, although they require advanced channel modeling and adaptive error correction techniques to mitigate environmental sensitivity. Careful spectrum management and coexistence strategies will be essential as \ac{HAPS} operate alongside \ac{TN} and other \ac{NTN} nodes in increasingly congested frequency spaces. 

Notably, recent \ac{3GPP} releases have incorporated \ac{NTN} support into the 5G \ac{NR} framework, standardizing the use of these bands for seamless terrestrial and non-terrestrial integration, particularly in the sub-7 GHz and \ac{mmWave} ranges.

A diverse range of application scenarios underscores the transformative potential of \ac{HAPS}. In connectivity enhancement, \ac{HAPS} extend coverage to remote and underserved areas and offer resilient infrastructure during natural disasters and unexpected traffic surges. In the domain of network infrastructure, \ac{HAPS} facilitate robust backhaul solutions and dynamic handover management, ensuring stable service delivery. \ac{HAPS} also serve as key enablers for emerging applications, supporting massive \ac{IoT} deployments, \ac{MEC} for latency-sensitive services, \ac{URLLC} for autonomous control, and aerial data centers capable of delivering cloud services and immersive \ac{AR}/\ac{VR} experiences. Notably, their quasi-stationary positioning and large coverage areas make \ac{HAPS} ideal for supporting structured \ac{UAV} corridors and future aerial highways.

Furthermore, \ac{ML} and \ac{AI} are increasingly integrated into \ac{HAPS} operations, enhancing network performance and automation. \ac{ML} techniques are pivotal for resource management, dynamic \ac{UE} association, mobility prediction, beamforming optimization, and interference mitigation. Intelligent algorithms enable \ac{HAPS} to adapt to dynamic environments, optimize energy consumption, and ensure service continuity even in challenging scenarios.

Despite these advancements, several challenges remain open. Scaling \ac{HAPS} networks to multi-\ac{UE}, multi-platform scenarios requires sophisticated coordination mechanisms. Achieving seamless \ac{TN}-\ac{NTN} integration mandates refined architectural strategies and flexible protocol designs. Energy efficiency, especially for long-endurance missions, remains a critical concern, as does the need for robust system designs to withstand stratospheric environmental conditions. Additionally, efficient interference management, advanced beamforming solutions, and spectrum-sharing frameworks enabling coexistence with terrestrial systems must be developed to fully exploit high-frequency bands.
Compared to previous surveys, this work distinguishes itself by providing a comprehensive and integrated analysis of \ac{HAPS} technologies, architectures, frequency challenges, use cases, and emerging machine learning applications. It offers a forward-looking perspective on deploying \ac{HAPS} as independent, intelligent network nodes rather than simple relays, solidifying their role in future \acp{SAGIN}.

In conclusion, \ac{HAPS} are envisioned as a cornerstone of 6G networks, enabling resilient, scalable, and sustainable global connectivity. Future research should prioritize:
\begin{itemize}
    \item Cross-layer optimization for \ac{TN}-\ac{NTN}-\ac{HAPS} integration.
    \item Scalable \ac{AI}-driven network management and adaptive control.
    \item Efficient multi-band operation strategies for dynamic spectrum access.
    \item Sustainable platform designs focused on energy harvesting and aerodynamic efficiency.
    \item Robust deployment frameworks to support diverse and evolving applications.
\end{itemize}

With continued innovation, \ac{HAPS} will significantly contribute to bridging the global digital divide and powering the next wave of wireless communication.

\bibliographystyle{IEEEtran}
\bibliography{references.bib}
\clearpage
% David López-Pérez
\begin{IEEEbiography}[{\includegraphics[width=1in,height=1.25in,clip,keepaspectratio]{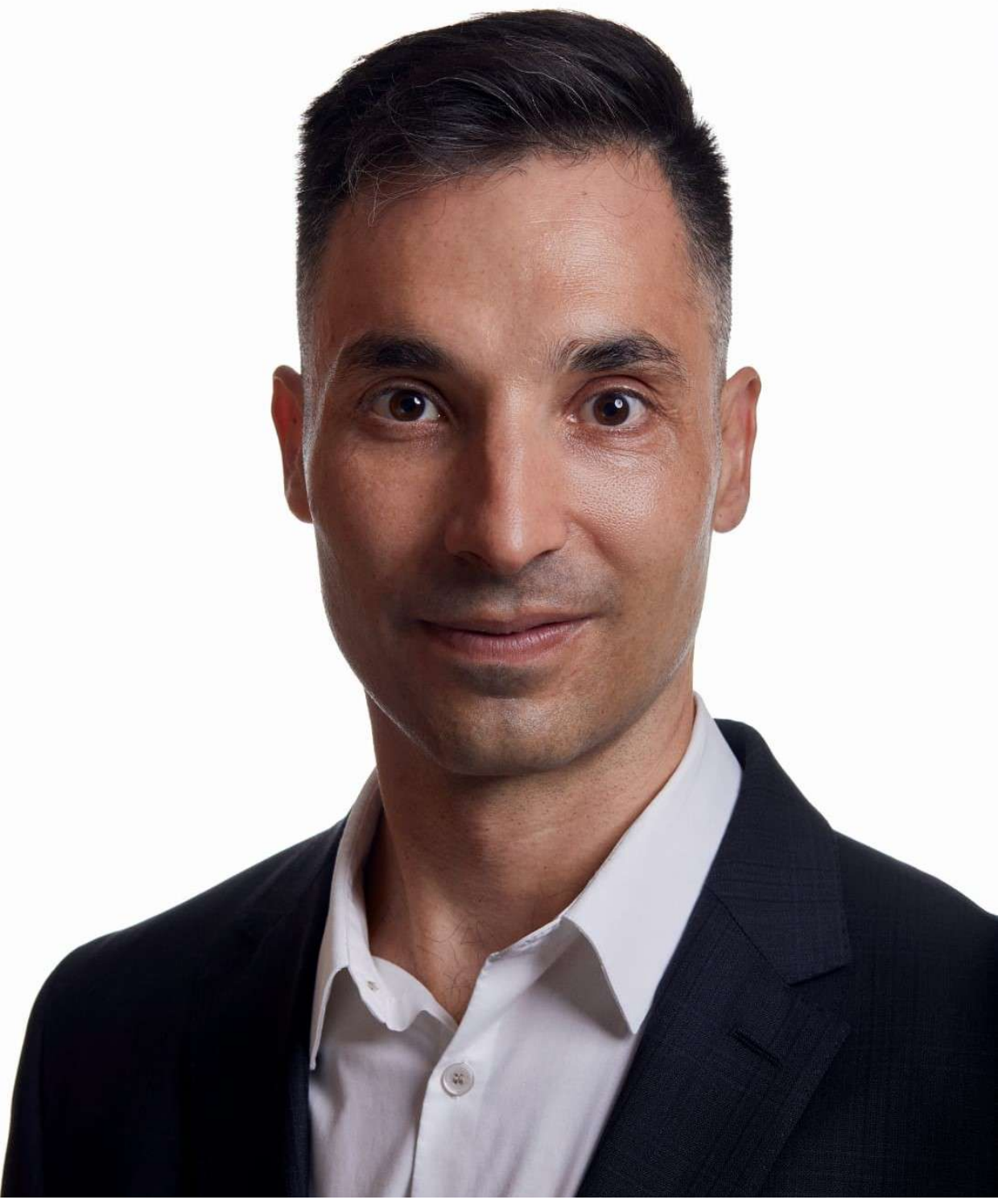}}]
{Dr. David López-Pérez} (Senior Member, IEEE) is a Distinguished Researcher at Universitat Politècnica de València. Before joining this position, David served as an Expert and Technical Leader at Huawei Technologies in Paris and held the title of Distinguished Member of Staff at Nokia Bell Labs in Dublin. Throughout his career, David has focused on the study of both cellular and Wi-Fi networks, with primary research interests in network performance analysis, network planning and optimization, heterogeneous networks,  NTNs, green networking, ML, and the development of new technologies and features. He has authored a book on small cells and another on ultra-dense networks, in addition to publishing over 150 research manuscripts covering a variety of related topics. He is also a prominent inventor with 60+ patents.
\end{IEEEbiography}
\begin{IEEEbiography}[{\includegraphics[width=1in,height=1.25in,clip,keepaspectratio]{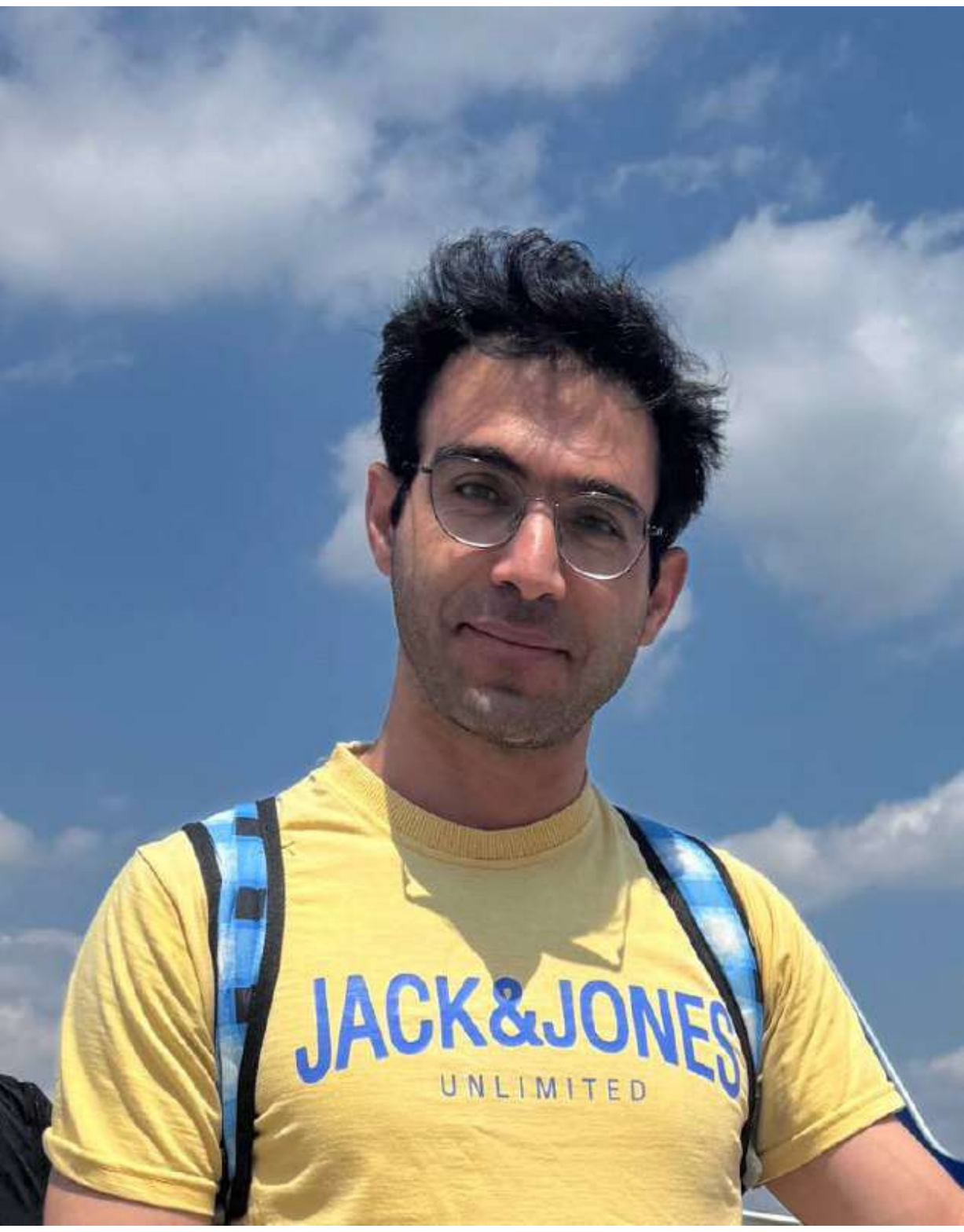}}]
{Azim Akhtarshenas} (Student Member, IEEE) is a Ph.D. candidate at the Universitat Politècnica de València (iTEAM), supervised by Dr. David López-Pérez. His current research focuses on integrated NTNs for UAV/HAPS communications, with emphasis on learning-based radio sensing via DRL approaches. 
He has co-authored work on a power-efficient CNN autoencoder for LoS/NLoS detection in MIMO-enabled UAV networks, DRL-based optimization of UAV aerial base-station flights, and federated learning for aerial and satellite platforms, alongside earlier ML research on open-set visual classification. 
His current projects develop derivative-free stochastic optimization and DRL methods for large-scale iNTN design, spectrum sharing, and coexistence across ground–air–space layers.
\end{IEEEbiography}
\begin{IEEEbiography}[{\includegraphics[width=1in,height=1.25in,clip,keepaspectratio]{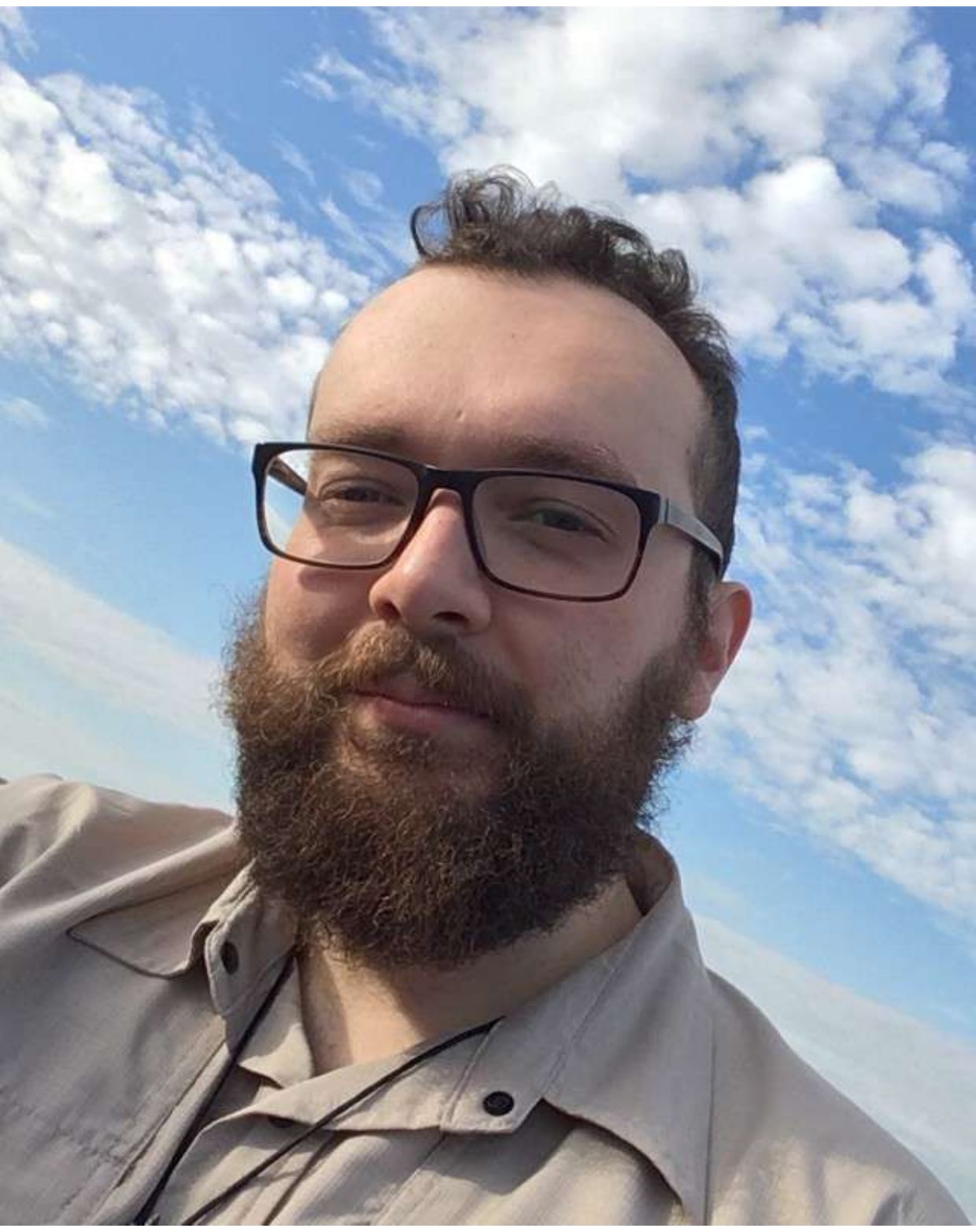}}]
{German Svistunov} (Student Member, IEEE) is a Ph.D. candidate at the Universitat Politècnica de València, supervised by Dr. David López-Pérez. His research focuses on the role of HAPS in TN–NTN integration and the design of HAPS-based network architectures for terrestrial and aerial users, including the potential replacement of terrestrial infrastructure with HAPS-based integrated networks.
Prior to his Ph.D., he worked as an R\&D Engineer at Huawei Technologies, where he conducted research on 5G-related algorithm design, error correction coding, post-5G communications, and ML applications in wireless systems. He is a co-author of several patents and IEEE conference papers.
\end{IEEEbiography}
\begin{IEEEbiography}[{\includegraphics[width=1in,height=1.25in,clip,keepaspectratio]{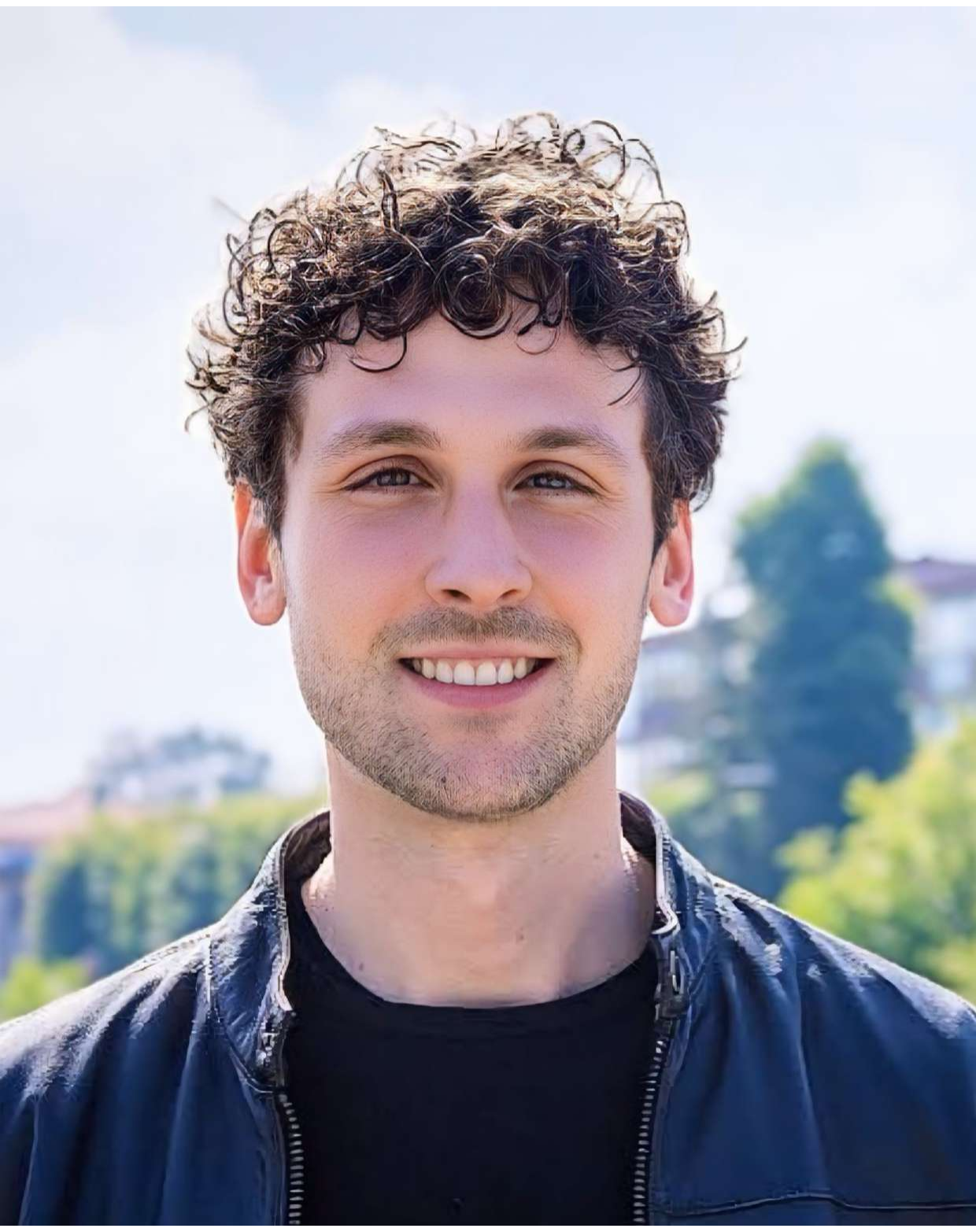}}]
{Marco Giordani} (Senior Member, IEEE) is an Associate Professor in Telecommunications at the Department of Information Engineering (DEI) of the University of Padova. During his Ph.D., he visited New York University (NYU), USA, and TOYOTA Infotechnology Center, Mountain View, CA, USA. He co-authored 80+ published articles in the area of wireless networks, three of which have received Best Paper Awards. He is a recipient of several awards, including the 2018 IEEE Daniel E. Noble Fellowship Award from the IEEE Vehicular Technology Society, the 2021 IEEE ComSoc Outstanding Young Researcher Award for EMEA, and the 2024 IEEE ComSoc Young Professional Award for Best Practitioner. Marco Giordani serves as Editor for the IEEE Transactions of Wireless Communications and the IEEE Transactions on Mobile Computing. He is the Director of the PhD Summer School of Information Engineering (SSIE), the Coordinator of the IEEE Italy Entrepreneurship Committee, and a Member of the same Committee for IEEE R8. His research focuses on 5G/6G cellular networks and PHY/MAC protocol design.
\end{IEEEbiography}
\begin{IEEEbiography}[{\includegraphics[width=1in,height=1.25in,clip,keepaspectratio]{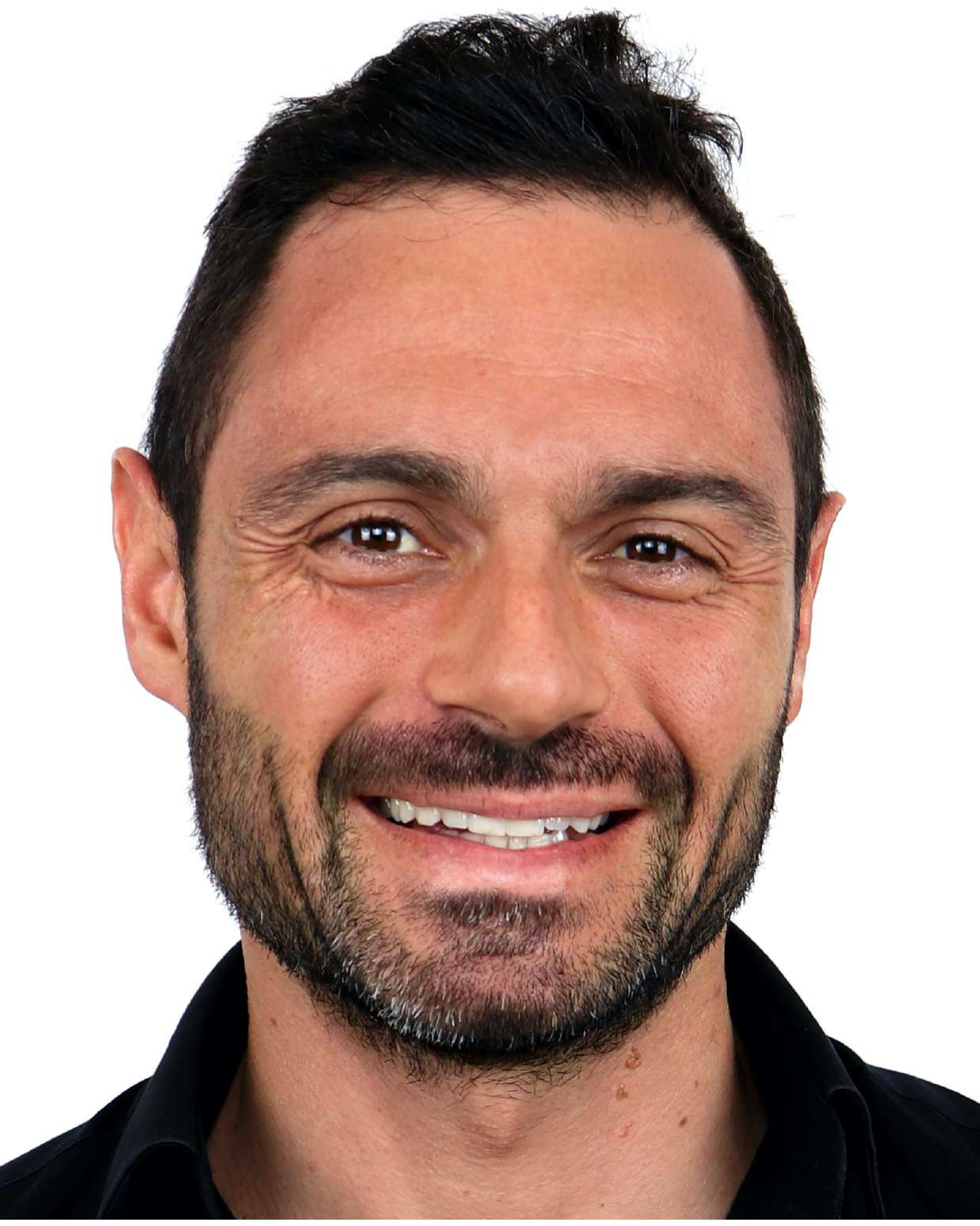}}]
{Giovanni Geraci} is a Senior Principal Research Lead at Nokia Standards and an Associate Professor at Universitat Pompeu Fabra in Barcelona. His current research focuses on AI/ML for wireless communications and the use of LLMs/agents in telecom. He served as an IEEE Distinguished Lecturer, has delivered 100+ international lectures, and received two IEEE early-career awards and three IEEE best paper awards.
\end{IEEEbiography}
\begin{IEEEbiography}[{\includegraphics[width=1in,height=1.25in,clip,keepaspectratio]{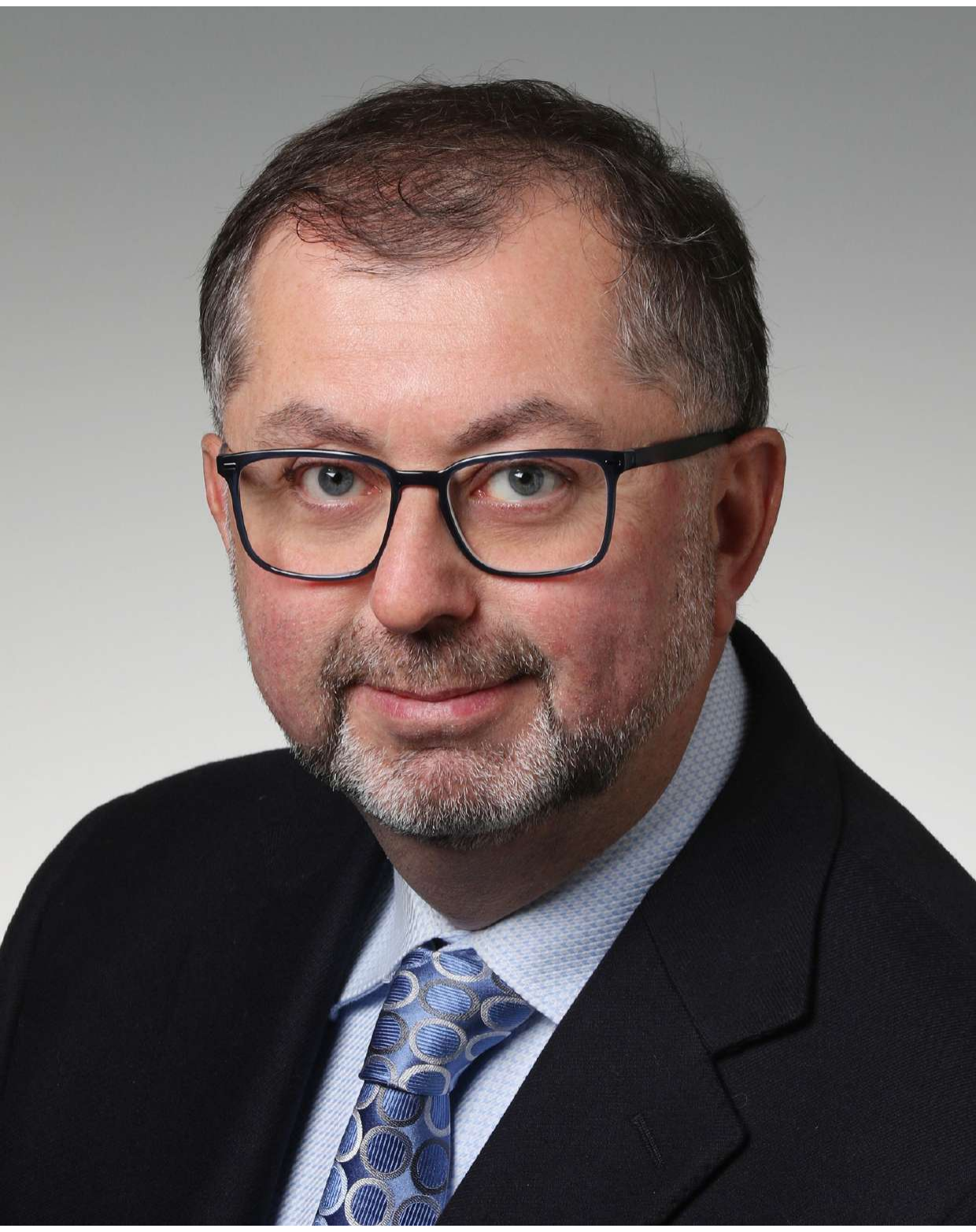}}]
{Halim Yanikomeroglu} is a Chancellor’s Professor in the Department of Systems and Computer Engineering at Carleton University, Canada; he is also the Founding Director of Carleton-NTN (Non-Terrestrial Networks) Lab. He is a Fellow of IEEE, Engineering Institute of Canada (EIC), Canadian Academy of Engineering (CAE), and Asia-Pacific Artificial Intelligence Association (AAIA). Dr. Yanikomeroglu has coauthored papers in 33 different IEEE journals; he also has 42 granted patents. He has supervised or hosted at Carleton 180 postgraduate researchers; several of his team members have become professors around the world. He gives around 25 invited seminars, keynotes, and panel talks every year. He has had a high number of leadership roles in IEEE. He also served as a Distinguished Lecturer for IEEE ComSoc and IEEE VTS. Dr. Yanikomeroglu received several awards for his research, teaching, and service. He holds a PhD degree in electrical and computer engineering from the University of Toronto. 
\end{IEEEbiography}
\begin{figure*}
    \centering
    \includegraphics[width=1\linewidth]{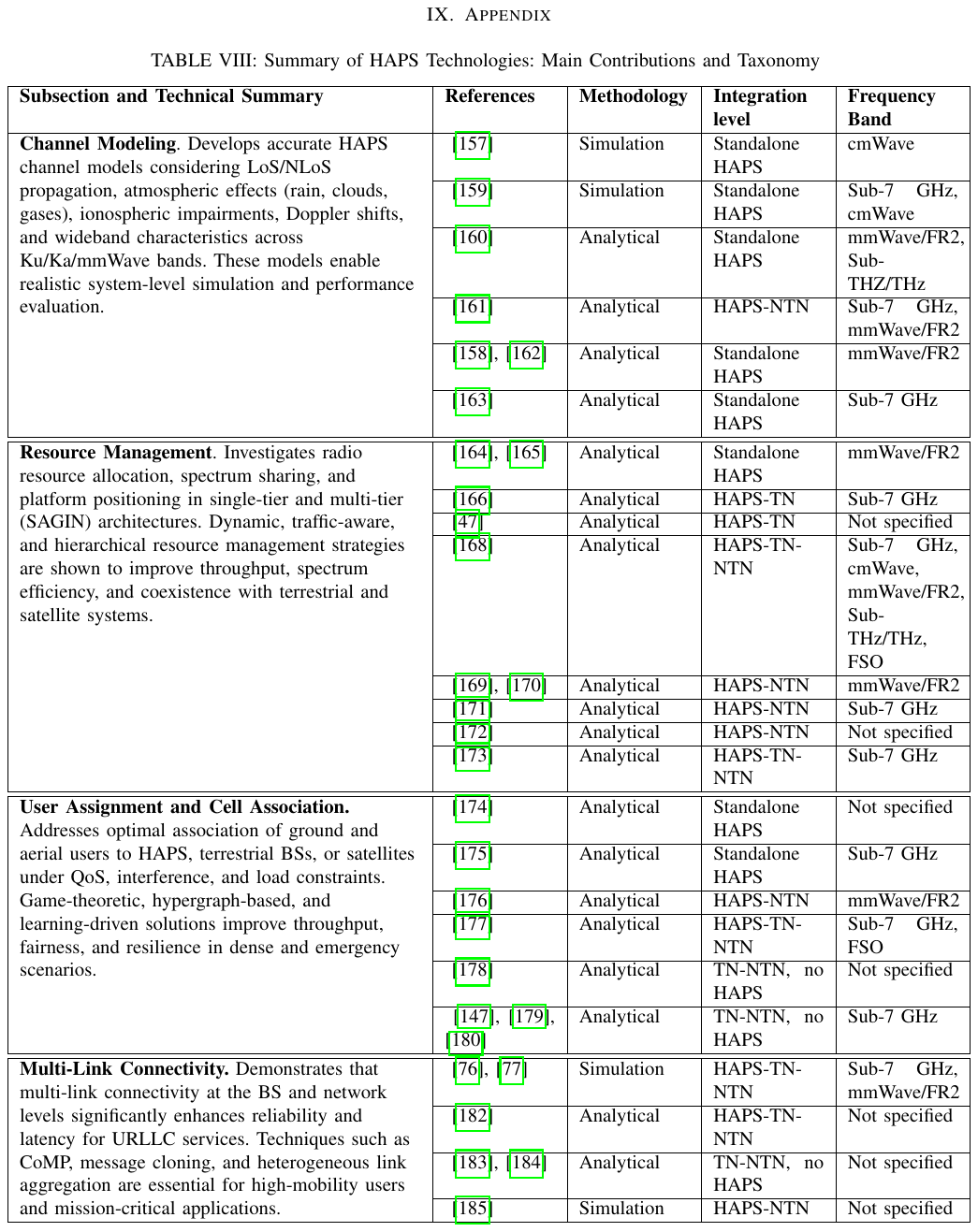}
    \label{fig:placeholder}
\end{figure*}
\begin{figure*}
    \centering
    \includegraphics[width=1\linewidth]{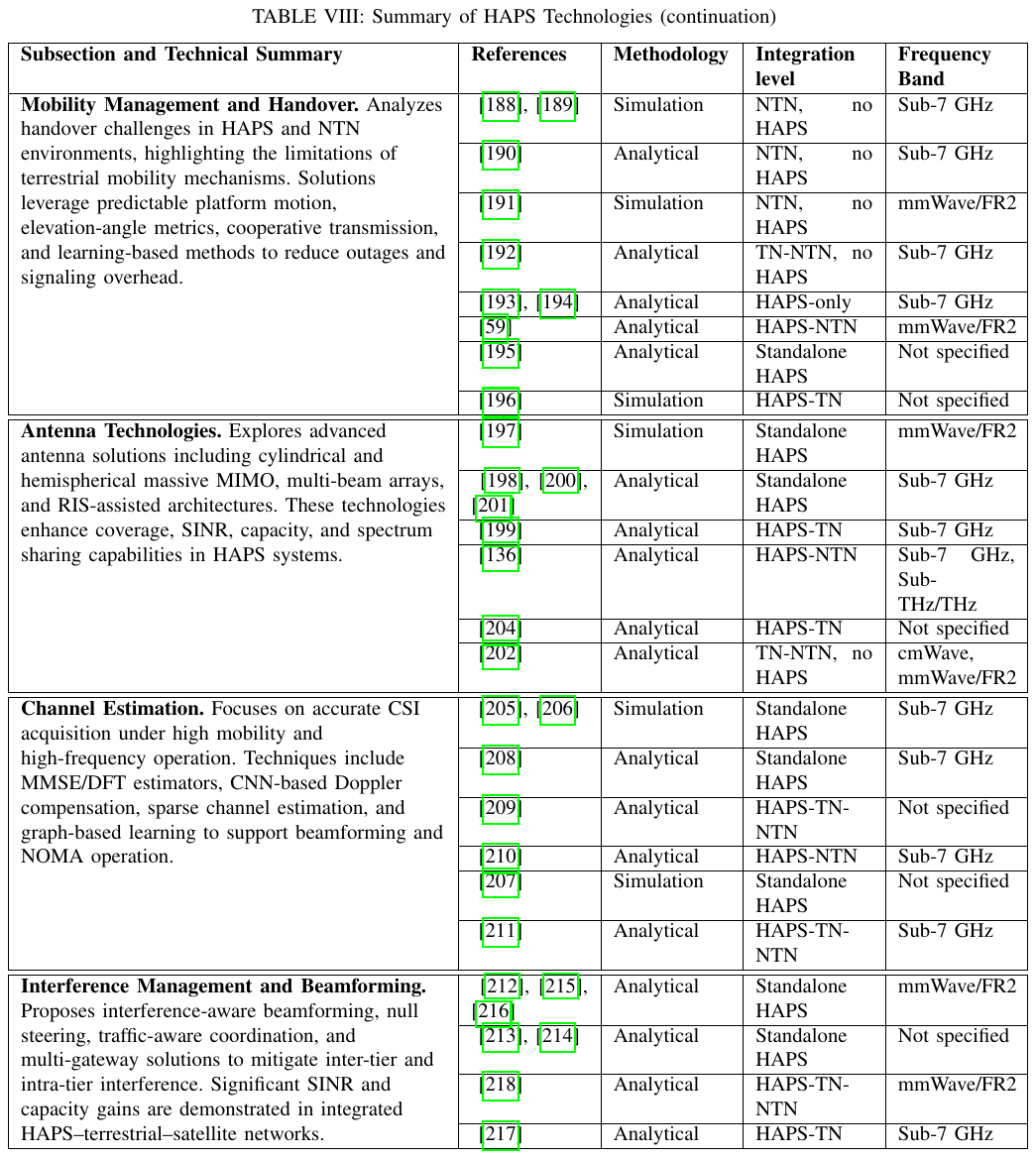}
    \label{fig:placeholder}
\end{figure*}
\begin{figure*}
    \centering
    \includegraphics[width=1\linewidth]{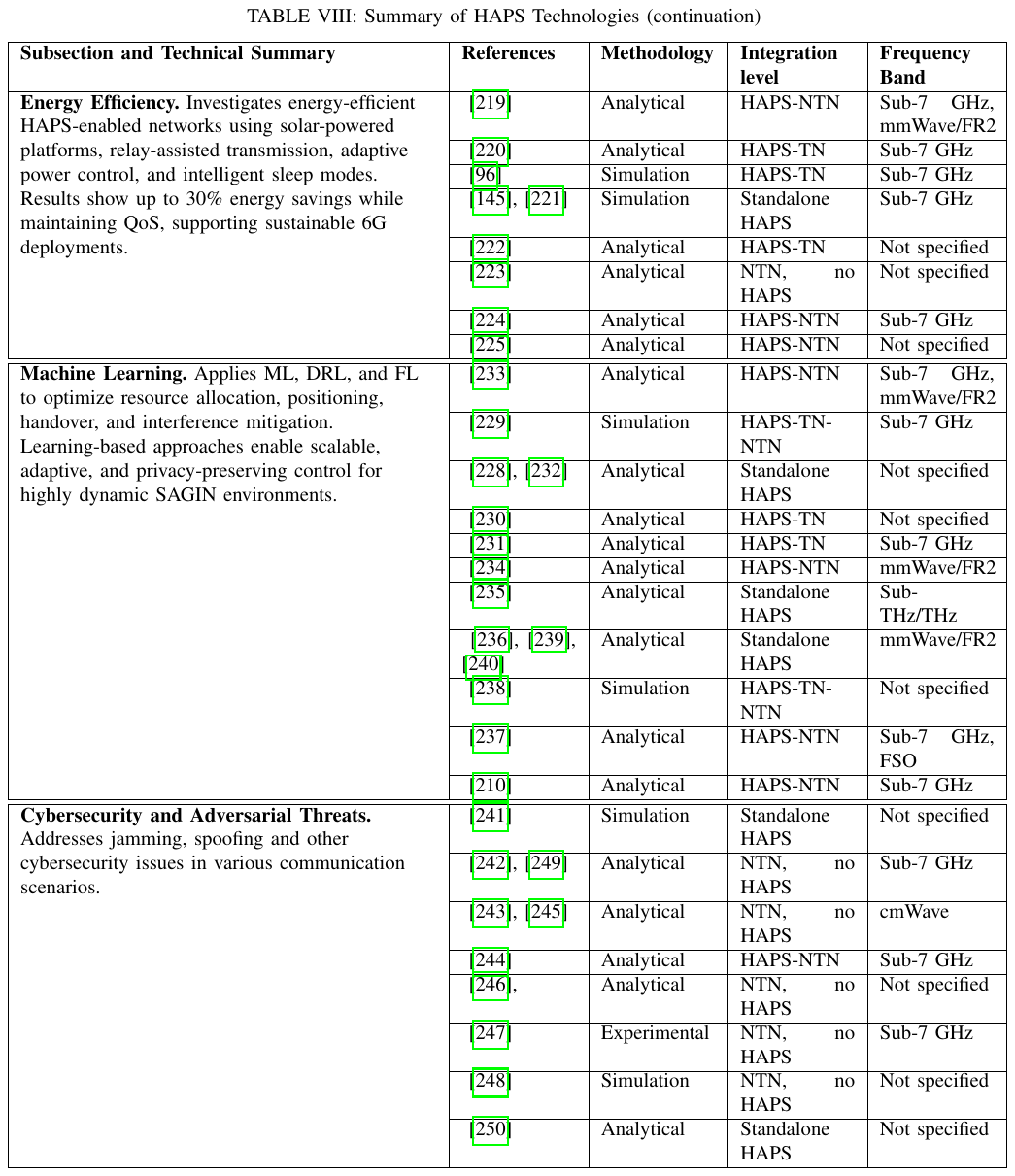}
    \label{fig:placeholder}
\end{figure*}
\end{document}